\documentclass[aps,prd,preprint,nofootinbib,amsmath,amssymb,showpacs,longbibliography,singlecolumn,superscriptaddress,10pt]{revtex4-1}

\usepackage{epsfig}
\usepackage{bm}
\usepackage{amssymb}
\usepackage{amsmath}
\usepackage{dsfont}
\usepackage{color}
\usepackage{subfigure}
\usepackage{dcolumn}
\usepackage{mathtools}%

\usepackage[utf8]{inputenc}

\usepackage{amsmath}
\usepackage{amssymb}
\usepackage{amsthm}
\usepackage{amscd, amsfonts, mathrsfs}
\usepackage{cases}
\usepackage{verbatim} 
\usepackage{epsf}
\usepackage[bookmarksnumbered,pdfpagelabels=true,plainpages=false,colorlinks=true,
            linkcolor=blue,citecolor=blue,urlcolor=blue]{hyperref}
\usepackage{xcolor}

\usepackage{amsmath}
\usepackage{graphicx}
\usepackage{dcolumn}
\usepackage{bm}

\colorlet{kigreen}{green!60!black}

\usepackage{dsfont}

\theoremstyle{plain}

\theoremstyle{definition}

\allowdisplaybreaks

\newcommand{\be}{\begin{equation}}
\newcommand{\ee}{\end{equation}}
\newcommand{\bea}{\begin{eqnarray}}
\newcommand{\eea}{\end{eqnarray}}

\newcommand{\bml}{\begin{subequations}}
\newcommand{\eml}{\end{subequations}}

\newcommand{\bbm}{\begin{bmatrix}}
\newcommand{\ebm}{\end{bmatrix}}
\newcommand{\bvm}{\begin{vmatrix}}
\newcommand{\evm}{\end{vmatrix}}

\begin{document}

\title{Modelling stochastic fluctuations in relativistic kinetic theory}

\author{Gabriel Soares Rocha}
\email{gabriel.soares.rocha@vanderbilt.edu}
\affiliation{Department of Physics and Astronomy, Vanderbilt University, 1221 Stevenson Center Lane,
Nashville, TN 37240, USA}

\author{Lorenzo Gavassino}
\email{lorenzo.gavassino@vanderbilt.edu}
\affiliation{Department of Mathematics, Vanderbilt University, Nashville, TN, USA}

\author{Nicki Mullins}
\email{nickim2@illinois.edu}
\affiliation{Illinois Center for Advanced Studies of the Universe Department of Physics, 
University of Illinois at Urbana-Champaign, Urbana, IL 61801, USA}

\begin{abstract}
Using the information current, we develop a Lorentz-covariant framework for modeling equilibrium fluctuations in relativistic kinetic theory in the grand-canonical ensemble. The resulting stochastic theory is proven to be causal and covariantly stable, and its predictions do not depend on the choice of spacetime foliation used to define the grand-canonical probabilities. As expected, in a box containing $N{>}5$ particles, Boltzmann's molecular chaos postulate is broken with (almost exact) probability $N^{-1/2}$, leading to a breakdown of the Boltzmann equation in small systems. We also verify that, in ultrarelativistic gases, transient hydrodynamics already accounts for at least 80\% of the equilibrium fluctuations of the stress-energy tensor at a given time. Finally, we compute the correlators at non-equal times for two selected collision kernels: That of a chemically active diluted solution, and that of ultrarelativistic scalar particles self-interacting via a quartic potential. For the former, we compute the density-density correlators analytically in real space, and dehydrodynamization of the stochastic theory is proven to occur whenever the mean free path diverges at high energy.
\end{abstract}

\maketitle

\tableofcontents

\section{Introduction}

The kinetic theory of gases is an insightful resource for all those who wish to model the statistical evolution of non-equilibrium systems. Although most real-world substances escape the ideal gas approximation (on which kinetic theory is based), several textbooks cite kinetic theory as the proof of principle upon which statistical mechanics and hydrodynamics are built \cite{huang_book,rezzolla_book,Hakim2011}. Indeed, together with the AdS/CFT correspondence \cite{FlorkowskiReview2018,Romatschke2010}, relativistic kinetic theory is currently our only means by which we can ``guess'' the analytical form of the dissipative terms in hydrodynamic models of the quark-gluon plasma \cite{Rocha:2023hts,Rocha:2023ilf}.

Perhaps the most famous feature of kinetic theory is its rigorous irreversibility. In fact, according to Boltzmann's celebrated H-theorem, we can define, within kinetic theory, an entropy functional $S$ that is non-decreasing in time \cite{DeGroot:1980dk,cercignani:90mathematical}. Historically, the H-theorem constituted the first derivation of the second law of thermodynamics from a microscopic (particle-based) model, and it set the foundations of modern statistical mechanics. At the time of Boltzmann, the H-theorem was not welcomed by mathematicians, who set out to prove the inconsistency of strict irreversibility with Hamiltonian mechanics \cite{huang_book}. The two most famous objections to the H-theorem, formulated at the time of Boltzmann, were Poincar\'{e}'s recurrence theorem, according to which any closed Hamiltonian system that starts out of equilibrium undergoes cycles in and out of equilibrium, and Loschmidt's ``reverse objection'', according to which, since the microscopic dynamics of a gas is time-reversible, a dynamical process is allowed if and only if its time-reversed is allowed, meaning that it should not be possible to argue irreversible dynamics from Hamiltonian mechanics alone.

Nowadays, we know that these objections are direct manifestations of the fluctuation-dissipation theorem, according to which every system that exhibits relaxation towards equilibrium must also undergo spontaneous fluctuations away from it. According to this understanding, the assumptions that led Boltzmann to prove the H-theorem are only an approximation, and the second law of thermodynamics is a property of the most probable macroscopic evolution, fixed some \textit{macroscopic} initial information \cite{Jaynes1965}. In fact, it is easy to show (see figure \ref{fig:Timereversal}) that the probability for a large system to transition from a macrostate $A$ to a macrostate $B$ is related to the probability of the time-reversed process (from $\Bar{B}$ to $\Bar{A}$) by the following identity \cite{Sevick2008}:
\begin{figure}
\begin{center}
\includegraphics[width=0.50\textwidth]{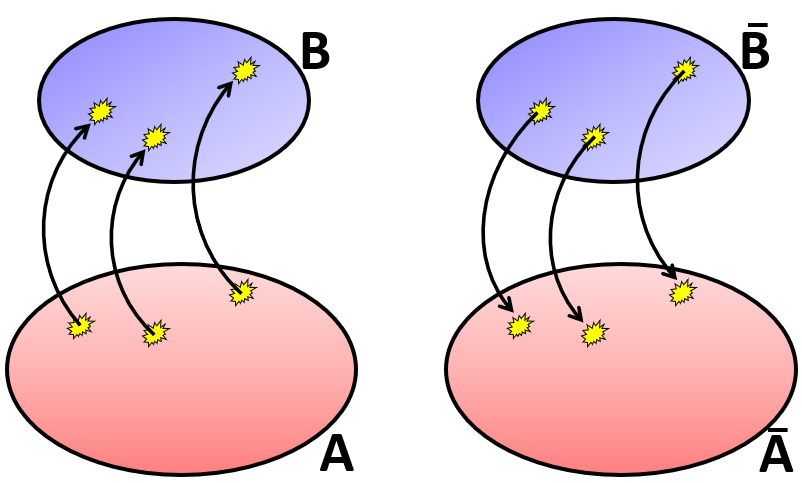}
\caption{Visual proof of equation \eqref{fluctuationssssssss}. A macroscopic state is the set of all microscopic states compatible with some large-scale measurement outcome. Let $N(A {\rightarrow} B)$ be the number of microscopic states in $A$ that evolve to $B$ in a time $\Delta t$. By time-reversal symmetry, this equals the number $N(\Bar{B}{\rightarrow} \Bar{A})$ of microscopic states in $\Bar{B}$ that evolve to $\Bar{A}$ during the same amount of time $\Delta t$. Assuming equal probability a priori of the microstates for a given macrostate, the probability of evolving to $B$ starting from $A$ is $\mathcal{P}(A{\rightarrow}B)=N(A{\rightarrow}B)/N(A)$, where $N(A)=e^{S_A}$ is the number of microscopic states in $A$. Analogously, the probability of evolving to $\Bar{A}$ starting from $\Bar{B}$ is $\mathcal{P}(\Bar{B}{\rightarrow}\Bar{A})=N(\Bar{B}{\rightarrow}\Bar{A})/N(\Bar{B})$, with $N(\Bar{B})=N(B)=e^{S_B}$. Equation \eqref{fluctuationssssssss} immediately follows.}
	\label{fig:Timereversal}
	\end{center}
\end{figure}
\begin{equation}\label{fluctuationssssssss}
    \dfrac{\mathcal{P}(\Bar{B}\rightarrow \Bar{A})}{\mathcal{P}(A \rightarrow B)}= e^{S_A-S_B} \, ,
\end{equation}
where $S_A$ and $S_B$ are the entropies of the respective states. If we assume that $S_B>S_A$, then, since the entropy is proportional to the particle number $N$, the right-hand side of \eqref{fluctuationssssssss} decays like $\sim e^{-N}$. Therefore, in the thermodynamic limit, the transition $\Bar{B}\rightarrow \Bar{A}$ is effectively forbidden, and the entropy cannot decrease. However, for finite values of $N$, both processes $\Bar{B}\rightarrow \Bar{A}$ and $A \rightarrow B$ are allowed\footnote{Indeed, Poincar\'{e}'s recurrence theorem is valid only for finite systems. In fact, the proof relies on the assumption that the phase space shell $E\leq \text{``Hamiltonian''} \leq E+\Delta E$ has finite volume \cite{huang_book}.}, and small violations of the second law of thermodynamics may occur. Such violations appear, to an observer who has access only to macroscopic data, as random fluctuations.

The mathematical framework that implements (seemingly) stochastic fluctuations into Boltzmann's transport equation is known as \textit{fluctuating kinetic theory}. In a non-relativistic setting, the theoretical foundations of such framework are well-understood, see e.g. \cite{landau10, fox1970contributions, Bixon:1969zz}. In relativity, the matter is less established, although some initial works exist \cite{Gavin:2016nir, Miron-Granese:2020mbf,Torrieri:2023thk}. Here, we provide, for the first time, a first-principle construction of relativistic fluctuating kinetic theory, within the grand-canonical ensemble. The key tool that will allow us to carry out a rigorous construction is the information current \cite{Gavassino:2021kjm}, whose existence automatically guarantees the consistency of the resulting theory with relativistic thermodynamics \cite{vanKampen1968,Noto_full,Israel_1981_review},  Lorentz-covariance, causality, and stability. Using the procedure developed in \cite{Mullins:2023tjg, Mullins:2023ott}, symmetrized correlation functions can be obtained using only the information current and the entropy production rate. 

Earlier studies of the correlation functions of relativistic kinetic theory, both symmetrized \cite{Gavin:2016nir, Miron-Granese:2020mbf} and retarded \cite{Romatschke:2015gic, Kurkela:2017xis, Bajec:2024jez}, were restricted to the relaxation time approximation. Furthermore, the main focus has always been on the correlation functions of conserved quantities (sometimes in the hydrodynamic regime), but there is no study of the structure of correlation functions of the kinetic distribution function itself. By using the information current, we are able to evaluate the equal-time correlation functions of arbitrary observables in any reference frame, without making any assumptions about the interactions. For dilute solutions in the relaxation time approximation, non-equal-time correlation functions of both the distribution function and the conserved currents are obtained \emph{analytically} in real space. This can be done with both a constant relaxation time, and a momentum-dependent relaxation time. We then show that correlation functions in the full linearized Boltzmann equation can be obtained if the spectrum of the linearized collision operator is known. Using the known spectrum for weakly interacting ultrarelativistic scalar particles \cite{Denicol:2022bsq}, we are able to obtain correlation functions in both the distribution function and conserved currents. This is the first time such correlation functions have been obtained in the relativistic regime.

Throughout the article, we adopt the metric signature $(+,-,-,-)$ and work in natural units $c=\hbar=k_B=1$. For an observable $f(x,p)$ that depends on both positions $x$ and momenta $p$ of the particles, we adopt the following Fourier convention:
\begin{equation}
\begin{aligned}
 &
 f(x,p) = \int \frac{d^{4}q}{(2 \pi)^{4}} e^{i q x} \widetilde{f}(q,p) \, .
\end{aligned}    
\end{equation}

\section{Constant-time theory}

In this section, we derive the probability distribution for fluctuations of an ideal relativistic gas at a given time (or across a generic spacelike surface). Then, we use it to compute equal-time correlators of physical observables.

\subsection{The information current}
\label{sec:info-current}

We consider an ideal gas of conserved particles in weak contact with a heat and particle bath. The (constant) intensive parameters that characterize the bath are its fugacity $\alpha^\star$ and its inverse temperature four-vector $\beta^\star_\mu$ \cite{Israel_1981_review,GavassinoTermometri}. In thermodynamic equilibrium, the gas is in the relativistic grand-canonical ensemble \cite{HakimBook}. Thus, the probability for a certain macroscopic state $\Psi$ to occur is
\begin{equation}\label{probabilitytuzzuz}
    \mathcal{P}(\Psi) = \dfrac{e^{S(\Psi)+\alpha^\star N(\Psi)-\beta_\nu^\star P^\nu(\Psi)}}{Z} \, ,
\end{equation}
where $S(\Psi)$, $N(\Psi)$, and $P^\mu(\Psi)$ are the entropy, particle number, and four-momentum of the macrostate $\Psi$, and $Z$ is the grand-canonical partition function.  In kinetic theory, the macroscopic state $\Psi$ is the profile of the one-particle distribution function $f_p(x^\mu)$ across a fixed Cauchy surface $\Sigma$. Thus, equation \eqref{probabilitytuzzuz} reads
\begin{equation}\label{PofF}
    \mathcal{P}[f_p] = Z^{-1} \exp \int_{\Sigma} \big(s^\mu+\alpha^\star J^\mu-\beta_\nu^\star T^{\mu \nu} \big)d\Sigma_\mu \, ,
\end{equation}
where $s^\mu$, $J^\mu$, and $T^{\mu \nu}$ are respectively the entropy current, the conserved particle current, and the (symmetric) stress-energy tensor, given by
\begin{equation}\label{Thedefinitions}
\begin{split}
    s^{\mu} ={}& - \int \frac{d^3p}{(2\pi)^3 p^0} \, p^{\mu} \left( f_p \ln f_p - f_p \right) , \\
    J^{\mu} ={}& \int \frac{d^3p}{(2\pi)^3 p^0} \, p^{\mu} f_p ,\\
    T^{\mu\nu} ={}& \int \frac{d^3p}{(2\pi)^3 p^0} \, p^{\mu} p^{\nu} f_p .\\
    \end{split}
\end{equation}
Here, we are assuming Maxwell-Boltzmann statistics, and we are neglecting possible spin degeneracies. The global equilibrium macrostate $f_{\text{eq},p}$ is, by definition, the most probable state. Any other state $f_p$ can be conveniently decomposed as $f_p=f_{\text{eq},p}(1+\phi_p)$, where $\phi_p$ is dimensionless, and it can be interpreted as the fractional displacement from equilibrium. Then, we have that
\begin{equation}\label{gringone}
\begin{split}
& \dfrac{\mathcal{P}[\phi_p]}{\mathcal{P}[\phi_p{=}0]} = e^{-\int_\Sigma E^\mu d\Sigma_\mu} \leq 1 \, ,\\
   & \text{with } \quad E^\mu = \int \dfrac{d^3 p}{(2\pi)^3 p^0} p^\mu f_{\text{eq},p} \bigg[ (\ln f_{\text{eq},p}{-}1{-}\alpha^\star{+}\beta^\star_\nu p^\nu)\phi_p {+}(1{+}\phi_p)\ln(1{+}\phi_p)\bigg] \, .
    \end{split}
\end{equation}
The vector field $E^\mu$ is the so-called ``information current''. The inequality in the first line (imposed on arbitrary $\Sigma$) implies that $\delta E^\mu/\delta \phi_p$ must vanish at $\phi_p=0$. This leads us to the expected formula for the equilibrium distribution function: $f_{\text{eq,p}}=e^{\alpha^\star{-}\beta^\star_\nu p^\nu}$. Therefore, the information current simplifies to
\begin{equation}\label{phuiummum}
    E^\mu = \int \dfrac{d^3 p}{(2\pi)^3 p^0} p^\mu f_{\text{eq},p}\bigg[ (1{+}\phi_p)\ln(1{+}\phi_p)-\phi_p\bigg] \, .
\end{equation}
This expression is valid for arbitrarily large fluctuations. Note that, while $\phi_p$ is allowed to become arbitrarily large in the positive direction, it cannot go below $-1$, because the distribution function $f_p$ is non-negative.

\subsection{Breakdown of the Boltzmann equation for small systems}

The Boltzmann equation is grounded on a delicate assumption, known as ``molecular chaos''. In a nutshell, molecular chaos assumes that the velocities of colliding particles are uncorrelated \cite{huang_book}. This allowed Boltzmann to derive an entirely deterministic equation of motion for $f_p$. Such equation predicts that, if a gas is initially in the equilibrium macrostate $f_{\text{eq},p}$, it cannot ``jump'' away from it spontaneously. On the other hand, equation \eqref{gringone} tells that $f_{\text{eq},p}$ is only the most probable state, and the system explores all macrostates with non-vanishing probability. It follows that the Boltzmann equation is only an approximation, and deviations from molecular chaos must occur, as predicted by \eqref{fluctuationssssssss}. Let us estimate the importance of these effects.

We set $\Sigma$ to be a constant-time Cauchy surface. Then, equation \eqref{gringone} reduces to
\begin{equation}\label{gavagringo}
  \dfrac{\mathcal{P}[\phi_p]}{\mathcal{P}[\phi_p{=}0]} = \exp \int_{\mathbb{R}^6} -\dfrac{d^3 x \, d^3p}{(2\pi)^3} f_{\text{eq},p}  \bigg[ (1{+}\phi_p)\ln(1{+}\phi_p)-\phi_p\bigg] \, .
\end{equation}
To understand the physical content of this distribution, let us restrict our attention to a specific type of fluctuation. Fixing an arbitrary phase-space region $\Gamma \subseteq \mathbb{R}^6$, we consider a perturbation of the form
\begin{equation}
\label{eq:box-pert}
    \phi_p(\textbf{x})= 
\begin{cases}
 \phi & \text{if } (\textbf{x},\textbf{p})\in \Gamma   \, ,\\
0 & \text{otherwise} \, . \\
\end{cases}
\end{equation}
Then, in equation \eqref{gavagringo}, the integral can be restricted to the region $\Gamma$, and the square bracket can be taken out from the integral. Introducing the dimensionless quantity
\begin{equation}
    N_{\text{eq}}:= \int_\Gamma \dfrac{d^3 x \, d^3p}{(2\pi)^3} f_{\text{eq},p} \, ,
\end{equation}
which is just the equilibrium number of particles in the region $\Gamma$, we finally obtain
\begin{figure}
\begin{center}
\includegraphics[width=0.60\textwidth]{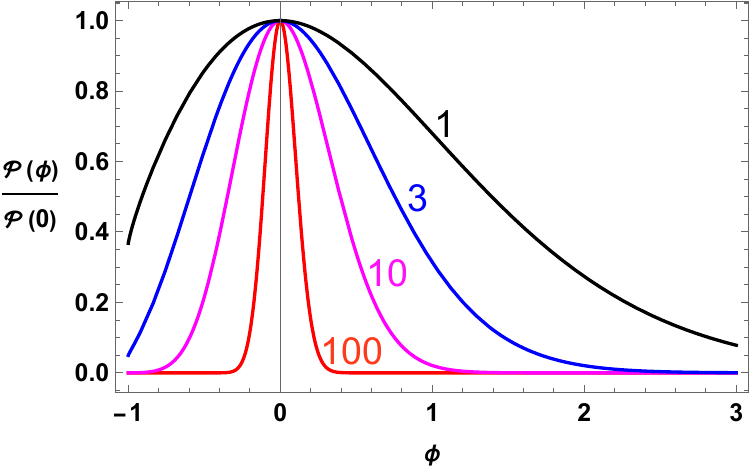}
	\caption{Grand-canonical probability \eqref{sao} that a fluctuation of relative size $\phi$ can occur in a region of phase space containing respectively $1$ (black), $3$ (blue), $10$ (magenta), and $100$ (red) particles at equilibrium. The standard deviation $\sqrt{\langle \phi^2\rangle }$ can be interpreted as the relative error that we commit if we assume that a volume element of gas containing (on average) the given number of particles obeys the molecular chaos assumption. Such error is 114\% for 1 particle, 60\% for 3 particles, 32\% for 10 particles, and 10\% for 100 particles. For one mole, it is $\sim 10^{-10}\,$\%.}
	\label{fig:Probabuz}
	\end{center}
\end{figure}
\begin{equation}\label{sao}
    \dfrac{\mathcal{P}(\phi)}{\mathcal{P}(\phi{=}0)}= e^{-N_{\text{eq}}\big[ (1{+}\phi)\ln(1{+}\phi)-\phi\big]}
\end{equation}
This can be interpreted as the probability for a fluctuation of size $\phi$ to occur in a portion of gas involving $N_{\text{eq}}$ particles. In figure \ref{fig:Probabuz}, we graph $\mathcal{P}(\phi)$ for some selected values of $N_{\text{eq}}$. The Gaussian approximation is already quite good for 5 particles, so that we can write
\begin{equation}
\label{eq:prob-dist-gauss-box}
    \mathcal{P}(\phi)\approx \sqrt{\dfrac{N_{\text{eq}}}{2\pi}} \, e^{-N_{\text{eq}}\phi^2/2} \, .
\end{equation}
The standard deviation of $\phi$ is, therefore, $N_{\text{eq}}^{-1/2}$, in agreement with section \S 113 of \cite{landau_statistical_1980}. Since molecular chaos entails that $\phi=0$ at all times in equilibrium, the standard deviation of $\phi$ can be interpreted as the relative error that we make if we assume that a fluid cell containing $N_{\text{eq}}$ particles fulfills molecular chaos exactly \cite{huang_book}. Taking 20\% as our maximum error bar for an acceptable description, we find that molecular chaos (and therefore the Boltzmann equation) necessarily breaks down when the fluid cell contains less than $\sim 25$ particles.

\subsection{Gaussian fluctuations}\label{TheGaussone}

As we saw in the previous section, if the system is not too small, the probability distribution of fluctuations can be approximated as Gaussian. This corresponds to truncating the information current \eqref{phuiummum} to second order in $\phi_p$. The result is
\begin{equation}\label{Eq:Kinetic_theory_E}
   E^\mu \approx \dfrac{1}{2} \int \dfrac{d^3 p}{(2\pi)^3 p^0} p^\mu f_{\text{eq},p} \phi^2_p\, .
\end{equation}
This vector field fulfills three properties. First, it is always timelike future-directed. This follows immediately from the fact that the difference $E^0-E^1$ is the integral of a non-negative definite quantity:
\begin{equation}
    E^0-E^1 = \dfrac{1}{2} \int \dfrac{d^3 p}{(2\pi)^3 p^0} (p^0{-}p^1) f_{\text{eq},p} \phi^2_p \geq 0 \, .
\end{equation}
Second, it is evident that $E^\mu=0$ if and only if $\phi=0$. Finally, due to the H-theorem, we have that $\partial_\mu E^\mu \leq 0$ along solutions of the linearised Boltzmann equation. It follows that the linearised Boltzmann equation is causal \cite{Gavassino:2021kjm}, and the equilibrium state is covariantly stable against fluctuations \cite{Gavassino:2021cli}, both deterministic and stochastic \cite{Gavassino:2024vyu}.

Plugging \eqref{Eq:Kinetic_theory_E} into the first line of \eqref{gringone}, and taking $\Sigma$ to be a constant-time hypersurface, we obtain the Gaussian approximation of \eqref{gavagringo}, namely
\begin{equation}\label{gavagringoGauss}
  \dfrac{\mathcal{P}[\phi_p]}{\mathcal{P}[\phi_p{=}0]} = \exp \int_{\mathbb{R}^6} - \dfrac{1}{2}\dfrac{d^3 x \, d^3p}{(2\pi)^3} f_{\text{eq},p}  \phi^2_p \, .
\end{equation}
The two-point correlator associated with this probability distribution can be evaluated using basic techniques for functional integration, giving the following:
\begin{equation}\label{equaltimecorreltor!}
    \langle \phi_p(\textbf{x})\phi_{p'}(\textbf{x}')\rangle = \dfrac{(2\pi)^3}{f_{\text{eq},p}} \delta^3(\textbf{x}-\textbf{x}')\delta^3(\textbf{p}-\textbf{p}') \, .
\end{equation}

\subsection{Equal-time correlators of conserved currents}

The equal-time correlator \eqref{equaltimecorreltor!} can be used to compute the size of fluctuations of the conserved particle current $J^\mu$ and of the stress-energy tensor $T^{\mu \nu}$. From the definitions \eqref{Thedefinitions}, we find that the deviation of the conserved currents from equilibrium is
\begin{equation}
\begin{split}\label{maladefinizioneequesta}
    \delta J^{\mu} ={}& \int \frac{d^3p}{(2\pi)^3 p^0} \, p^{\mu} f_{\text{eq},p}\phi_p ,\\
    \delta T^{\mu\nu} ={}& \int \frac{d^3p}{(2\pi)^3 p^0} \, p^{\mu} p^{\nu} f_{\text{eq},p}\phi_p .\\
    \end{split}
\end{equation}
Straightforward integration of \eqref{equaltimecorreltor!} then gives
\begin{equation}
\begin{split}
\langle \delta J^\mu(\textbf{x})\delta J^\nu(\textbf{x}')\rangle ={}& \delta^3(\textbf{x}{-}\textbf{x}') \int \dfrac{d^3p}{(2\pi)^3(p^0)^2} p^\mu p^\nu f_{\text{eq},p} \, , \\
\langle \delta J^\mu(\textbf{x})\delta T^{\nu\rho}(\textbf{x}')\rangle ={}& \delta^3(\textbf{x}{-}\textbf{x}') \int \dfrac{d^3p}{(2\pi)^3(p^0)^2} p^\mu p^\nu p^\rho f_{\text{eq},p} \, , \\
\langle \delta T^{\mu\nu}(\textbf{x})\delta T^{\rho \lambda}(\textbf{x}')\rangle ={}& \delta^3(\textbf{x}{-}\textbf{x}') \int \dfrac{d^3p}{(2\pi)^3(p^0)^2} p^\mu p^\nu p^\rho p^\lambda f_{\text{eq},p} \, . \\
\end{split}
\end{equation}
Note that, in the denominator of the momentum-volume form, the energy $p^0$ appears to the \textit{second} power, and not to the first as usual. This implies that, if at least one of the indices in the correlators is time, we recover the usual moments of the equilibrium distribution function, namely:
\begin{equation}
\begin{split}
&\langle \delta J^0(\textbf{x})\delta J^\nu(\textbf{x}')\rangle = \delta^3(\textbf{x}{-}\textbf{x}') J_{\text{eq}}^\nu \, , \\
& \langle \delta J^0(\textbf{x})\delta T^{\nu\rho}(\textbf{x}')\rangle = \langle \delta J^\nu(\textbf{x})\delta T^{0\rho}(\textbf{x}')\rangle = \delta^3(\textbf{x}{-}\textbf{x}') T_{\text{eq}}^{\nu \rho} \, , \\
&\langle \delta T^{0\nu}(\textbf{x})\delta T^{\rho \lambda}(\textbf{x}')\rangle =\delta^3(\textbf{x}{-}\textbf{x}') A_{\text{eq}}^{\nu \rho \lambda} \, , \\
\end{split}
\end{equation}
where $A^{\nu \rho \lambda}$ is the third moment of the distribution function. The other components of the correlators do not have such a simple interpretation, and one needs to evaluate all the integrals explicitly.

\subsection{Evaluation of the equal-time correlators in the ultrarelativistic limit}\label{UUUultra}

Let us compute all the components of the equal-time correlators in the ultrarelativistic limit, where all the integrals admit an analytic solution. We will work in the equilibrium rest frame, so that $f_{\text{eq},p}=e^{\frac{\mu-p}{T}}$, where $\mu$ is the chemical potential, $T$ is the temperature, and $p=p^0$ is the magnitude of the three-momentum. In this limit, the equilibrium particle density $n$, pressure $P$, and energy density $\varepsilon$ are, respectively,
\begin{equation}\label{edaicisono}
    n= \dfrac{T^3}{\pi^2} \, e^{\mu/T}\, , \quad \quad  \quad P=nT \, , \quad \quad \quad \varepsilon=3P \, .
\end{equation}
To lighten the notation, we express all correlators in the form $\langle \delta A(\textbf{x})\delta B(\textbf{x}')\rangle = \delta^3(\textbf{x}-\textbf{x}')\overline{AB}$, and we just provide the formula for the (constant) correlation amplitude $\overline{AB}$. For the current-current correlator, we have
\begin{equation}
    \overline{J^\mu J^\nu}=
    \begin{bmatrix}
n & 0 & 0 & 0 \\
0 & n/3 & 0 & 0 \\
0 & 0 & n/3 & 0 \\
0 & 0 & 0 & n/3 \\
    \end{bmatrix} \, .
\end{equation}
For the mixed current-energy-momentum correlator, we find
\begin{equation}
    \overline{J^0 T^{\nu \rho}}=
    \begin{bmatrix}
\varepsilon & 0 & 0 & 0 \\
0 & P & 0 & 0 \\
0 & 0 & P & 0 \\
0 & 0 & 0 & P \\
    \end{bmatrix} \, , \quad \quad \quad 
    \overline{J^1 T^{\nu \rho}}=
    \begin{bmatrix}
0 & P & 0 & 0 \\
P & 0 & 0 & 0 \\
0 & 0 & 0 & 0 \\
0 & 0 & 0 & 0 \\
    \end{bmatrix} \, .  
\end{equation}
The correlators involving $J^2$ and $J^3$ are analogous to that involving $J^1$, by isotropy. For the energy-momentum correlators, we find
\begin{equation}
\begin{split}
    \overline{T^{00} T^{\nu \rho}}=4T
    \begin{bmatrix}
\varepsilon & 0 & 0 & 0 \\
0 & P & 0 & 0 \\
0 & 0 & P & 0 \\
0 & 0 & 0 & P \\
    \end{bmatrix} \, , & \quad \quad \quad 
    \overline{T^{01} T^{\nu \rho}}=4T
    \begin{bmatrix}
0 & P & 0 & 0 \\
P & 0 & 0 & 0 \\
0 & 0 & 0 & 0 \\
0 & 0 & 0 & 0 \\
    \end{bmatrix} \, , \\ 
\overline{T^{11} T^{\nu \rho}}=4T
    \begin{bmatrix}
P & 0 & 0 & 0 \\
0 & 3P/5 & 0 & 0 \\
0 & 0 & P/5 & 0 \\
0 & 0 & 0 & P/5 \\
    \end{bmatrix} \, , & \quad \quad \quad 
    \overline{T^{12} T^{\nu \rho}}=4T
    \begin{bmatrix}
0 & 0 & 0 & 0 \\
0 & 0 & P/5 & 0 \\
0 & P/5 & 0 & 0 \\
0 & 0 & 0 & 0 \\
    \end{bmatrix} \, . \\ 
    \end{split}
\end{equation}
All the others can be recovered from the above ones using isotropy. For example, if we apply a $90^\text{o}$ rotation around the $x^2$ axis, we find that $\overline{T^{23} T^{23}}=\overline{T^{12} T^{12}}=4TP/5$, and $\overline{T^{33} T^{33}}=\overline{T^{11} T^{11}}=4T\varepsilon/5$. The correlators in general flow configurations and general statistics are discussed in Appendix \ref{apn:information-current-FD-BE}.

\subsection{Comparison with the correlators of the Israel-Stewart theory}

Since we have exact formulas for the equal-time correlators in kinetic theory, we can compare them with the hydrodynamic correlators of relativistic transient hydrodynamics. The most widespread hydrodynamic theory of this kind is the Israel-Stewart theory, whose information current is well-known \cite{OLSON199018,GavassinoSymmetricQuasi2022roi}. If we work in the equilibrium rest frame, and consider an ultrarelativistic gas as above, the information density (in the Landau frame) reads
\begin{equation}\label{infoofIS}
    E^0 = \dfrac{1}{2} \bigg[ \dfrac{(\delta n)^2}{n} + \dfrac{3n}{T^2} (\delta T)^2 -4n \,\delta u^j \delta u_j -b_1 \delta \nu^j \delta \nu_j+b_2 \delta \pi^{jk}\delta \pi_{jk} \bigg] \, ,
\end{equation}
where $\delta u^j$, $\delta \nu^j$, and $\delta \pi^{jk}$ are the fluctuations to the (Landau-frame) flow velocity, to the charge-diffusion flux, and to the shear stress tensor, respectively. There is no bulk viscous stress because ultrarelativistic gases are conformal. The background constants $b_1$ and $b_2$ are two second-order transport coefficients, whose exact value depends on the procedure that one uses to derive hydrodynamics from kinetic theory [e.g., Denicol-Niemi-Molnar-Rischke (DNMR) \cite{Denicol:2012cn}, or inverse Reynolds dominance (IReD) \cite{Wagner:2022ayd}]. In fluid dynamics, the macroscopic state is the profile of $\Psi=\{\delta n,\delta T, \delta u^j,\delta \nu^j,\delta \pi^{jk} \}$ across a constant-time hypersurface, and the grand-canonical probability distribution is $\mathcal{P}(\Psi)=e^{-\int E^0 d^3x}$ \cite{Gavassino:2021kjm}. All the equal-time correlators can be evaluated by functional integration, and we find
\begin{equation}\label{fluxium}
    \overline{nn}=n \, , \quad \quad \overline{TT}=\dfrac{T^2}{3n} \, , \quad \quad \overline{u^j u^k}=\dfrac{\delta^{jk}}{4n} \, , \quad \quad \overline{\nu^j \nu^k}=\dfrac{\delta^{jk}}{b_1} \, .
\end{equation}
For the shear stress tensor, one needs to have extra care in performing the functional integration, due to the constraints of symmetry and tracelessness, $\delta \pi^{[jk]}=\delta \pi^j_{\, \, j}=0$. Applying the same technique as in Appendix B of \cite{Gavassino:2022roi}, we find
\begin{equation}\label{fluxium2}
    \overline{\pi^{jk}\pi^{lm}}= \dfrac{1}{2b_2} \bigg[\delta^{jl}\delta^{km}+\delta^{jm}\delta^{kl}-\dfrac{2}{3} \delta^{jk}\delta^{lm} \bigg] \, . 
\end{equation}
Since in the information density \eqref{infoofIS} all the perturbation fields are decoupled (i.e. statistically independent), all the mixed correlators, e.g. $\overline{nT}$ and $\overline{u^j\nu^k}$, vanish identically. Note that the equal-time correlators of the ordinary fluid fields $\delta n$, $\delta T$, and $\delta u^j$ agree with the predictions of the standard theory of hydrodynamic fluctuations \cite{landau_statistical_part_II}.

We can use the correlators \eqref{fluxium} and \eqref{fluxium2} to evaluate the correlators of $\delta J^\mu$ and $\delta T^{\mu \nu}$, which are expressed in terms of the fluid variables $\Psi$ as follows:
\begin{equation}
    \delta J^0=\delta n \, , \quad \quad \delta J^j=n\delta u^j +\delta \nu^j \, , \quad \quad \delta T^{00}=3\delta P \, , \quad \quad \delta T^{0j}=4P\delta u^j \, , \quad \quad \delta T^{jk}=\delta P\, \delta^{jk}+\delta \pi^{jk} \, ,
\end{equation}
where the pressure fluctation is $\delta P=T\delta n+n\delta T$, and its correlator amplitude is $\overline{PP}=4TP/3$. The amplitude of the current correlator turns out to be
\begin{equation}
    \overline{J^\mu J^\nu}=
    \begin{bmatrix}
n & 0 & 0 & 0 \\
0 & n/4{+}b_1^{-1} & 0 & 0 \\
0 & 0 & n/4{+}b_1^{-1} & 0 \\
0 & 0 & 0 & n/4{+}b_1^{-1} \\
    \end{bmatrix} \, .
\end{equation}
The amplitude of the mixed correlator is
\begin{equation}
    \overline{J^0 T^{\nu \rho}}=
    \begin{bmatrix}
\varepsilon & 0 & 0 & 0 \\
0 & P & 0 & 0 \\
0 & 0 & P & 0 \\
0 & 0 & 0 & P \\
    \end{bmatrix} \, , \quad \quad \quad 
    \overline{J^1 T^{\nu \rho}}=
    \begin{bmatrix}
0 & P & 0 & 0 \\
P & 0 & 0 & 0 \\
0 & 0 & 0 & 0 \\
0 & 0 & 0 & 0 \\
    \end{bmatrix} \, .  
\end{equation}
Finally, the amplitude of the energy-momentum correlator is
\begin{equation}
\begin{split}
    \overline{T^{00} T^{\nu \rho}}=4T
    \begin{bmatrix}
\varepsilon & 0 & 0 & 0 \\
0 & P & 0 & 0 \\
0 & 0 & P & 0 \\
0 & 0 & 0 & P \\
    \end{bmatrix} \, , & \quad \quad \quad 
    \overline{T^{01} T^{\nu \rho}}=4T
    \begin{bmatrix}
0 & P & 0 & 0 \\
P & 0 & 0 & 0 \\
0 & 0 & 0 & 0 \\
0 & 0 & 0 & 0 \\
    \end{bmatrix} \, , \\ 
\overline{T^{11} T^{\nu \rho}}=\dfrac{1}{3}
    \begin{bmatrix}
12TP & 0 & 0 & 0 \\
0 & 4TP{+}2b_2^{-1} & 0 & 0 \\
0 & 0 & 4TP{-}b_2^{-1} & 0 \\
0 & 0 & 0 & 4TP{-}b_2^{-1} \\
    \end{bmatrix} \, , & \quad \quad \quad 
    \overline{T^{12} T^{\nu \rho}}=\dfrac{1}{2b_2}
    \begin{bmatrix}
0 & 0 & 0 & 0 \\
0 & 0 & 1 & 0 \\
0 & 1 & 0 & 0 \\
0 & 0 & 0 & 0 \\
    \end{bmatrix} \, . \\ 
    \end{split}
\end{equation}
Comparing the kinetic correlation amplitudes with the hydrodynamic correlation amplitudes we find that the two coincide if and only if
\begin{equation}\label{contrucztion}
    b_1=\dfrac{12}{n} \, , \quad \quad \quad b_2 = \dfrac{5}{8TP} \, .
\end{equation}
On the other hand, in the Israel-Stewart theory, $b_1$ and $b_2$ are related to the dynamical transport coefficients. In particular, one has that $b_1=\tau_n/\kappa$, where $\kappa$ is the charge diffusivity, and $\tau_n$ is the diffusion relaxation time, while $b_2=\tau_\pi/(2T\eta)$, where $\eta$ is the shear viscosity, $\tau_\pi$ is the shear relaxation time. Thus, equation \eqref{contrucztion}, if taken at face value, translates into a constraint relating the values of the relaxation times to the first-order transport coefficients:
\begin{equation}\label{tritoncino}
    \tau_n = \dfrac{12 \kappa}{n} \, , \quad \quad \quad \tau_\pi = \dfrac{5\eta}{4P} \, .
\end{equation}
If the transport coefficients are computed from the Boltzmann equation in the Anderson-Witting Relaxation Time Approximation (RTA), the above relations are recovered \textit{exactly}, see equations (168) and (169) of \cite{Ambrus:2022vif}. This is not surprising, since the Israel-Stewart theory and the RTA describe the same off-equilibrium dynamics. If, instead, we compute the transport coefficients using the full Boltzmann collision kernel, the relations in \eqref{tritoncino} can still be recovered exactly, both within the DNMR \cite{Denicol:2012cn} and the IReD \cite{Wagner:2022ayd} framework (for a constant cross-section gas), but only if we limit ourselves to the 14-moment approximation. The reason is that the Israel-Stewart theory itself can be viewed as a sort of ``kinetic theory'' with only 14 moments.

\subsection{Fluctuations away from transient hydrodynamics}
\label{subsec:fluct-trans-hydro}

In general, one should not expect the equal-time correlators of a hydrodynamic theory to coincide with those of a more microscopic description. In fact, hydrodynamics is only supposed to reproduce the low-frequency behavior of the system, while the equal-time correlators contain all frequencies, since
\begin{equation}
    \langle \delta A(t) \, \delta B(t')\rangle =\int_{\mathbb{R}} \dfrac{d\omega}{2\pi} e^{-i\omega (t-t')}  (\delta A \, \delta B)_\omega \, .
\end{equation}
However, the Israel-Stewart theory is not a conventional infrared theory. When interpreted as a transient hydrodynamic theory, Israel-Stewart is expected to reproduce also a portion of the high-frequency spectrum through the relaxation times, see \cite{Wagner:2023jgq}, interpretation (iii). Clearly, a fluctuating theory with 14 degrees of freedom (the Israel-Stewart theory) cannot reproduce all the complexity of a fluctuating theory with infinite degrees of freedom (full kinetic theory). Hence, in the high-frequency regime, kinetic theory can ``fluctuate away'' from the Israel-Stewart states, leading to a violation of equation \eqref{tritoncino}. It is, therefore, instructive to consider the relative errors (subscripts: kt=``kinetic theory'', IS=``Israel-Stewart''),
\begin{equation}
\begin{split}
    z_n={}& \dfrac{(\overline{J^1J^1})_{\text{kt}}-(\overline{J^1J^1})_{\text{IS}}}{(\overline{J^1J^1})_{\text{kt}}} = \dfrac{1}{4}-\dfrac{3\kappa}{n\tau_n} \, , \\
    z_\pi ={}& \dfrac{(\overline{T^{12}T^{12}})_{\text{kt}}-(\overline{T^{12}T^{12}})_{\text{IS}}}{(\overline{T^{12}T^{12}})_{\text{kt}}} =1-\dfrac{5\eta}{4P\tau_\pi} \, ,\\
\end{split}
\end{equation}
which measure the discrepancy between Israel-Stewart and kinetic theory fluctuations.
In appendix \ref{thezetaarecosi}, we prove that, if the Israel-Stewart theory is consistent with kinetic theory, then both $z_n$ and $z_\pi$ fall in the interval $[0,1]$ (i.e., kinetic theory fluctuates more than hydrodynamics). For this reason, we can interpret $z_n$ and $z_\pi$ as the percentage contributions to the fluctuations of the corresponding fluxes ($\delta \nu^j$ for $z_n$, and $\delta \pi^{jk}$ for $z_\pi$) coming from those degrees of freedom that are neglected by transient hydrodynamics. Different derivations of Israel-Stewart dynamics from kinetic theory adopt different interpretations for which fluctuations are ``hydrodynamic'' and which are ``not hydrodynamic''\footnote{The state-space of linearised kinetic theory is the Hilbert space $L^2(\mathbb{R}^6)$, whose elements are the functions $\phi_p(\textbf{x})$ at a given time. Each derivation of Israel-Stewart (e.g. DNMR) assumes a specific regime of applicability within kinetic theory. The regime of applicability is a projector $\mathbb{P}$ on $L^2(\mathbb{R}^6)$, which returns $1$ if the dynamics of the (conserved currents of the) state is captured correctly by Israel-Stewart hydrodynamics, and $0$ otherwise. If two theories $a$ and $b$ have different regimes of applicability, they have different projectors, $\mathbb{P}_a$ and $\mathbb{P}_b$. The behaviour of $E^0$ across the image of $P_a$ differs from that across the image of $\mathbb{P}_b$. Thus, different interpretations of transient hydrodynamics have different $z_n$ and $z_\pi$.}. Thus, $z_n$ and $z_\pi$ are theory-dependent. 

For a constant-cross-section gas, the DNMR \cite{Denicol:2012cn} values of $z_n$ and $z_\pi$ (computed using 41 moments) are respectively 6\% and 20\%. This means that DNMR states are responsible for $4/5$ of the fluctuations of $\pi^{jk}$, and the remaining $1/5$ comes from high-frequency degrees of freedom. Not surprisingly \cite{Wagner:2023jgq}, IReD \cite{Wagner:2022ayd} does significantly better in reproducing the high-frequency regime of kinetic theory, with $z_n=2$\% and $z_\pi=4$\%. This means that roughly $24/25$ of the fluctuations of $\pi^{jk}$ come from IReD states, and only $1/25$ escapes hydrodynamics. It should be kept in mind that the entirety of the non-hydrodynamic contribution to the fluctuations of $\nu^j$ and $\pi^{jk}$ is sensitive to the details of the interaction. If, instead of postulating a constant cross-section, we compute the cross-section from scalar quantum field theory with a $\lambda\varphi^4$ interaction, non-hydrodynamic contributions are more important, with $z_n=10$\% and $z_\pi=17$\% (according to IReD \cite{Rocha:2023hts}). Overall, we see that Israel-Stewart thermodynamics already accounts for most of the kinetic fluctuations.

\section{Time-evolution of linear fluctuations}

Thermodynamics alone can only give us correlators between spacelike-separated events. If two events are causally connected, their correlation depends also on the details of the dynamics of the system. In this section, we discuss how to compute the two-point correlators of fluctuating relativistic kinetic theory at non-equal times.

\subsection{Causal theory of time-dependent fluctuations}\label{jarjar}

It is useful to formulate the problem in an abstract way.
Let the column vector $\Psi(x)$ be a (possibly infinite) list of non-equilibrium fields, which fully characterize the state of the gas at a given spacetime event $x$. In the kinetic theory case, $\Psi(x)=\{\phi^1(x),\phi^2(x),... \}$ is the list of linear combination coefficients of any expansion $\phi_p=\sum_A \phi^A \, g_A(p)$ on some basis $\{g_1(p),g_2(p),...\}$ of $L^2(\mathbb{R}^3)$. In the presence of random fluctuations, $\Psi$ is a solution of a linear stochastic differential equation of the form
\begin{equation}\label{stoquo}
    \mathcal{L}(\partial_\mu)\Psi = \xi \, ,
\end{equation}
where $\mathcal{L}(\partial_\mu)$ is a linear differential operator, and $\xi(x)$ is a stochastic noise term with zero average, i.e. $\langle \xi \rangle=0$. In kinetic theory, equation \eqref{stoquo} can be interpreted as the Langevin generalization of the Boltzmann equation, expanded on the fields $\phi^A(x)$. The solution of \eqref{stoquo} takes the form
\begin{equation}
    \Psi(x)=\int d^4x' \mathcal{G}(x{-}{x'})\xi({x'}) \, ,
\end{equation}
where the matrix $\mathcal{G}$ is a linear-response retarded Green function, solution to the Cauchy problem
\begin{equation}\label{cauchyyy}
   \begin{cases}
    \mathcal{L}(\partial_\mu)\mathcal{G}(x)=\mathbb{I} \, \delta^4(x) \, , \\
    \mathcal{G}(x^0<0)=0 \, .
  \end{cases}  
\end{equation}
In the case of Gaussian fluctuations, all the relevant information is encoded in two-point correlators. Using invariance under spacetime translations, we are, therefore, interested in the infinite matrix $\langle \Psi(x) \Psi^T(0)\rangle$. Thus, let us right-multiply both sides of \eqref{stoquo} by $\Psi^T(0)$ and average. The result is
\begin{equation}
    \mathcal{L}(\partial_\mu)\langle \Psi(x) \Psi^T(0) \rangle =\int d^4 x' \, \langle \xi(x)\xi^T(x')\rangle \mathcal{G}^T(-x') \, .
\end{equation}
Now, let us assume that the noise is covariantly Markovian, i.e. $\langle \xi(x)\xi^T(x')\rangle=\mathcal{Q} \, \delta^4(x{-}x')$, for some constant matrix $\mathcal{Q}$ (note that the Markovian assumption is an approximation, especially in relativity \cite{Petrosyan:2021lqi}). Then we have the following:
\begin{equation}\label{gabubbo}
\mathcal{L}(\partial_\mu) \langle \Psi(x)\Psi^T(0)\rangle =  \mathcal{Q} \, \mathcal{G}^T(-x) \, .  
\end{equation}
Finally, we can make a useful observation: If the evolution operator $\mathcal{L}(\partial_\mu)$ is causal (which is guaranteed if there is an information current), then $\mathcal{G}^T(-x)$ is supported inside the past lightcone $J^-(0)$, see the system \eqref{cauchyyy}. Therefore, if we evaluate \eqref{gabubbo} on a point $x$ that falls strictly outside $J^-(0)$, the right-hand side vanishes, so that
\begin{equation}\label{thirtyseven}
    \mathcal{L}(\partial_\mu) \langle \Psi(x)\Psi^T(0)\rangle =0  \quad \quad \text{if }x \notin J^-(0) \, .
\end{equation}
This tells us that, to evaluate the correlator $\langle \Psi(x)\Psi^T(0)\rangle$ for $x^0>0$, we only need to solve the Boltzmann equation \textit{without} noise (i.e. \eqref{stoquo} with $\xi=0$), using the equal-time correlator as our initial data.

\subsection{Time-dependent fluctuations in kinetic theory}\label{conniunzo}

Combining the reasoning of the above subsection, with the results of subsection \ref{TheGaussone}, we can conclude that the two-point correlator $\langle \phi_p(x)\phi_k(0) \rangle$ is the (unique) solution of the following Cauchy problem:
\begin{equation}\label{cauchyyyCorr}
   \begin{cases}
    p^\mu \partial_\mu \langle \phi_p(x)\phi_k(0) \rangle-{\displaystyle \int} \dfrac{d^3 \Bar{p}}{(2\pi)^3 \Bar{p}^0} K_{p\Bar{p}} \,  \langle \phi_{\Bar{p}}(x)\phi_k(0) \rangle=0 \, , & \text{ for }t>0 \, , \\
    \\
    \langle \phi_p(\textbf{x})\phi_k(0) \rangle= \dfrac{(2\pi)^3}{f_{\text{eq},p}} \delta^3(\textbf{x})\delta^3(\textbf{p}-\textbf{k}) \, , & \text{ at }t=0 \, ,
  \end{cases}  
\end{equation}
where $K_{p\Bar{p}}$ is the kernel of the corresponding linearised collision operator model. For negative times, one can just invoke time-translation invariance: $\langle \phi_p(x)\phi_k(0) \rangle=\langle \phi_p(0)\phi_k(-x) \rangle=\langle \phi_k(-x)\phi_p(0) \rangle$. We note that, while the first equation of \eqref{cauchyyyCorr} is manifestly Lorentz-covariant, the second seems to pick a preferred frame. Also, the distinction between positive and negative times depends on the frame of reference (outside the lightcone). However, the full fluctuating theory is Lorentz-covariant. The proof is provided below.

Let us focus on the retarded correlator $\mathcal{C}_{pk}(x)=\Theta(x^0-0^+)\langle \phi_p(x)\phi_k(0)\rangle$. For both $x^0<0^+$ and $x^0>0^+$, such correlator is a solution of the linearised Boltzmann equation without noise. In fact, for negative $x^0-0^+$, it vanishes identically, and the Boltzmann equation holds as an identity ``$0=0$''. For positive $x^0-0^+$, the point $x$ falls outside the past lightcone $J^-(0)$, and we can invoke equation \eqref{thirtyseven}. At $x^0=0^+$, the derivative of $\Theta(x^0-0^+)$ contributes to $\partial_t \mathcal{C}_{pk}$. Hence, recalling equation \eqref{equaltimecorreltor!}, we obtain
\begin{equation}\label{unpodilavoro}
    p^\mu \partial_\mu \mathcal{C}_{pk}(x)-\int \dfrac{d^3 \Bar{p}}{(2\pi)^3 \Bar{p}^0} K_{p\Bar{p}} \,  \mathcal{C}_{\Bar{p}k}(x)= \dfrac{(2\pi)^3}{f_{\text{eq},p}} \delta^4(x) p^0 \delta^3(\textbf{p}-\textbf{k}) \, ,
\end{equation}
To derive the above equation, the infinitesimal positive number $0^+$ in the Heaviside step function is crucial. The reason is that, since the event $x=0$ belongs to $J^-(0)$, the right-hand side of \eqref{gabubbo} does not vanish at $x=0$, while it vanishes for $x^0=0^+$. 

Equation \eqref{unpodilavoro} clearly shows us that, although we have picked a reference frame to construct the theory, no physical prediction depends on this choice. In fact, recalling that the product $p^0 \delta^3(\textbf{p}-\textbf{k})$ is a Lorentz scalar \cite{Peskin_book}, we see that both sides of \eqref{unpodilavoro} are indeed Lorentz scalars. This implies that, although we need to choose a hypersurface $\Sigma$ to define the probability distribution \eqref{PofF}, all the physical predictions, including the second line of \eqref{cauchyyyCorr} are ``$\Sigma$-independent''. Additionally, since the linear operator on the left-hand side of \eqref{unpodilavoro} is causal (as it admits an information current), the support of $\mathcal{C}_{pk}(x)$ cannot exit the future lightcone $J^+(0)$. Thus relativity of simultaneity does not introduce any time-ordering ambiguity in \eqref{cauchyyyCorr}, because $\langle \phi_p(x)\phi_k(0) \rangle$ is supported in the two-folded lightcone $J^+(0)\cup J^-(0)$.

\subsection{Application: Diluted solutions in the relaxation-time approximation}\label{diluzzio}

Consider a highly diluted solution, and treat the solute species as the gas of interest, and the solvent species as the heat bath. In this setting, binary collisions between particles of the solute (i.e. ``gas-gas'' interactions) are negligibly rare, and all collisions in the gas are mediated by the solvent (i.e., the relaxation of the gas is governed by ``gas-bath'' interactions). We also assume that the collisions give rise to chemical conversion between the solvent and the solute, so that $\partial_\mu J^\mu_{\text{gas}} =-\partial_\mu J^\mu_{\text{bath}} \neq 0$. Then, working in the relaxation-time approximation, the Boltzmann equation for the solute reads, in the rest frame of the solvent, 
\begin{equation}\label{relaxtakeiteasy}
    p^\mu \partial_\mu f_p = p^0 \,  \dfrac{f_{\text{eq},p}(\alpha^\star,\beta^\star_\nu)-f_p}{\tau_p} \, ,
\end{equation}
where $\tau_p$ is a (possibly momentum-dependent) relaxation time. This is \textit{not} the usual (Anderson-Witting) relaxation-time approximation for one-component gases, because $f_{\text{eq},p}(\alpha^\star,\beta^\star_\nu)$ is not a function of $f_p$ itself. Instead, $f_{\text{eq},p}=e^{\alpha^\star-\beta_\nu^\star p^\nu}$ depends solely on the intensive parameters $\{\alpha^\star,\beta^\star_\nu \}$ of the solvent. This models the fact that the relaxation is due uniquely to interactions with the bath, and particles in the gas are not ``aware'' of each other. Indeed, equation \eqref{relaxtakeiteasy} enables direct transfusion of energy, momentum, and particle number between the solute and the solvent, since $\partial_\mu T^{\mu \nu}_{\text{gas}}\neq 0$ and $\partial_\mu J^\mu_{\text{gas}} \neq 0$\footnote{In a relaxation time model where the relaxation time depends on particle momenta and local conservation laws are obeyed, the ansatz for the collision term has to be modified \cite{Rocha:2021zcw, Rocha:2022fqz}.}.

With the above choice of collision integral, the first equation of \eqref{cauchyyyCorr} becomes
\begin{equation}\label{unpodilavoro2}
    \bigg[ p^\mu \partial_\mu+ \dfrac{p^0}{\tau_p} \bigg] \langle \phi_p(x)\phi_k(0) \rangle= 0 \, , \quad \quad \text{for }t>0
\end{equation}
This can be viewed as an ordinary differential equation along the parametric geodesic $x^\mu(\lambda)=x^\mu+p^\mu \lambda/p^0$, which can be solved analytically with the initial data in \eqref{cauchyyyCorr}. The result is
\begin{equation}\label{quinquin}
    \langle \phi_p(x)\phi_k(0) \rangle=e^{-|t|/\tau_p} \dfrac{(2\pi)^3}{f_{\text{eq},p}}\delta^3(\textbf{x}-\textbf{v}t)\delta^3(\textbf{p}-\textbf{k}) \, ,
\end{equation}
where $v^j=p^j/p^0$ is the three-velocity\footnote{Note that the same result can be obtained using the covariant approach of \cite{Mullins:2023tjg, Mullins:2023ott}, guaranteeing that this correlator does not depend on the choice of foliation used to define equal-time.}. The above formula has been extended also to negative times using the symmetry $\langle \phi_p(x)\phi_k(0) \rangle=\langle \phi_k(-x)\phi_p(0) \rangle$. This gives rise to the absolute value in the exponent.

\subsection{Density fluctuations in a (chemically active) diluted solution with constant relaxation time}

We can use the correlator \eqref{quinquin} to evaluate the density-density correlator of the solute at non-equal times. Recalling equation \eqref{maladefinizioneequesta}, we obtain (in the rest frame of the solvent)
\begin{equation}\label{lesolve}
    \langle \delta J^0(x) \, \delta J^0(0)\rangle = \int \dfrac{d^3p}{(2\pi)^3}  \, f_{\text{eq},p} \, e^{-|t|/\tau_p} \delta^3(\textbf{x}-\textbf{v}t) \, .
\end{equation}
To evaluate the integral explicitly, let us first assume that $\tau_p=\tau$ does not depend on the momentum. If we work in the ultrarelativistic limit, as in section \ref{UUUultra}, we obtain
\begin{equation}
  \langle \delta J^0(x) \, \delta J^0(0)\rangle =  \dfrac{ne^{-|t|/\tau}}{4\pi |\textbf{x}|^2} \delta(|\textbf{x}|-|t|) \, .
\end{equation}
This describes a spherical wavefront that propagates at the speed of light, and decays exponentially over a timescale $\tau$. If, instead, we work in the non-relativistic limit (still with constant $\tau_p=\tau$), we obtain ($m$ is the mass of the constituents)
\begin{equation}
    \langle \delta J^0(x) \, \delta J^0(0)\rangle =\dfrac{n e^{-|t|/\tau}}{(2\pi T m^{-1} t^2)^{3/2}} \, \exp \bigg[-\dfrac{m \textbf{x}^2}{2Tt^2} \bigg] \, ,
\end{equation}
which describes a Gaussian profile that expands with standard deviation proportional to $t$, and decays over a timescale $\tau$. In the intermediate regime, the result is more cumbersome:
\begin{figure}
\begin{center}
\includegraphics[width=0.46\textwidth]{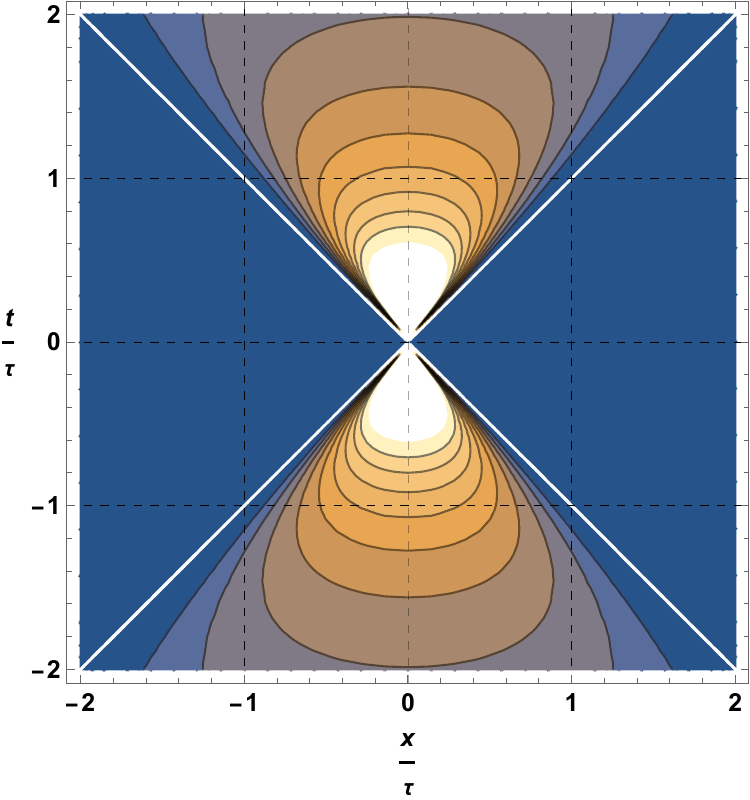}
\includegraphics[width=0.46\textwidth]{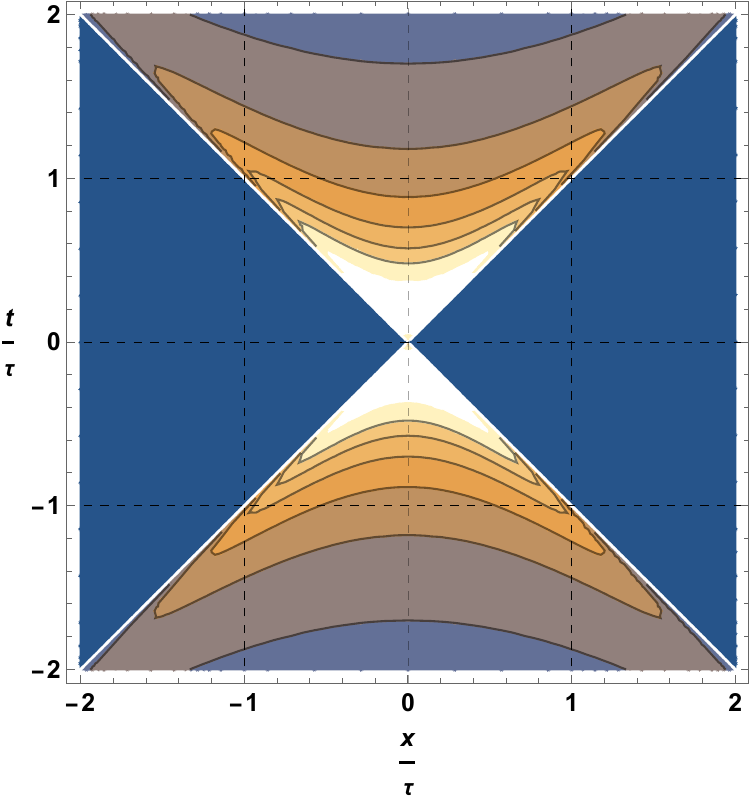}
	\caption{Minkowski diagram of the density-density correlator \eqref{fivofourthreetwoone} on the two-dimensional spacetime plane $y=z=0$ for mildly relativistic gases. The left panel refers to a solvent with $T/m=0.1$, while the right panel refers to a solvent with $T/m=0.5$ (where $m$ is the mass of the constituents of the solute). The colormap marks the intensity of the correlator (white=large, blue=small). The lightcone has been highlighted in white.}
	\label{fig:Temperare}
	\end{center}
\end{figure}
\begin{equation}\label{fivofourthreetwoone}
   \langle \delta J^0(x) \, \delta J^0(0)\rangle = n_{\text{ur}}e^{-|t|/\tau} \,  \dfrac{t^2 \Theta(t^2{-}\textbf{x}^2)}{8\pi (t^2{-}\textbf{x}^2)^{5/2}} \, \dfrac{m^3}{T^3}  \exp \bigg[- \dfrac{m \, |t|}{T\sqrt{t^2{-}\textbf{x}^2}} \bigg] \, ,  
\end{equation}
where $n_{\text{ur}}$ is the equilibrium density in the ultrarelativistic limit, as given in equation \eqref{edaicisono}. In figure \ref{fig:Temperare}, we sketch the Minkowski diagram of $\langle \delta J^0(x) \, \delta J^0(0)\rangle$ for mildly relativistic gases.

\subsection{The case of $\lambda \varphi^4$ relaxation time}

In the previous subsection, we evaluated the density correlators assuming that $\tau_p$ is independent of momentum. However, that is not a realistic assumption within a QFT framework. Let us consider, instead, a more interesting setting, where $\tau_p{=}\tau p^0/T$ (with $\tau{=}\text{const}$), which is the relaxation time of a gas with $\lambda \varphi^4$ cross-section. Then, our diluted solution reduces to the radiation hydrodynamic model considered in \cite{Gavassino:2024pgl}, and equation \eqref{lesolve} becomes
\begin{equation}\label{lesolve39}
    \langle \delta J^0(x) \, \delta J^0(0)\rangle = \int \dfrac{d^3p}{(2\pi)^3}  \, \exp \bigg[\dfrac{\mu}{T}{-}\dfrac{p^0}{T}{-}\dfrac{|t|T}{\tau p^0}\bigg] \delta^3(\textbf{x}-\textbf{v}t) \, .
\end{equation}
Working in the ultrarelativistic limit, the integral can still be solved analytically. The result is
\begin{equation}
\langle \delta J^0(x) \, \delta J^0(0)\rangle = \dfrac{n}{4\pi |\textbf{x}|^2} (|t|/\tau)^{3/2} K_3(2\sqrt{|t|/\tau}) \delta(|\textbf{x}|-|t|) \, ,
\end{equation}
where $K_3$ is the modified Bessel function of the second kind. Taking the long-time limit of this correlator, we obtain the following decay law:
\begin{equation}\label{sgridiamolo}
\langle \delta J^0(x) \, \delta J^0(0)\rangle \sim  e^{-2\sqrt{|t|/\tau}} \, .
\end{equation}
As can be seen, the thermal correlations decay subexponentially, which is a manifestation of the dehydrodynamization mechanism discussed in \cite{Kurkela:2017xis}. This phenomenon arises due to the existence of non-hydrodynamic states with arbitrarily long equilibration timescales, as discussed in \cite{Gavassino:2024rck,Rocha:2024cge}. Indeed, equation \eqref{sgridiamolo} shows that, although these non-thermal long-lived states have a low probability of occurring, they are not fully suppressed within the grand-canonical ensemble. This also explains why the existence of these states leads to a divergence of the gradient expansion discussed in \cite{Gavassino:2024pgl}.

\section{Evaluating the correlators from the information current}

The full Boltzmann equation for a single-component gas is more complicated than that of infinitely diluted solutions considered in section \ref{diluzzio}. For the former, there is in general no hope of obtaining an exact formula for the $\phi_p$ correlator in real space. Instead, we need to work in Fourier space. For practical purposes, this is not a strong limitation, since most applications involve the Fourier component of the correlators from the start. Here, we discuss how one can derive the formula for the correlators directly from the knowledge of the information current. Our strategy is a relativistically-covariant generalization \cite{Mullins:2023tjg, Mullins:2023ott} of the Fox-Uhlenbeck approach \cite{doi:10.1063/1.1693183, fox1970contributions}.

\subsection{Correlators in Fourier space}
\label{sec:corr-fourier-RTA}

First, a bit of notation. By translation invariance, the correlator in Fourier space has the form
\begin{equation}
\label{eq:phi-corr-notation}
    \langle \phi_p(q) \phi_k^*(q') \rangle =(2\pi)^4 (\phi_p \phi_k)_q \, \delta^4(q-q') \, ,  
\end{equation}
where $\phi_p(q)$ is the Fourier transform of $\phi_p(x)$, and we are adopting the notation of \cite{landau_statistical_1980}, according to which
\begin{equation}
    (\phi_p\phi_k)_q := \int d^4 x \, e^{-iqx} \langle \phi_p(x) \phi_k(0) \rangle \, .
\end{equation}
We know that the integral is supported within the two-sided lightcone $J^+(0) \cup J^-(0)$ (recall the discussion in section \ref{conniunzo}). Thus, we can decompose it as the sum of two integrals, over $J^+(0)$ and $J^-(0)$ respectively. Using the symmetry $\langle \phi_p(x)\phi_k(0)\rangle = \langle \phi_k(-x)\phi_p(0)\rangle$, we therefore obtain
\begin{equation}\label{qwertyuiop}
    (\phi_p\phi_k)_q = \mathcal{C}_{pk}(q)+\mathcal{C}_{kp}(-q) \, ,
\end{equation}
where $\mathcal{C}_{pk}(q)$ is the Fourier transform of the retarded correlator introduced in section \ref{conniunzo}. If we Fourier transform equation \eqref{unpodilavoro}, we find that $\mathcal{C}_{pk}(q)$ solves the following equation:
\begin{equation}\label{asdfghjkl}
f_{\text{eq},p} \, ip^\mu q_\mu \, \mathcal{C}_{pk}(q)-\int \dfrac{d^3 \Bar{p}}{(2\pi)^3 \Bar{p}^0} f_{\text{eq},p} K_{p\Bar{p}} \,  \mathcal{C}_{\Bar{p}k}(q)= (2\pi)^3 p^0 \delta^3(\textbf{p}-\textbf{k}) \, .
\end{equation}
Equations \eqref{qwertyuiop} and \eqref{asdfghjkl} fully determine the value of $(\phi_p\phi_k)_q$. For example, setting the collision kernel to be
\begin{equation}
\begin{aligned}
&
K_{p\Bar{p}}=-\frac{(2\pi)^3}{\tau_p}(p^0)^2 \delta^3(\textbf{p}-\Bar{\textbf{p}}),
\end{aligned}    
\end{equation}
as in equation \eqref{unpodilavoro2}, we (correctly) recover the Fourier tranform of \eqref{quinquin}, namely
\begin{equation}
\label{eq:phi-corr-RTA-final}
    (\phi_p\phi_k)_q = \dfrac{(2\pi)^3}{f_{\text{eq},p}} \, \dfrac{2\tau_p \delta^3(\textbf{p}-\textbf{k})}{1+\tau^2_p(q^0-\textbf{v}\cdot \textbf{q})^2}  \, .
\end{equation}

\subsection{The covariant Fox-Uhlenbeck approach}

Let us go back to the abstract framework of section \ref{jarjar}. It can be rigorously shown \cite{Gavassino:2023odx,Gavassino:2023qwl} that, if the information current $E^\mu$ is timelike future-directed in all states, and the entropy production rate $\sigma=-\partial_\mu E^\mu$ is non-negative definite, and we have $E^\mu = \Psi^T \mathbb{E}^\mu \Psi/2$, and $\sigma=\Psi^T \sigma \! \! \! \sigma \Psi$ (where $\mathbb{E}^\mu$ and $\sigma \! \! \! \sigma$ are symmetric matrices), then the linear operator on the left-hand side of equation \eqref{stoquo} can be taken to be 
\begin{equation}
   \mathcal{L}(\partial_\mu)=\mathbb{E}^\mu \partial_\mu + \sigma \! \! \! \sigma \, . 
\end{equation}
Under these conditions, the covariantly Markovian noise on the right-hand side of \eqref{stoquo} has variance \cite{Mullins:2023tjg, Mullins:2023ott}
\begin{equation}\label{covuzko}
    \langle \xi(x) \xi^T(x')\rangle = 2 \sigma \! \! \! \sigma \delta^4(x-x') \, .
\end{equation}
This is the covariant generalization of Fox and Uhlenbeck's formulation of the fluctuation-dissipation theorem \cite{fox1970contributions}, as it states that systems with stronger dissipation have larger noise (which is responsible for fluctuations). We remark that \eqref{covuzko} is manifestly Lorentz-invariant. 

In the kinetic theory case, we recall that the array of discrete variables $\Psi=\{\phi^1,\phi^2,... \}$ is the (infinite) list of linear-combination coefficients of any given expansion $\phi_p=\sum_A \phi^A g_A(p)$, see section \ref{jarjar}. Thus, it is straightforward to verify that the matrix components of $\mathbb{E}^\mu$ and $\sigma \! \! \! \sigma$ are, respectively,
\begin{equation}
\begin{split}
E^\mu_{AB}={}& \int \dfrac{d^3 p}{(2\pi)^3 p^0} p^\mu f_{\text{eq},p} g_A(p)g_B(p)   \, , \\
\sigma_{AB}={}& -\int \dfrac{d^3 p}{(2\pi)^3 p^0} \dfrac{d^3 \Bar{p}}{(2\pi)^3 \Bar{p}^0} f_{\text{eq},p} \, g_A(p) K_{p\Bar{p}} \, g_B(\Bar{p}) \, .
\end{split}
\end{equation}
Hence, the stochastic equation \eqref{stoquo}, expressed in components, reads $E^\mu_{AB} \partial_\mu \phi^B {+}\sigma_{AB}\phi^B = \xi_A$, where the summation over $B$ is understood, and $\xi_A$ is a stochastic noise with $\langle \xi_A(x)\xi_B(x')\rangle = 2\sigma_{AB}\delta^4(x{-}x')$, as required by equation \eqref{covuzko}. If we express $\xi_A$ as the integral of a noise distribution in momentum, i.e.
\begin{equation}
    \xi_A = \int \dfrac{d^3 p}{(2\pi)^3 p^0} g_A(p) \xi_p \, ,
\end{equation}
the discrete system of equations $E^\mu_{AB} \partial_\mu \phi^B {+}\sigma_{AB}\phi^B = \xi_A$ is equivalent to the stochastic Boltzmann equation
\begin{equation}\label{sprintone}
\begin{cases}
 f_{\text{eq},p}p^\mu \partial_\mu \phi_p - f_{\text{eq},p} \int \dfrac{d^3 \Bar{p}}{(2\pi)^3 \Bar{p}^0} K_{p\Bar{p}} \phi_{\Bar{p}} = \xi_p \, , \\  
 \langle \xi_p(x)\xi_{\Bar{p}}(x')\rangle = -2f_{\text{eq},p}K_{p\Bar{p}} \delta^4(x-x') \, .\\
\end{cases}
\end{equation}
Note that both sides of the equations above are Lorentz scalars, confirming that the resulting fluctuating version of kinetic theory is manifestly Lorentz-covariant. Indeed, it was shown in \cite{Mullins:2023tjg} that if the linear equations of motion \eqref{stoquo} are obtained from the information current (as discussed above), the resulting correlators of the dynamical variables, in this case $\phi_p$, do not depend on the choice of spacetime foliation.

The system \eqref{sprintone}, expressed in Fourier space, will be used in section \ref{The full boltz} to evaluate the correlators of the full Boltzmann equation.

\section{The full fluctuating Boltzmann equation}\label{The full boltz}

Building on the knowledge acquired in the previous sections, we will now turn our attention to the evolution of fluctuations arising from more realistic models for the collisions. To this end, we consider a system in which only two-to-two scatterings occur, and quantum statistics effects can be neglected. In this case, the relativistic Boltzmann equation in the full non-linear regime reads
\begin{equation}
\label{eq:Boltzmann}
p^{\mu} \partial_{\mu} f_{p}  
=
\frac{1}{2}
\int\frac{d^{3}k}{(2 \pi)^{3} k^{0}} \ \frac{d^{3}k'}{(2 \pi)^{3} k'^{0}} \ \frac{d^{3}p'}{(2 \pi)^{3} p'^{0}} W_{pp' \leftrightarrow kk'} (f_{k}f_{k'}  -  f_{p}f_{p'}), 
\end{equation}
where we introduce the transition rate  $W_{pp' \leftrightarrow qq'}  = (2\pi)^{6} s \sigma(s,\Theta) \delta^{4}(p+p'-k-k')$, $\sigma(s,\Theta)$ is the differential cross section, $s \equiv (k^{\mu}+k'^{\mu})(k_{\mu}+k'_{\mu}) = (p^{\mu}+p'^{\mu})(p_{\mu}+p'_{\mu})$ is the square of the total energy in the zero-total-momentum frame and we define $\Theta$ as the angle between the 3-momentum vectors $\mathbf{p}$ and $\mathbf{k}$ in the zero-total-momentum frame. The cross-section contains information about the specific microscopic theory of interest and its specific functional form greatly influences the non-equilibrium dynamics \cite{DeGroot:1980dk,cercignani:02relativistic,Gavassino:2024rck,Rocha:2024cge}. 

In the linear regime, perturbations to the global equilibrium distribution for this system evolve through the linearized Boltzmann equation
\begin{equation}
\label{Lhat}
\begin{aligned}
&
f_{\mathrm{eq},p}
p^{\mu}\partial_{\mu}\phi_p
=
f_{\mathrm{eq},p}
\hat{L}[\phi_{p}] \\
&
\equiv
\frac{1}{2}\int \frac{d^{3}k}{(2 \pi)^{3} k^{0}} \ \frac{d^{3}k'}{(2 \pi)^{3} k'^{0}} \ \frac{d^{3}p'}{(2 \pi)^{3} p'^{0}}  W_{pp' \leftrightarrow kk'} f_{\mathrm{eq},p} f_{\mathrm{eq},p'} (  \phi_{k} + \phi_{k'} - \phi_{p} - \phi_{p'}  ),
\end{aligned}    
\end{equation}
where we defined the linearized collision operator $\hat{L}$, which is a linear functional of the relative deviation function $\phi_{p}$. In order to express the linearized Boltzmann equation in the form of Eq.~\eqref{sprintone}, we must find an integral kernel such that 
\begin{equation}
\label{eq:fund-K-prop}
\begin{aligned}
&
f_{\text{eq},p} \int \dfrac{d^3 \Bar{p}}{(2\pi)^3 \Bar{p}^0} K_{p\Bar{p}} \phi_{\Bar{p}}
=
f_{\text{eq},p}
\Hat{L}[\phi_{p}].
\end{aligned}    
\end{equation}
This is indeed achieved by the following manifestly symmetric integral kernel 
 (details in Appendix \ref{eq:K-kernel})
\begin{equation}
\label{eq:kernel-K}
\begin{aligned}
f_{\mathrm{eq},p}
K_{p \Bar{p}} 
&
= 
\frac{(2\pi)^3}{4} \int \frac{d^3k}{(2\pi)^3 k^0}  \ \frac{d^{3}k'}{(2 \pi)^{3} k'^{0}} \ \frac{d^{3}p'}{(2 \pi)^{3} p'^{0}} \,  
\\
&
\times
\left\{ 
\tilde{p}^0   W_{pp' \rightarrow kk'} f_{\mathrm{eq},p} f_{\mathrm{eq},p'}
\left[ \delta^{3}(\textbf{k}-\Bar{\textbf{p}}) + \delta^{3}(\textbf{k}'-\Bar{\textbf{p}}) 
- \delta^{3}(\textbf{p}-\Bar{\textbf{p}}) - \delta^{3}(\textbf{p}'-\Bar{\textbf{p}}) \right]
\right.
\\
&
\left.
+
p^{0} W_{\Bar{p}p' \leftrightarrow kk'} f_{\mathrm{eq},\widetilde{p}} f_{\mathrm{eq},p'}  
\left[\delta^{3}(\textbf{k} - \textbf{p} ) 
+ 
 \delta^{3}(\textbf{k}' - \textbf{p}) 
- 
\delta^{3}(\Bar{\textbf{p}} - \textbf{p}) 
- 
\delta^{3}(\textbf{p}' - \textbf{p})  \right]
\right\}.
\end{aligned}
\end{equation}
Then, in Fourier space, the Boltzmann equation can be expressed as 
\begin{equation}
\label{eq:lin-BEq-Fou}
f_{\mathrm{eq},p} [i \Omega E_p  \Bar{\phi}_{p} + i p^{\mu} q_{\langle \mu \rangle} \Bar{\phi}_{p}]  
-
\int \dfrac{d^3 \Bar{p}}{(2\pi)^3 \Bar{p}^0} f_{\text{eq},p} K_{p\Bar{p}} \Bar{\phi}_{\Bar{p}} =  \Bar{\xi}_p ,
\end{equation}
where we define $\Omega = q_{\mu} u^{\mu}$,  $E_p = p_{\mu} u^{\mu}$, and $q^{\langle \mu \rangle} = \Delta^{\mu}_{\ \nu}q^{\nu}$, which is given in terms of the fundamental projector $\Delta^{\mu \nu} = g^{\mu \nu} - u^{\mu} u^{\nu}$. In the rest frame of the background fluid, $\Omega = q^{0}$, $E_{p} = p^{0}$, $u^{\mu} = (1,0,0,0)$, $(q^{\langle \mu \rangle}) = (0, \mathbf{q}) $.  From Eq.~\eqref{sprintone}, the Fourier-transformed stochastic noise correlator reads 
\begin{equation}
\label{eq:Fourier-correlators-Xi}
\begin{aligned}
&
\langle \Bar{\xi}_{p}(q) \Bar{\xi}^{*}_{\tilde{p}}(q') \rangle = - 2 (2 \pi)^{4} f_{\text{eq},p} K_{p \Bar{p}} \delta^{4}(q-q').
\end{aligned}    
\end{equation}
Equation \eqref{eq:lin-BEq-Fou} is an inhomogeneous linear integral equation for $\Bar{\phi}_{p}$, whose solution can be found by expanding $\Bar{\phi}_{p}$ in the eigenfunctions of $\Hat{L}$. In the remainder of the section, we shall analyze a particular system, for which the eigensystem of $\Hat{L}$ (the set of eigenvalues and eigenvectors) is known analytically.  In this case, one can also expand $\xi_{p}$ in this basis and express the collision kernel $K_{p \Bar{p}}$ by a spectral expansion. This was already recognized in the non-relativistic limit \cite{fox1970contributions}. The obtention of the eigenfunctions, however, can be achieved only in very rare instances. In the non-relativistic regime, this is the case of the so-called Maxwell particles. Then, the eigenfunctions are expressible in terms of Hermite polynomials and spherical harmonics, but there is not a closed expression for the eigenvalues \cite{chang1952propagation,cercignani:90mathematical}. In relativistic systems, this was only recently achieved in Ref.~\cite{Denicol:2022bsq} for a system of weakly interacting massless scalar particles. In the remainder of this section, we shall analyze the dynamics of fluctuations in Fourier space for the latter system.

\subsection{Fluctuations for weakly interacting ultrarelativistic scalar particles}

For a system of scalar particles described by the Lagrangian density $\mathcal{L} = \frac{1}{2} \partial_{\mu} \varphi \ \partial^{\mu} \varphi
-
(\lambda/4!) \varphi^{4}$, the differential cross-section at leading order in the coupling is, $\sigma(s, \Theta) = g/s$, with $g = \lambda^{2}/(32 \pi)$. This specific functional form for $\sigma(s, \Theta)$ enables the computation of the eigensystem of the linearized collision term (and by extension of the kernel $K_{p \Bar{p}}$) in exact form, in contrast to what was described for the non-relativistic case. Indeed, it was shown that the eigensystem is so that  \cite{Denicol:2022bsq} 
\begin{equation}
\label{eq:eigenvalues-lin-col}
\begin{aligned}
&
\hat{L}\left[L^{(2\ell+1)}_{n,p} p^{\langle \mu_{1}} \cdots p^{\mu_{\ell} \rangle} \right] 
=
\chi_{n \ell} 
L^{(2\ell+1)}_{n,p} p^{\langle \mu_{1}} \cdots p^{\mu_{\ell} \rangle},
 \\
&
\chi_{n \ell} = - \frac{g n_{\mathrm{eq}} \beta}{4}  \left(\frac{n+\ell-1}{n+\ell+1} + \delta_{n0}\delta_{\ell 0}\right),
\end{aligned}    
\end{equation}
where $L^{(2\ell+1)}_{n,p} = L^{(2\ell+1)}_{n}(\beta E_p)$ denotes associated Laguerre polynomials \cite{NIST:DLMF,gradshteyn2014table}, $n_{\mathrm{eq}} = e^{\alpha}/(\pi^{2} \beta^{3})$ is the equilibrium particle density, $\alpha = \mu/T$ is the thermal potential, and $p^{\langle \mu_{1}} \cdots p^{\mu_{\ell} \rangle}$ is the irreducible projection of products of momentum. It is given in terms of projection tensors, $p^{\langle \mu_{1}} \cdots p^{\mu_{\ell} \rangle} \equiv \Delta^{\mu_{1} \cdots \mu_{\ell}}_{\nu_{1} \cdots \nu_{\ell}} p^{\nu_{1}} \cdots p^{\nu_{\ell}}$. The $2 \ell$-rank tensor $\Delta^{\mu_{1} \cdots \mu_{\ell}}_{\nu_{1} \cdots \nu_{\ell}}$ is constructed from the $\Delta^{\mu \nu} = g^{\mu \nu} - u^{\mu} u^{\nu}$ projectors in such a way that it is completely symmetric with respect to permutations in any of the indices $\mu_{1} \cdots \mu_{\ell}$ and $\nu_{1} \cdots \nu_{\ell}$, separately, and also traceless within each subset of indices \cite{DeGroot:1980dk,Denicol:2021} (see also Appendix \ref{sec:F-moms}).  One of the main properties of these irreducible tensors is 
\begin{equation}
\label{eq:main-property-irred-tens}
\begin{aligned}
\int \dfrac{d^3 p}{(2\pi)^3 p^0}\,
p^{\langle \mu_{1}} \cdots p^{\mu_{\ell} \rangle}
p_{\langle \nu_{1}} \cdots p_{\nu_{m} \rangle}
H(E_{\mathbf{p}})
& = \frac{\ell!}{(2 \ell + 1)!!} \Delta^{\mu_{1} \cdots \mu_{\ell}}_{\nu_{1} \cdots \nu_{\ell} } \delta_{\ell m} 
\int \dfrac{d^3 p}{(2\pi)^3 p^0}
\left(\Delta^{\mu \nu} p_{\mu} p_{\nu}\right)^{\ell}
H(E_p), 
\end{aligned}
\end{equation}
where $H(E_{p})$ is an arbitrary weight function assumed to be sufficiently regular so that the integral converges. As we shall see in the remainder of section, the property above allows the separation of different tensor components from a given equation by integration. The polynomials $L_{n}^{(\alpha)}$ form a complete set of orthogonal functions, indeed, they can be normalized so that they are orthogonal with respect to the global equilibrium distribution,
\begin{equation}
\label{eq:orth-laguerre}
\begin{aligned}
&
\frac{\ell!}{(2 \ell + 1)!!} \int \dfrac{d^3 p}{(2\pi)^3 p^0} \left( \Delta^{\mu \nu} p_{\mu} p_{\nu} \right)^{\ell} L_{n,p}^{(2 \ell + 1)} L_{m,p}^{(2 \ell + 1)} f_{\mathrm{eq}, p}
=
A^{(\ell)}_{n} \delta_{nm} 
\equiv
(-1)^{\ell} \frac{\ell !}{(2 \ell + 1)!!}\frac{n_{\mathrm{eq}}}{2 \beta^{2 \ell -1}}
\frac{(n+2\ell+1)!}{n!}
\delta_{nm}.
\end{aligned}    
\end{equation}
It is noted that the spectrum \eqref{eq:eigenvalues-lin-col} contains basic properties of the linearized collision term: all eigenvalues are non-positive\footnote{This is expected from the property that, given an arbitrary function of momentum $A_{p}$, $\int \dfrac{d^3 p}{(2\pi)^3 p^0} f_{\mathrm{eq},p} A_{p} \hat{L}[A_{p}]\leq 0$, which is crucial to prove the non-negativeness of entropy production in the linear regime.} and the zero-modes (i.e.~the eigenfunctions $\Hat{L}$ with zero eigenvalue) correspond to linear combinations of $1$, $p^{\mu}$ which are associated, respectively, to the conservation of particle number\footnote{The particle number conservation is a consequence of the employment of the cross-section at leading order in the coupling $\lambda$. Beyond this order, further processes which do not conserve the particle number emerge \cite{Denicol:2022bsq}.} and of 4-momentum in the collisions. Namely, the zero modes correspond to the values $(n, \ell) = (0, 0), (1, 0)$ and $(0, 1)$, since $L_{0,p}^{(1)} = 1$  and $p^{\mu} = E_p u^{\mu} + p^{\langle \mu \rangle} = (2 L^{(1)}_{0,p} - L^{(1)}_{1,p}) u^{\mu} + L^{(3)}_{0,p} p^{\langle \mu \rangle}$. Another interesting feature of the spectrum is that it is bounded from below and above,
\begin{equation}
\label{eq:bounded-spectrum}
\begin{aligned}
&
- \frac{g n_{\mathrm{eq}} \beta}{4} < \chi_{n \ell} \leq 0.
\end{aligned}   
\end{equation}

When written in the eigenbasis \eqref{eq:eigenvalues-lin-col}, the problem of solving the integral equation \eqref{eq:lin-BEq-Fou} is simplified. To explicitly see this, we shall expand the deviation function and the noise in the eigenfunction basis, 
\begin{subequations}
\label{eq:phi-xi-expn}
\begin{align}
&
\label{eq:phi-xi-expn-1}
 \Bar{\phi}_p = \sum_{n,\ell = 0}^{\infty} \Phi_{n}^{\mu_{1} \cdots \mu_{\ell}} L^{(2 \ell + 1)}_{n p} p_{\langle \mu_{1}} \cdots p_{ \mu_{\ell} \rangle}
, 
\\
&
\label{eq:phi-xi-expn-1b}
\Bar{\xi}_p = f_{\mathrm{eq}, p} \sum_{\ell, n = 0}^{\infty} \Xi_{n}^{\mu_{1} \cdots \mu_{\ell}} L^{(2 \ell + 1)}_{n p} p_{\langle \mu_{1}} \cdots p_{\mu_{\ell} \rangle},
\\
&
\label{eq:phi-xi-expn-2}
A_{n}^{(\ell)}\Xi_{n}^{\mu_{1} \cdots \mu_{\ell}}(q)
=
\int \dfrac{d^3 p}{(2\pi)^3 p^0} L^{(2 \ell + 1)}_{n p} p^{\langle \mu_{1}} \cdots p^{ \mu_{\ell} \rangle} \Bar{\xi}_p
\equiv
\widetilde{\Xi}_{n}^{\mu_{1} \cdots \mu_{\ell}},
\\
&
\label{eq:phi-xi-expn-3}
A_{n}^{(\ell)}\Phi_{n}^{\mu_{1} \cdots \mu_{\ell}}(q)
=
\int \dfrac{d^3 p}{(2\pi)^3 p^0} f_{\mathrm{eq},p}  L^{(2 \ell + 1)}_{n p} p^{\langle \mu_{1}} \cdots p^{ \mu_{\ell} \rangle} \Bar{\phi}_p
\equiv
\widetilde{\Phi}_{n}^{\mu_{1} \cdots \mu_{\ell}},
\end{align}    
\end{subequations}
which conveniently separates the dependence on the Fourier variable $q^{\mu}$ ($\Phi_{n}^{\mu_{1} \cdots \mu_{\ell}}$, $\Xi_{n}^{\mu_{1} \cdots \mu_{\ell}}$, or equivalently $\widetilde\Phi_{n}^{\mu_{1} \cdots \mu_{\ell}}$, $\widetilde\Xi_{n}^{\mu_{1} \cdots \mu_{\ell}}$, which we shall refer to as the \textit{Fourier-Laguerre components} of $\Bar{\phi}_{p}$ and $\Bar{\xi}_{p}$, respectively) from the dependence on the momentum, $p^{\mu}$ ($L^{(2 \ell + 1)}_{n, p} p_{\langle \mu_{1}} \cdots p_{ \mu_{\ell} \rangle} $) . In terms of the $\Phi$-components, the Fourier-space correlators read
\begin{equation}
\label{eq:phi-phi-correlators}
\begin{aligned}
&
\langle \Bar{\phi}_{p}(q) \Bar{\phi}^{*}_{k}(q')  \rangle 
=
\sum_{n,\ell, n',m = 0}^{\infty} \frac{1}{A_{n}^{(\ell)} A_{n'}^{(m)} } 
\left\langle \widetilde{\Phi}_{n}^{\mu_{1} \cdots \mu_{\ell}}(q) \widetilde{\Phi}_{n'}^{*\nu_{1} \cdots \nu_{m}}(q') \right\rangle
L^{(2 \ell + 1)}_{n, p}
L^{(2 \ell + 1)}_{n', k}
 p_{\langle \mu_{1}} \cdots p_{ \mu_{\ell}
 \rangle}
  k_{\langle \mu_{1}} \cdots k_{ \mu_{m}
 \rangle}.
\end{aligned}    
\end{equation}
Some of the $\widetilde{\Phi}$-components possess physical interpretation in terms of stochastic fluctuations around the global equilibrium particle four-current and the energy-momentum tensor. In general, deviations from these quantities can be expressed in Kinetic theory as
\begin{equation}
\begin{aligned}
 &
 \delta J^{\mu} = \int \dfrac{d^3 p}{(2\pi)^3 p^0} p^{\mu} f_{\mathrm{eq},p} \Bar{\phi}_{p}, \\
 &
 \delta T^{\mu \nu} = \int \dfrac{d^3 p}{(2\pi)^3 p^0} p^{\mu} p^{\nu} f_{\mathrm{eq},p} \Bar{\phi}_{p}.
\end{aligned}    
\end{equation}
Then, shear-stress deviations can be identified as $\pi^{\mu \nu} = \delta T^{\langle \mu \nu \rangle}$, and, one can then compute from Eq.~\eqref{eq:phi-xi-expn-3} that, since $L_{0,p}^{(5)} = 1$,
\begin{equation}
\label{eq:shear-as-phi}
\begin{aligned}
&
\widetilde{\Phi}_{0}^{\mu \nu}
= \pi^{\mu \nu}.
\end{aligned}    
\end{equation}
Besides, from the fact that  $L^{(3)}_{0, p} = 1$, and $L^{(3)}_{1, p} = 4 - \beta E_{p}$, it is seen that $\widetilde{\Phi}_{0}^{\mu}$ and $\widetilde{\Phi}_{1}^{\mu}$ are related to the particle diffusion, $\nu^{\mu} \equiv \delta J^{\langle \mu \rangle} $, and to energy diffusion, $h^{\mu} = u_{\nu} \delta T^{\langle \mu \rangle \nu}$,  associated with the fluctuations and, indeed,
\begin{equation}
\label{eq:vectors-as-phi}
\begin{aligned}
&    
\widetilde{\Phi}_{0}^{\mu} = \nu^{\mu},
\quad
\widetilde{\Phi}_{1}^{\mu} = 4 \nu^{\mu} - \beta h^{\mu}.
\end{aligned}    
\end{equation}
Similarly, since $L^{(1)}_{0, p} = 1$, and $L^{(1)}_{1, p} = 2 - \beta E_{p}$, and $L^{(1)}_{2, p} = (1/2)[6 - 6 \beta E_{p} + (\beta E_{p})^{2}]$, the scalar components
$\widetilde{\Phi}_{1}$ and $\widetilde{\Phi}_{2}$ can be related to corrections to the particle density, $\delta n = u_{\mu} \delta J^{\mu}$, energy density $\delta \varepsilon = u_{\mu} u_{\nu} \delta T^{\mu \nu}$, associated with the fluctuations, 
\begin{equation}
\label{eq:relations-Phi-phys}
\begin{aligned}
&
\widetilde{\Phi}_{1} = 2 \widetilde{\Phi}_{0} - \beta \delta n,
\quad
\widetilde{\Phi}_{2} = \frac{1}{2} 
\left(6 \widetilde{\Phi}_{0}
-
6 \beta \delta n
+
\beta^{2} \delta \varepsilon
\right). 
\end{aligned}    
\end{equation}
In case the particles of the system had mass $m$,  $\widetilde{\Phi}_{0}$ would be related to deviations to the trace of the energy momentum tensor in equilibrium, $g_{\mu\nu} \delta T^{\mu \nu} = m^{2} \widetilde{\Phi}_{0}$. Besides, it is noted, from expansion \eqref{eq:phi-xi-expn-1} and Eq.~\eqref{eq:eigenvalues-lin-col}, that $\Phi_{0}$, $\Phi_{1}$, and $\Phi_{0}^{\mu}$ are the components associated with the zero-modes, $L^{(1)}_{0,p} = 1$, $L^{(1)}_{1,p} = 2 - \beta E_{p}$, and $L^{(3)}_{0,p} p^{\langle \mu \rangle} = p^{\langle \mu \rangle}$, respectively. 

Exact equations of motion for a generic Fourier-Laguerre component  $\widetilde{\Phi}_{n}^{\mu_{1} \cdots \mu_{\ell}}$ can be derived by integrating Eq.~\eqref{eq:lin-BEq-Fou}, with the eigenmoment basis, $\int d^3 p/[(2\pi)^3 p^0] (\cdots) L_{n,p}^{(2 \ell +1)} f_{\mathrm{eq},p} p_{\langle \mu_{1}} \cdots p_{ \mu_{\ell} \rangle}$, and performing manipulations involving the Laguerre polynomials and the irreducible tensors. Then, we can derive the following linear system of equations for $\widetilde{\Phi}_{n}^{\mu_{1} \cdots \mu_{\ell}}$ (for details, see Appendix \ref{sec:F-moms}),
\begin{equation}
\label{eq:moments-lag-main}
\begin{aligned}
&
i \frac{\Omega}{\beta} \left[
-(n+1) \widetilde{\Phi}_{n+1}^{\mu_{1} \cdots \mu_{\ell}}
+
2(n+\ell+1)
\widetilde{\Phi}_{n}^{\mu_{1} \cdots \mu_{\ell}}
-
(n+2\ell+1)
\widetilde{\Phi}_{n-1}^{\mu_{1} \cdots \mu_{\ell}}
\right]
+ 
i q_{\langle \mu \rangle} \left[ \widetilde{\Phi}_{n}^{\mu_{1} \cdots \mu_{\ell} \mu}
-
2 \widetilde{\Phi}_{n-1}^{\mu_{1} \cdots \mu_{\ell} \mu}
+
\widetilde{\Phi}_{n-2}^{\mu_{1} \cdots \mu_{\ell} \mu}
\right]
\\
&
-
\frac{i}{\beta^{2}}  
 \frac{\ell}{2 \ell + 1}
q^{\langle \mu_{1}} 
\left[
(n+1)(n+2)
\widetilde{\Phi}_{n+2}^{\mu_{2} \cdots \mu_{\ell} \rangle}
-
2(n+1)(n+2\ell+1)
\widetilde{\Phi}_{n+1}^{\mu_{2} \cdots \mu_{\ell} \rangle}
+
(n+2\ell+1)(n+2\ell)
\widetilde{\Phi}_{n}^{\mu_{2} \cdots \mu_{\ell} \rangle}
\right]
\\
&
-
\chi_{n\ell}
\widetilde{\Phi}_{n}^{\mu_{1} \cdots \mu_{\ell}}
   = 
\widetilde{\Xi}_{n}^{\mu_{1} \cdots \mu_{\ell}},
\quad n = 0, 1, \cdots, \quad \ell = 0, 1, \cdots, 
\end{aligned}
\end{equation}
where we consider that $\widetilde{\Phi}_{n}^{\mu_{1} \cdots \mu_{\ell}} \equiv 0$, for $n < 0$. Besides, the stochastic noise Fourier-Laguerre components, $\widetilde{\Xi}_{n}^{\mu_{1} \cdots \mu_{\ell}}$, can be shown to be correlated as (see also Appendix \ref{sec:F-moms})
\begin{equation}
\label{eq:XiXi-corr}
\begin{aligned}
&
\left\langle \widetilde{\Xi}_{n}^{\mu_{1} \cdots \mu_{\ell}}(q) \widetilde{\Xi}_{n'}^{*\nu_{1} \cdots \nu_{m}}(q')
\right\rangle
= 
-
2 (2 \pi)^{4} A_{n}^{(\ell)} \chi_{n,\ell}  \delta^{4}(q-q') \delta_{\ell m} \delta_{nn'} \Delta^{\mu_{1} \cdots \mu_{\ell}\nu_{1} \cdots \nu_{\ell}}.
\end{aligned}    
\end{equation}
We note from Eq.~\eqref{eq:moments-lag-main} that the term proportional to $\chi_{n\ell}$ does not contribute for $(n,\ell) = (0,0), (0,1), (1,0)$, for which $\chi_{n\ell} = 0$. Besides, from that equation, it is seen  that there is only coupling among Fourier-Laguerre components with rank $\ell$ and ranks $\ell \mp 1$ and among components with Laguerre indices $n$, $n \pm 1$ and $n \pm 2$. We note that the tensor-rank coupling is erased in the limit where $q^{\langle \mu \rangle} \to 0$. This limit encompasses fluctuations which are spatially homogeneous with respect to background fluid, which is what we shall consider for the reminder of this section.

In the the limit where $q^{\langle \mu \rangle} \to 0$,    Eqs.~\eqref{eq:moments-lag-main} become
\begin{equation}
\label{eq:moments-lag-homog}
\begin{aligned}
&
i \frac{\Omega}{\beta} \left[
-(n+1) \widetilde{\Phi}_{n+1}^{\mu_{1} \cdots \mu_{\ell}}
+
2(n+\ell+1)
\widetilde{\Phi}_{n}^{\mu_{1} \cdots \mu_{\ell}}
-
(n+2\ell+1)
\widetilde{\Phi}_{n-1}^{\mu_{1} \cdots \mu_{\ell}}
\right]
-
\chi_{n,\ell}
\widetilde{\Phi}_{n}^{\mu_{1} \cdots \mu_{\ell}}
   = 
\widetilde{\Xi}_{n}^{\mu_{1} \cdots \mu_{\ell}},
\\
&
\quad n = 0, 1, \cdots, \quad \ell = 0, 1, \cdots,
\end{aligned}
\end{equation}
It is readily seen that Eqs.~\eqref{eq:moments-lag-homog} form a system of algebraic equations that are linear in $\widetilde{\Phi}_{n}^{\mu_{1} \cdots \mu_{\ell}}$ in which the different tensor ranks are decoupled and that can be expressed in terms of a dimensionless tridiagonal matrix in the Laguerre indices $n$, $n'$,
\begin{equation}
\label{eq:homogeneous-main}
\begin{aligned}
&
\sum_{n'=0}^{\infty}\mathcal{M}_{n n'}^{(\ell)} \widetilde{\Phi}_{n'}^{\mu_{1} \cdots \mu_{\ell}}
=
\widehat{\Xi}_{n}^{\mu_{1} \cdots \mu_{\ell}},
\quad
n = 0,1,2,3, \cdots,
\\
&
\mathcal{M}_{n n'}^{(\ell)} = 
b_{n}^{(\ell)}(\widehat{\Omega})
\delta_{n',n+1}
+
a_{n}^{(\ell)}(\widehat{\Omega})
\delta_{n',n}
+
c_{n}^{(\ell)}(\widehat{\Omega})
\delta_{n',n-1}, \\
&
a_{n}^{(\ell)}(\widehat{\Omega}) = 2 i \widehat{\Omega} (n+\ell+1)
-
\widehat{\chi}_{n\ell},
\quad
b_{n}^{(\ell)}(\widehat{\Omega}) = - i \widehat{\Omega} (n+1),
\quad
c_{n}^{(\ell)}(\widehat{\Omega}) = - i \widehat{\Omega} (n+2\ell+1), 
\end{aligned}    
\end{equation}
where, in the above equation, we have employed the  variables 
\begin{equation}
\label{eq:Omega-Q2-defs-homog}
\begin{aligned}
&
\widehat{\Omega} = \frac{\Omega}{g n_{\mathrm{eq}} \beta^{2}},
\quad
\widehat{\chi}_{n \ell} = \frac{\chi_{n \ell}}{g n_{\mathrm{eq}} \beta} = -\frac{1}{4}\left(\frac{n + \ell -1}{n+\ell+1} + \delta_{n 0} \delta_{\ell 0}\right),
\quad
\widehat{\Xi}_{n}^{\mu_{1} \cdots \mu_{\ell}} = \frac{\widetilde{\Xi}_{n}^{\mu_{1} \cdots \mu_{\ell}}}{g n_{\mathrm{eq}} \beta},
\end{aligned}    
\end{equation}
in which $\widehat{\Omega}$ and $\widehat{\chi}_{n 0}$ are dimensionless and $\widehat{\Xi}_{n}^{\mu_{1} \cdots \mu_{\ell}}$ has dimensions of $T^{\ell-2}$. We also define the dimensionless quantity
\begin{equation}
\label{eq:A-adimens-def-homog}
\begin{aligned}
&
\widehat{A}_{n}^{(\ell)} \equiv A_{n}^{(\ell)} \frac{\beta^{2 \ell-1}}{n_{\mathrm{eq}}} = (-1)^{\ell} \frac{\ell !}{(2 \ell + 1)!!}\frac{1}{2}
\frac{(n+2\ell+1)!}{n!}.
\end{aligned}   
\end{equation}
The correlators for the stochastic source components $\widehat{\Xi}_{n}$ can be obtained from Eq.~\eqref{eq:XiXi-corr} for $\ell =0$ and definitions \eqref{eq:phi-xi-expn-2} and \eqref{eq:Omega-Q2-defs-homog} and they read
\begin{equation}
\begin{aligned}
&
\label{eq:moments-lag-homog-b}
\left\langle \widehat{\Xi}_{n}^{\mu_{1} \cdots \mu_{\ell}}(\Omega) \widehat{\Xi}_{n'}^{*\mu_{1} \cdots \mu_{m}}(\Omega')
\right\rangle
= 
- 2 \frac{(2 \pi)^{4}}{g \beta^{2 \ell}} \widehat{A}_{n}^{(\ell)} \widehat\chi_{n\ell}  \delta(\Omega-\Omega') \delta_{nn'} \delta_{\ell m} \Delta^{\mu_{1} \cdots \mu_{\ell}\nu_{1} \cdots \nu_{\ell}}.
\end{aligned}    
\end{equation}

For tridiagonal matrices, an analytic formula for the inverse is known \cite{huang1997analytical}, and, indeed, we shall employ for a matrix dimension truncation of rank $N$ (or, equivalently, employing $0 \leq n,n' \leq N-1$ in Eq.~\eqref{eq:homogeneous-main}),
\begin{equation}
\begin{aligned}
& 
[(\mathcal{M}^{(\ell)})^{-1}]_{ij} =
\begin{cases}
(-1)^{i+j}b_{i} \cdots b_{j-1} \theta_{i-1} \varphi_{j+1}/\theta_{N}, \quad \text{if $i < j$}
\\
\theta_{i-1} \varphi_{j+1}/\theta_{N}, \quad \text{if $i = j$}     
\\
(-1)^{i+j}c_{j} \cdots c_{i-1} \theta_{j-1} \varphi_{i+1}/\theta_{N}, \quad \text{if $i > j$}
\end{cases}
,
\end{aligned}    
\end{equation}
where the coefficients $\{\varphi_{i}\}_{i=1, \cdots, N}$ and $\{\theta_{i}\}_{i=1, \cdots, N}$  are obtained, respectively, through the following recursion relations
\begin{equation}
\begin{aligned}
&
\varphi_{N+1} = 1, \quad \varphi_{N} = a_{N-1}, \quad \theta_{0} = 1, \quad \theta_{1} = a_{0},  
\\
&
\varphi_{j} = a_{j-1} \varphi_{j+1} - b_{j-1} c_{j} \varphi_{j+2}, \quad j = N-1, N-2, \cdots, 2, 1, \\
&
\theta_{j+1} = a_{j-1} \theta_{j-1} - b_{j-2} c_{j-1} \theta_{j-2}, \quad j = 2, 3,  \cdots N. 
\end{aligned}    
\end{equation}
Once this inversion is performed, the $\widetilde{\Phi}$-$\widetilde{\Phi}$ correlators can be expressed in terms of $\widehat{\Xi}$-$\widehat{\Xi}$ correlators as
\begin{equation}
\label{eq:lag-fou-phi4-corr-final}
\begin{aligned}
&
\left\langle \widetilde{\Phi}^{\mu_{1} \cdots \mu_{\ell}}_{n}(\Omega)
\widetilde{\Phi}_{n'}^{*\mu_{1} \cdots \mu_{m}}(\Omega')
 \right\rangle
= 
\lim_{N \to \infty}
\sum_{a,a'=0}^{N-1}
[(\mathcal{M}^{(\ell)})^{-1}(\widehat{\Omega})]_{na} [(\mathcal{M}^{(m)})^{-1}(\widehat{\Omega}')]^{*}_{n'a'}
 \left\langle
\widehat{\Xi}_{a}^{\mu_{1} \cdots \mu_{\ell}}
(\Omega)
\widehat{\Xi}^{*\mu_{1} \cdots \mu_{m}}_{a'}
(\Omega')
 \right\rangle
 \\
 &
= 
-2\frac{(2 \pi)^{4}}{g\beta^{2\ell}}
\lim_{N \to \infty}
\sum_{a=0}^{N-1}
[(\mathcal{M}^{(\ell)})^{-1}(\widehat{\Omega})]_{na} [(\mathcal{M}^{(\ell)})^{-1}(\widehat{\Omega})]^{\dagger}_{an'}
\widehat{A}_{a}^{(\ell)} \widehat{\chi}_{a \ell} \delta(\Omega - \Omega')
\delta_{\ell m} \Delta^{\mu_{1} \cdots \mu_{\ell}\nu_{1} \cdots \nu_{\ell}}
\\
&
\equiv
(2 \pi)^{4} \frac{(-1)^{\ell}}{\beta^{2\ell}} \left( \widetilde{\Phi}_{n\ell}
\widetilde{\Phi}_{n' \ell}^{*}
 \right)_{\Omega} \delta(\Omega - \Omega')  \delta_{\ell m} \Delta^{\mu_{1} \cdots \mu_{\ell}\nu_{1} \cdots \nu_{\ell}}
,
 \end{aligned}    
\end{equation}
where we have employed the quantities defined in Eqs.~\eqref{eq:Omega-Q2-defs-homog} and \eqref{eq:A-adimens-def-homog} and defined the dimensionless correlator amplitudes $\left( \widetilde{\Phi}_{n \ell}
\widetilde{\Phi}_{n' \ell}^{*}
 \right)_{\Omega}$. They have the property that $\left( \widetilde{\Phi}_{n \ell}
\widetilde{\Phi}_{n' \ell}^{*}
 \right)_{\Omega} =  \left( \widetilde{\Phi}_{n' \ell}^{*}
\widetilde{\Phi}_{n \ell}
 \right)_{\Omega} =  \left( \widetilde{\Phi}_{n'\ell}
\widetilde{\Phi}_{n\ell}^{*}
 \right)_{\Omega}$ and are always real. We also note that, since the above correlator is proportional to $\delta_{\ell m}$, Fourier-Laguerre modes with different tensor ranks are decorrelated, as expected.

\begin{figure}[!h]
\begin{center}
\includegraphics[width=0.43\textwidth]{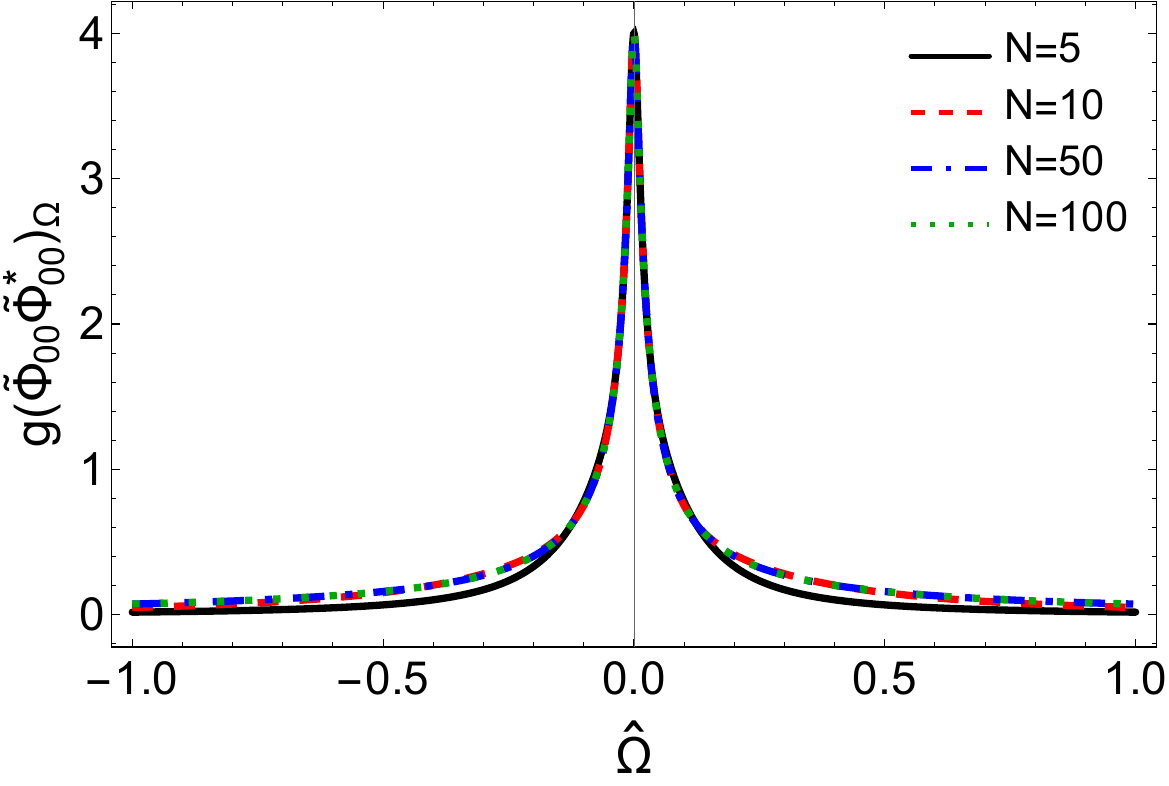}
\includegraphics[width=0.43\textwidth]{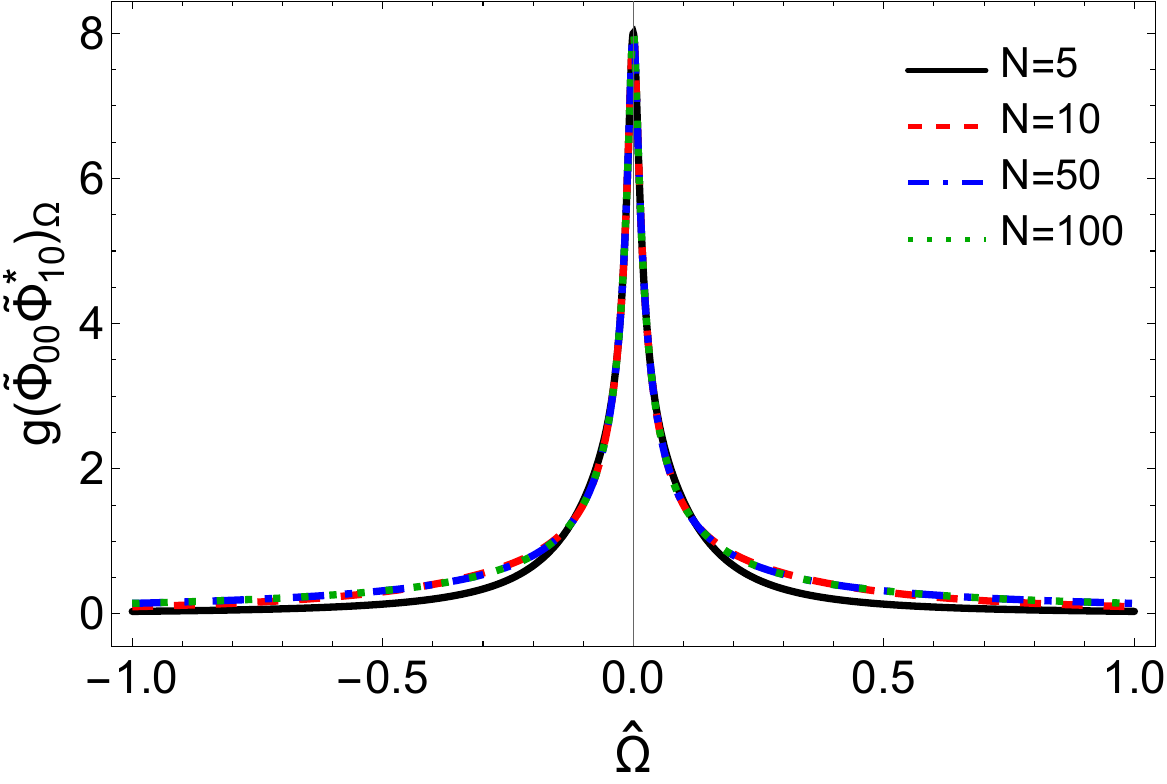}
\\
\includegraphics[width=0.43\textwidth]{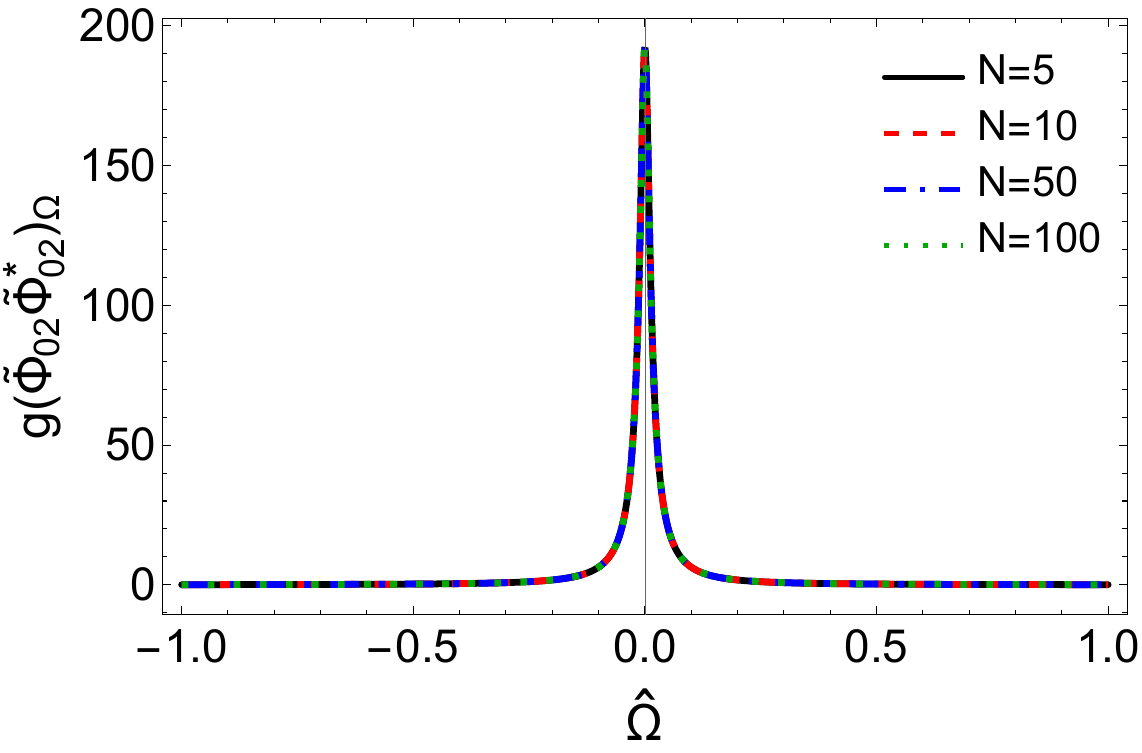}
\includegraphics[width=0.43\textwidth]{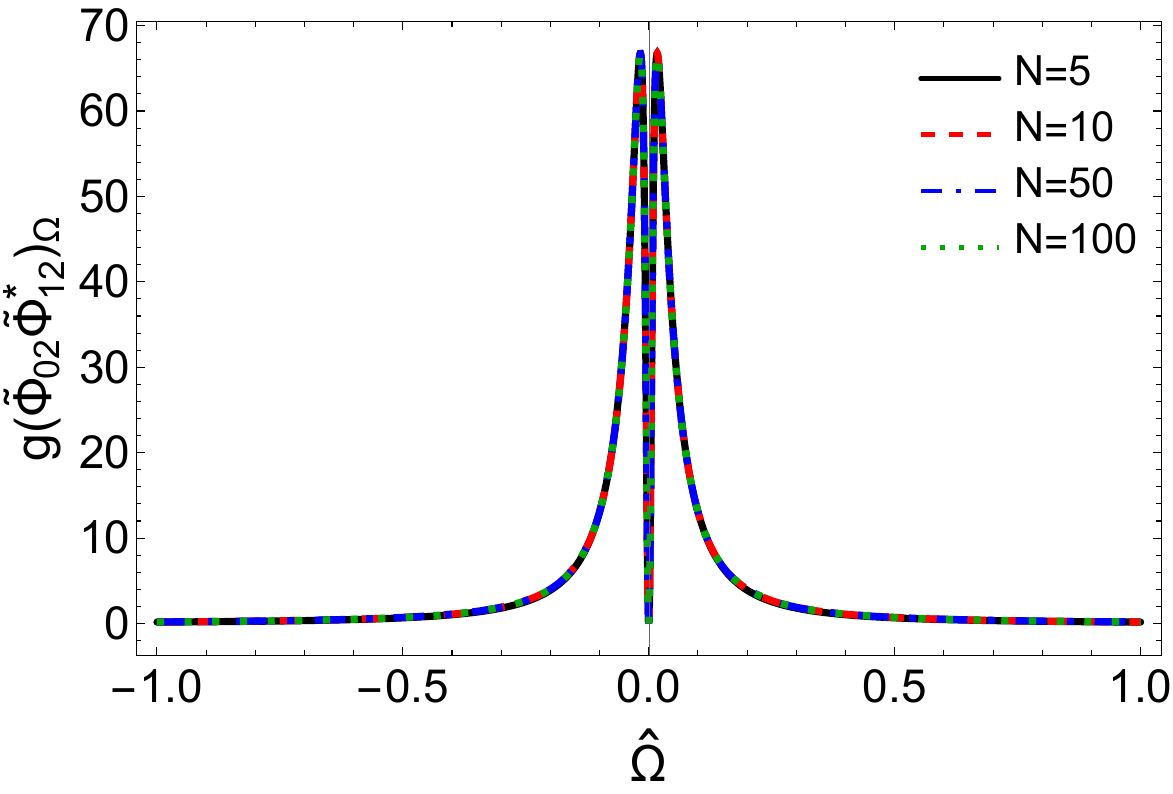}
 \caption{Convergence of the correlator amplitudes $\left( \widetilde{\Phi}_{n \ell} \widetilde{\Phi}_{n' \ell}  \right)_{\Omega}$ for successive truncations and with different values of $n$ and $n'$ in each panel. (Left-hand panels) Diagonal correlators. (Right-hand panels) Off-diagonal correlators.}
	\label{fig:convergence}
	\end{center}
\end{figure}

In order to check the convergence of the procedure outlined above, in Fig.~\ref{fig:convergence}, we display some $\left( \widetilde{\Phi}_{n \ell} \widetilde{\Phi}^{*}_{n'\ell}  \right)_{\Omega}$ correlator amplitudes as a function of $\widehat{\Omega}$ for various truncation orders, $N$. In general, we see that already for $N = 5$ the procedure yields results which are quantitatively close to the higher truncation orders. In the upper panels of Fig.~\ref{fig:convergence}, we display the scalar correlator amplitudes $\left( \widetilde{\Phi}_{00}
\widetilde{\Phi}_{00}^{*}
 \right)_{\Omega}$ and $\left( \widetilde{\Phi}_{00}
\widetilde{\Phi}_{10}^{*}
 \right)_{\Omega}$, which are related to the zero-mode Fourier-Laguerre components $\widetilde{\Phi}_{0}$ and $\widetilde{\Phi}_{1}$ of the linearized collision term and to the fluctuations in conserved currents (see Eqs.~\eqref{eq:relations-Phi-phys} and \eqref{eq:lag-fou-phi4-corr-final}). In the lower panels of Fig.~\ref{fig:convergence}, we display the rank-2 tensor correlator amplitudes $\left( \widetilde{\Phi}_{02}
\widetilde{\Phi}_{02}^{*}
 \right)_{\Omega}$ and $\left( \widetilde{\Phi}_{02}
\widetilde{\Phi}_{12}^{*}
 \right)_{\Omega}$, which are related to the Fourier-Laguerre components $\widetilde{\Phi}_{0}^{\mu \nu}$ and $\widetilde{\Phi}_{1}^{\mu \nu}$. We note that the former is related to shear fluctuations (see Eq.~\eqref{eq:shear-as-phi}) and the latter is a non-hydrodynamic fluctuation.  We see that the convergence of the scalar correlators is slightly slower with respect to the rank-2 tensor correlator amplitudes. Additionally, the diagonal ($n = n'$) correlators (left panels of Fig.~\ref{fig:convergence}) converge to a single peak structure centered at $\widehat{\Omega} = 0$, whereas the off-diagonal amplitudes (right panels of Fig.~\ref{fig:convergence}) can have a more complex structure. In fact, the amplitude $\left( \widetilde{\Phi}_{02}
\widetilde{\Phi}_{12}^{*}
 \right)_{\Omega}$ (lower-right panel) possesses a double-peak structure with a sharp valley near $\widehat{\Omega} = 0$. Moreover, off-diagonal correlators can become negative in certain regions, as it will be seen below.

\begin{figure}[!h]
\begin{center}
\includegraphics[width=0.48\textwidth]{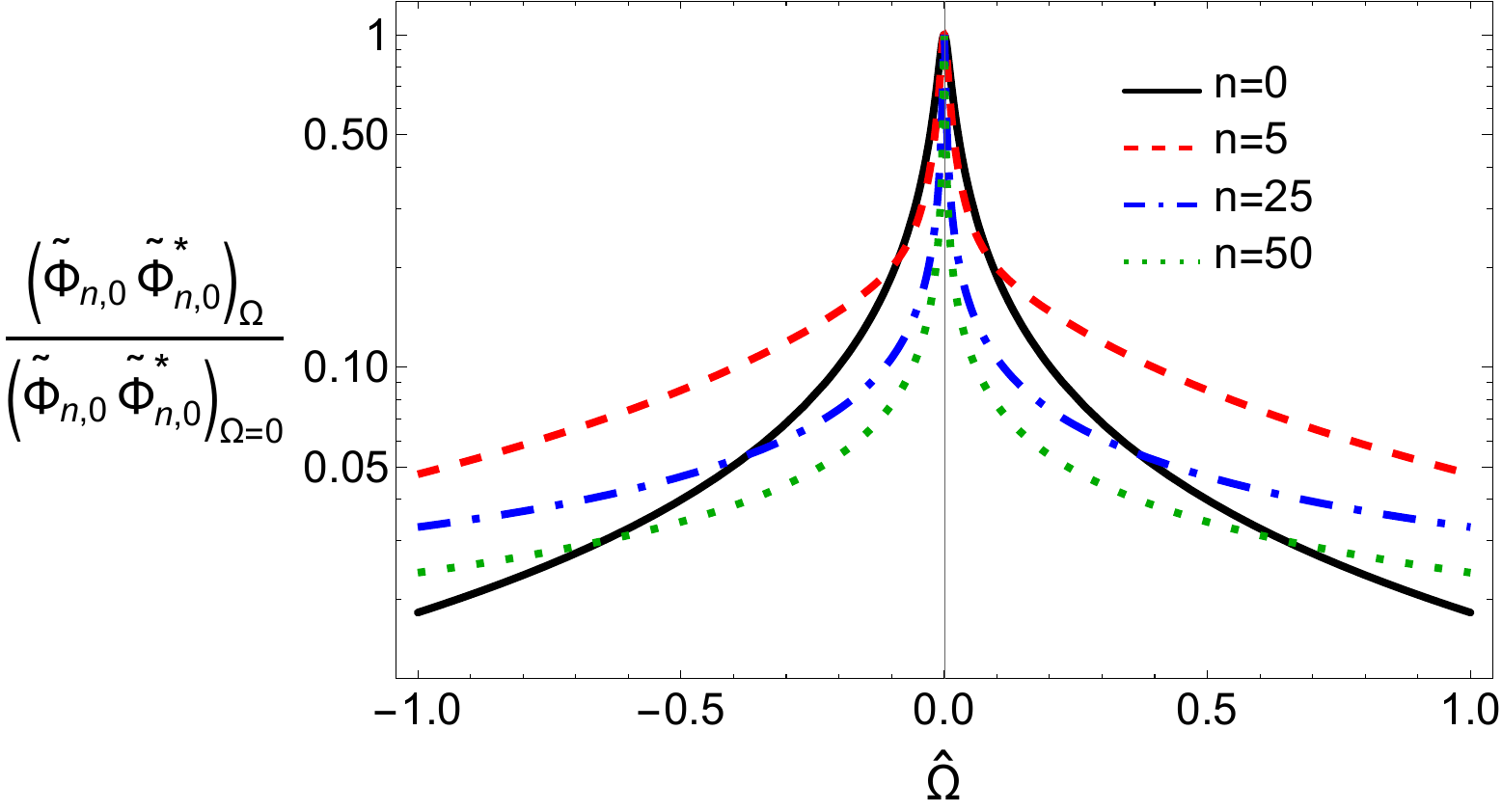}
\includegraphics[width=0.49\textwidth]{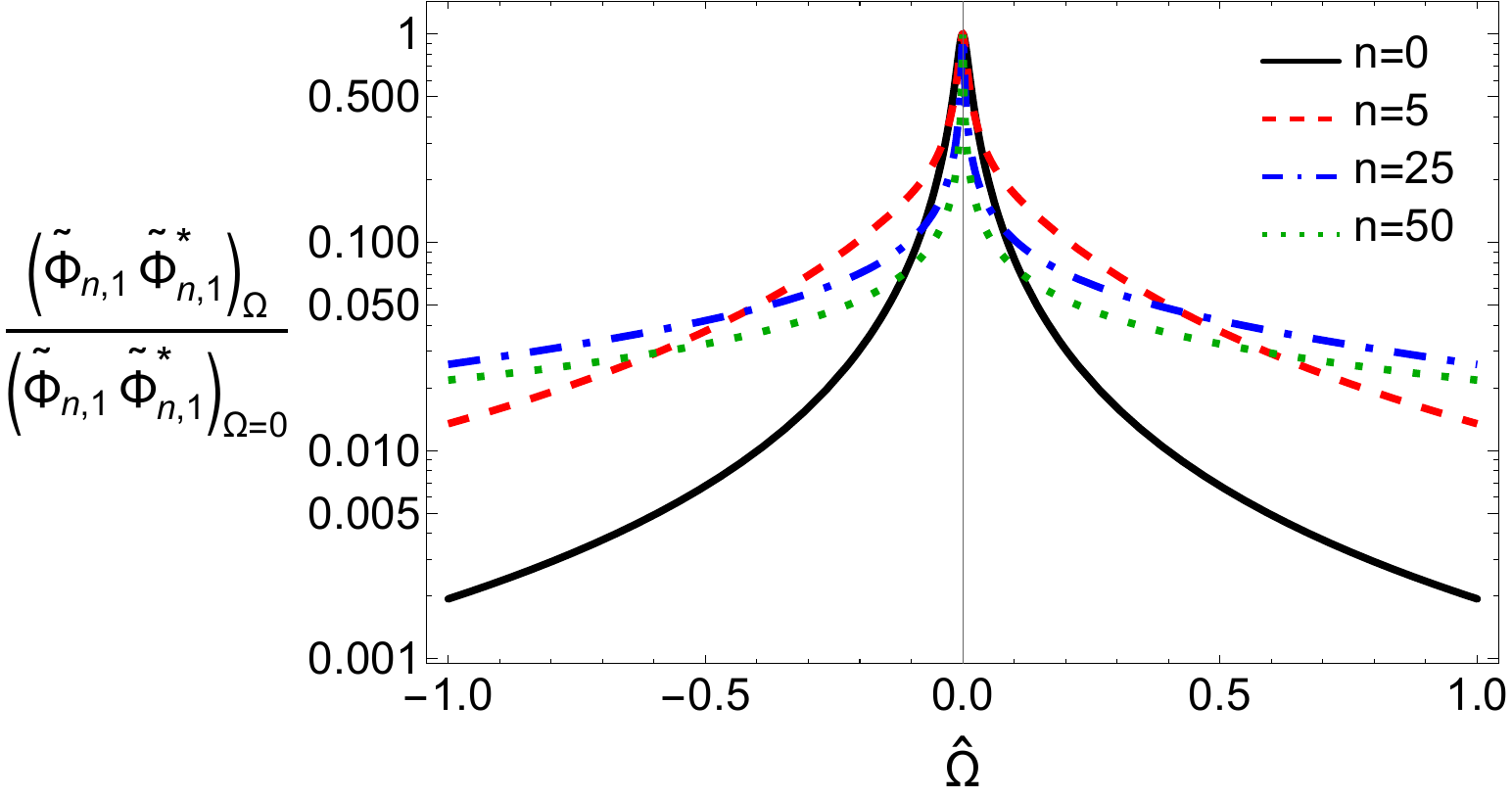}
 \caption{Diagonal eigenmode correlator amplitudes for  $N=125$. (Left-hand panel) $\ell = 0$ correlator amplitudes. (Right-hand panel) $\ell = 1$ correlator amplitudes.}
	\label{fig:various-corrs-diag}
	\end{center}
\end{figure}

\begin{figure}[!h]
\begin{center}
\includegraphics[width=0.45\textwidth]{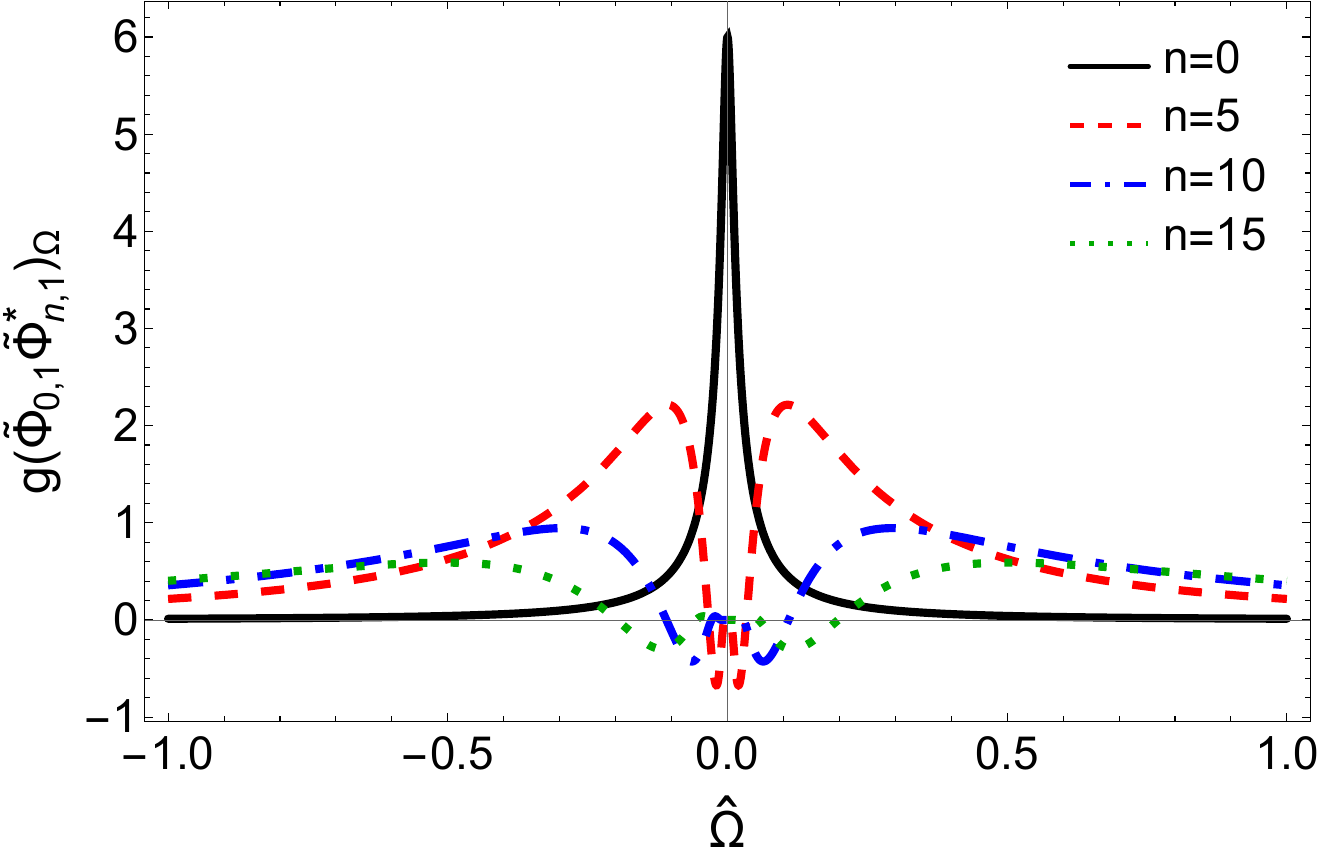}
\includegraphics[width=0.45\textwidth]{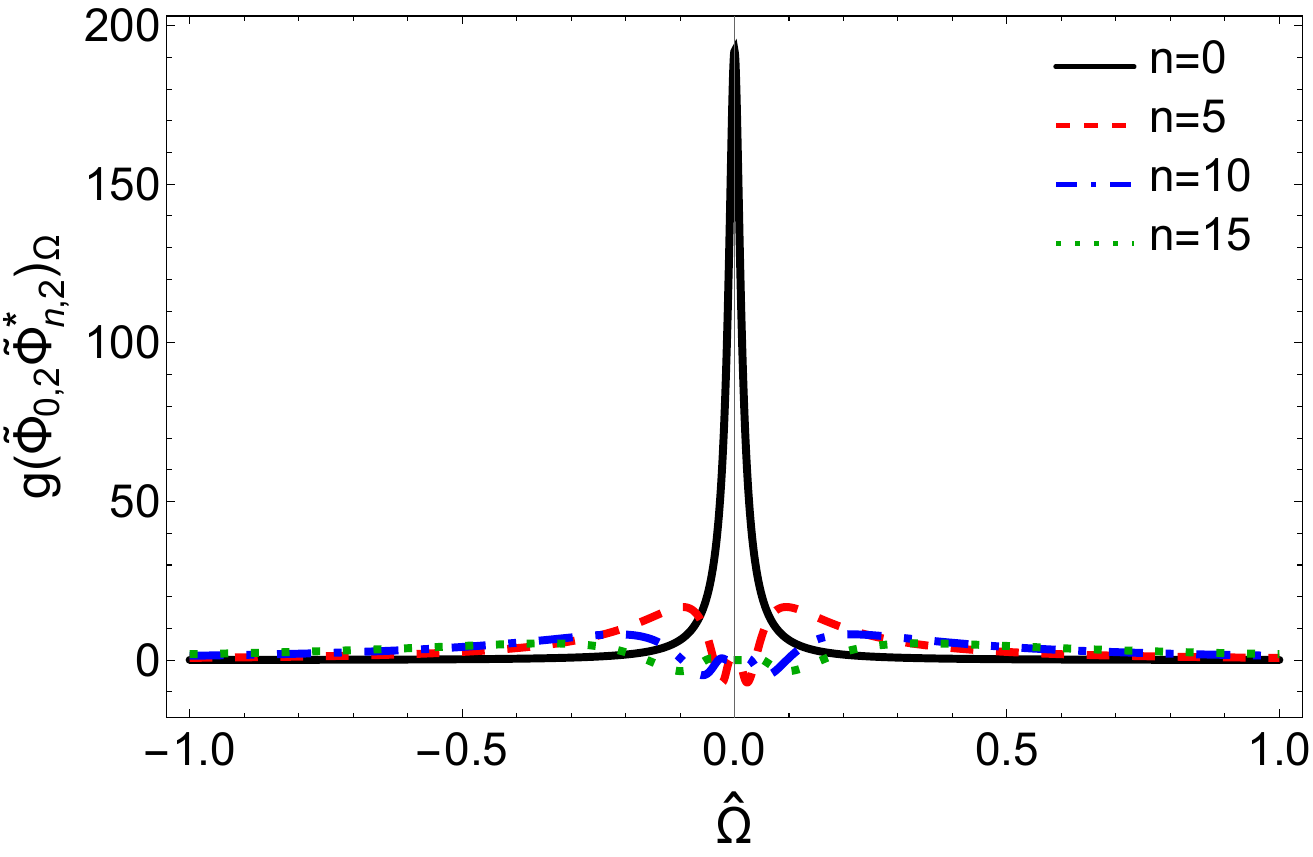}
 \caption{Eigenmode correlator amplitudes for  $N=125$. (Left-hand panel) $\ell = 1$ correlator amplitudes. (Right-hand panel) $\ell = 2$ correlator amplitudes.}
	\label{fig:various-corrs-off}
	\end{center}
\end{figure}

In Figures \ref{fig:various-corrs-diag} and \ref{fig:various-corrs-off} general aspects of the shape of the correlator amplitudes are assessed at fixed but large ($N=125$) truncation order. In Fig.~\ref{fig:various-corrs-diag}, we display some diagonal scalar ($\ell = 0$) and vector ($\ell = 1$) correlator amplitudes normalized with their values at $\widehat\Omega = 0$. It is seen that for small $|\widehat\Omega|$ the distribution becomes more sharply peaked as $n$ increases and it broadens for large $|\widehat\Omega|$.  For $\ell =1$, however, we see that the small $|\widehat\Omega|$ region where the amplitudes $\left( \widetilde{\Phi}_{n,1}
\widetilde{\Phi}_{n,1}^{*}
 \right)_{\Omega}$ become sharply peaked is much smaller than the $\ell = 0$ case. In Fig.~\ref{fig:various-corrs-off}, we see the behavior of the off-diagonal amplitudes $\left( \widetilde{\Phi}_{0,1}
\widetilde{\Phi}_{n,1}^{*}
 \right)_{\Omega}$ and $\left( \widetilde{\Phi}_{0,2}
\widetilde{\Phi}_{n,2}^{*}
 \right)_{\Omega}$. We remind that $\widetilde{\Phi}_{0,1}$ and $\widetilde{\Phi}_{0,2}$ can both be related to vector and rank-2 tensor fluctuations in conserved currents (see Eqs.~\eqref{eq:vectors-as-phi} and \eqref{eq:shear-as-phi}, respectively), whereas the remaining modes $\widetilde{\Phi}_{n,1}$ and $\widetilde{\Phi}_{n,2}$ are, in general, related non-hydrodynamic modes. We note that, overall, the higher the value of $n$, the smaller the typical value of the amplitude is. This is sensible because the more distant the modes are in the hierarchy of Laguerre moment equations \eqref{eq:moments-lag-main} at fixed $\ell$, the less correlated these modes are expected to be, since these equations couple modes with Laguerre index $n$ with $n \pm 1$. We also note that the typical value of $\left( \widetilde{\Phi}_{0,1}
\widetilde{\Phi}_{n,1}^{*}
 \right)_{\Omega}$ decays with $n$ much slower than $\left( \widetilde{\Phi}_{0,2}
\widetilde{\Phi}_{n,2}^{*}
 \right)_{\Omega}$. 

\newpage

Now, we turn our attention to correlators related to the deviations from the global equilibrium particle four-current and energy momentum tensor. As a matter of fact, inverting Eqs.~\eqref{eq:shear-as-phi}--\eqref{eq:relations-Phi-phys}  these correlators can be readily expressed in terms of the Fourier-Laguerre correlators $\left\langle \widetilde{\Phi}_{n}^{\mu_{1} \cdots \mu_{\ell}}
\widetilde{\Phi}_{n'}^{*\mu_{1} \cdots \mu_{m}}
 \right\rangle$. Indeed,
\begin{equation}
\label{eq:scalar-Tmunu-Nmu-corrs}
\begin{aligned}
\beta^{2} \langle \delta n (\Omega) \delta n^{*}(\Omega')\rangle 
& 
= 
4
\left\langle \widetilde{\Phi}_{0}
\widetilde{\Phi}_{0}^{*}
 \right\rangle
-
4
\left\langle \widetilde{\Phi}_{0}
\widetilde{\Phi}_{1}^{*}\right\rangle
+
\left\langle \widetilde{\Phi}_{1}
\widetilde{\Phi}_{1}^{*}
 \right\rangle,
 \\
\beta^{3} \langle \delta n (\Omega) \delta \varepsilon^{*}(\Omega')\rangle 
& =
4
\left\langle \widetilde{\Phi}_{0}
\widetilde{\Phi}_{2}^{*}
 \right\rangle
 +
12 
\left\langle \widetilde{\Phi}_{0}
\widetilde{\Phi}_{0}^{*}
 \right\rangle
-
18 \left\langle \widetilde{\Phi}_{0}
\widetilde{\Phi}_{1}^{*}
 \right\rangle
-2 \left\langle \widetilde{\Phi}_{1}
\widetilde{\Phi}_{2}^{*}  \right\rangle
+ 
6 \left\langle \widetilde{\Phi}_{1}
\widetilde{\Phi}_{1}^{*}
 \right\rangle,
 \\
\beta^{4} \langle \delta \varepsilon (\Omega) \delta \varepsilon^{*}(\Omega')\rangle 
&=
36 \left\langle \widetilde{\Phi}_{0}
\widetilde{\Phi}_{0}^{*}
 \right\rangle
+
36 \left\langle \widetilde{\Phi}_{1}
\widetilde{\Phi}_{1}^{*} \right\rangle
+
4
\left\langle \widetilde{\Phi}_{2}
\widetilde{\Phi}_{2}^{*} \right\rangle
+
24
\left\langle \widetilde{\Phi}_{2}
\widetilde{\Phi}_{0}^{*} \right\rangle
-
24
\left\langle \widetilde{\Phi}_{2}
\widetilde{\Phi}_{1}^{*} \right\rangle
-
72
\left\langle \widetilde{\Phi}_{0}
\widetilde{\Phi}_{1}^{*} \right\rangle, \\
\left\langle \nu^{\alpha}
\nu^{*\beta} \right\rangle
&=
\left\langle \widetilde{\Phi}_{0}^{\alpha} 
\widetilde{\Phi}_{0}^{*\beta}
\right\rangle,
\\
\beta\left\langle \nu^{\alpha}
h^{*\beta} \right\rangle
&
=
4
\left\langle \widetilde{\Phi}_{0}^{\alpha} 
\widetilde{\Phi}_{0}^{*\beta}
\right\rangle
-
\left\langle \widetilde{\Phi}_{0}^{\alpha} 
\widetilde{\Phi}_{1}^{*\beta}
\right\rangle
,
\\
\beta^{2}\left\langle h^{\alpha}
h^{*\beta} \right\rangle
&
=
16
\left\langle \widetilde{\Phi}_{0}^{\alpha} 
\widetilde{\Phi}_{0}^{*\beta}
\right\rangle
-
8
\left\langle \widetilde{\Phi}_{0}^{\alpha} 
\widetilde{\Phi}_{1}^{*\beta}
\right\rangle
+
\left\langle \widetilde{\Phi}_{1}^{\alpha} 
\widetilde{\Phi}_{1}^{*\beta}
\right\rangle
,
\\
\left\langle \pi^{\mu \nu} \pi^{*\alpha\beta} \right\rangle
&=
\left\langle  \widetilde{\Phi}_{0}^{\mu \nu}
\widetilde{\Phi}_{0}^{*\alpha \beta}
\right\rangle,
\end{aligned}    
\end{equation}
where we note that many of the above correlators vanish. First, as a consequence of Eq.~\eqref{eq:moments-lag-homog-b}, correlators between currents with different tensor ranks ($\ell \neq m$ in Eq.~\eqref{eq:moments-lag-homog-b}) are zero. Second, some correlators are zero due to the local conservation laws, expressed in Fourier space as $i q_{\mu} \delta J^{\mu} = 0$ and $i q_{\nu} \delta T^{\mu \nu}u_{\nu} = 0$. Then, since $q_{\mu} = i \Omega u_{\mu} + q_{\langle \mu \rangle}$, we have that $i \Omega \delta n = 0$ and $i \Omega (\delta \varepsilon u^{\mu} + h^{\mu}) = 0$, in the homogeneous limit. Contracting the latter with $u_{\nu}$ and $\Delta_{\nu}^{\alpha}$, we have, respectively, $\delta \varepsilon = 0 = h^{\alpha}$. Thus, correlators involving $\delta n$, $\delta \varepsilon$ and $h^{\mu}$ vanish in the homogeneous limit, and this fact was confirmed numerically. For the non-vanishing correlators in Eqs.~\eqref{eq:scalar-Tmunu-Nmu-corrs}, we have
\begin{equation}
\label{eq:diffusion-shear-amplitude}
\begin{aligned}
\left\langle \nu^{\alpha}
\nu^{*\beta} \right\rangle
&\equiv
-\frac{(2 \pi)^{4}}{\beta^{2}}\left( \nu
\nu^{*} \right)_{\Omega, \mathrm{kt}}
\delta(\Omega-\Omega') \Delta^{\alpha \beta},
\\
\left\langle \pi^{\mu \nu} \pi^{*\alpha\beta} \right\rangle
&
\equiv
\frac{(2 \pi)^{4}}{\beta^{4}}
\left( \pi \pi^{*} \right)_{\Omega, \mathrm{kt}}
\delta(\Omega-\Omega') \Delta^{\mu \nu \alpha \beta},
\end{aligned}    
\end{equation}
where the subscript $\mathrm{kt}$ denotes that the amplitudes stem from kinetic theory. This notation will be useful for the comparison with fluctuating Israel-Stewart below. Besides, from Eqs.~\eqref{eq:lag-fou-phi4-corr-final} and \eqref{eq:scalar-Tmunu-Nmu-corrs}, we find that $\left( \nu
\nu^{*} \right)_{\Omega, \mathrm{kt}} = \left( \widetilde{\Phi}_{0,1}
\widetilde{\Phi}_{0,1}^{*}
 \right)_{\Omega}$ and  
$\left( \pi \pi^{*} \right)_{\Omega, \mathrm{kt}} = \left( \widetilde{\Phi}_{0,2}
\widetilde{\Phi}_{0,2}^{*}
 \right)_{\Omega}$. And thus the behavior of these amplitudes with $\widehat{\Omega}$ can be seen in Figs.~\ref{fig:various-corrs-diag} and \ref{fig:various-corrs-off}.

\begin{figure}[!h]
\begin{center}
\includegraphics[width=0.45\textwidth]{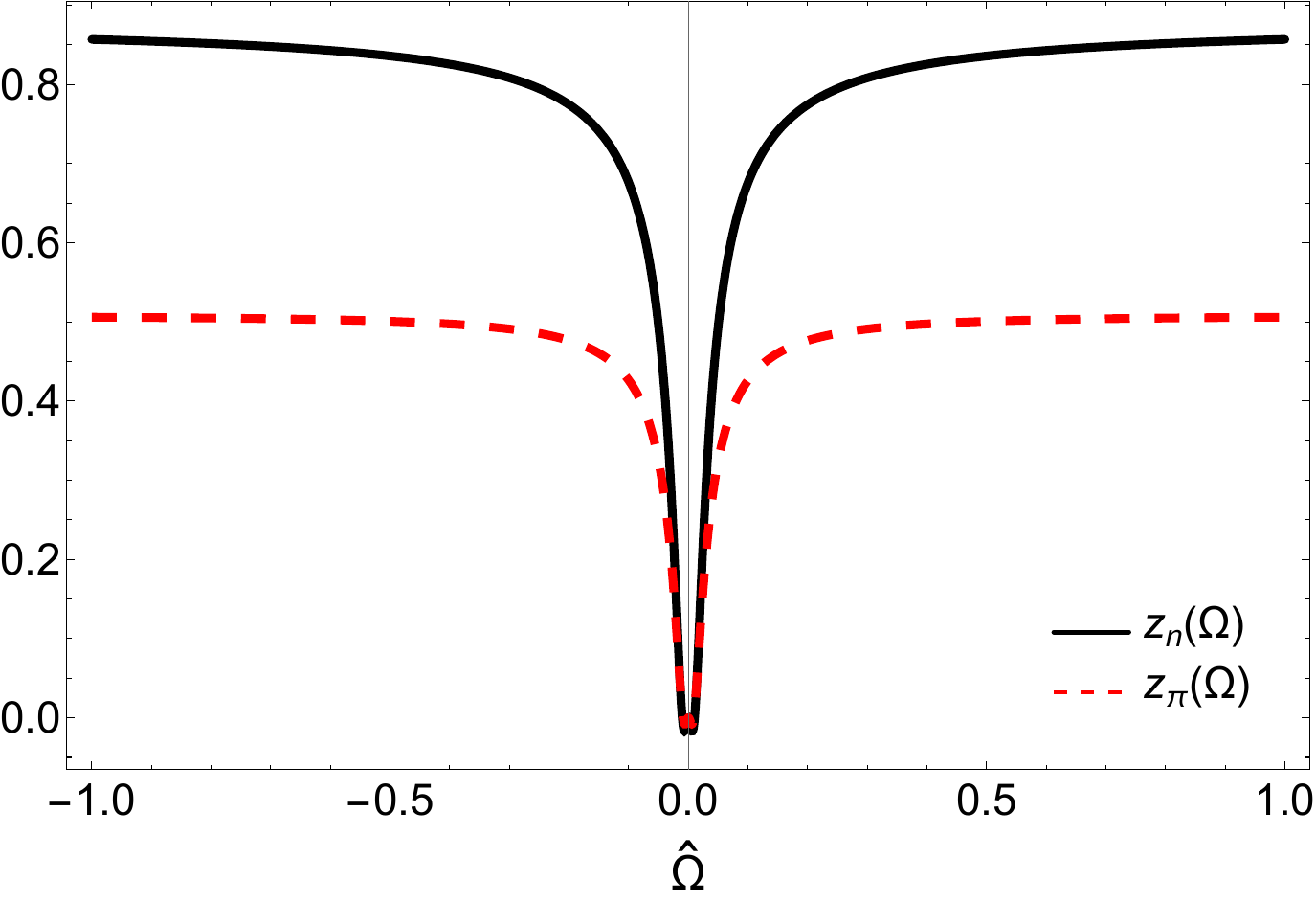}
 \caption{Relative difference between hydrodynamic and kinetic theory fluctuation amplitudes for particle diffusion and the shear-stress tensor (see definition in Eq.~\eqref{eq:relat-corr-ampl}).}
	\label{fig:phi4-IS-vs-kin}
	\end{center}
\end{figure}

Now we compare the amplitudes in Eq.~\eqref{eq:diffusion-shear-amplitude} with their counterparts computed within Israel-Stewart hydrodynamics. Then, analogously with subsection ~\ref{subsec:fluct-trans-hydro}, we define
\begin{equation}
\label{eq:relat-corr-ampl}
\begin{aligned}
&
z_{n}(\Omega) = \frac{\left( \nu \nu^{*} \right)_{\Omega, \mathrm{kt}} - \left( \nu \nu^{*} \right)_{\Omega, \mathrm{IS}}}{\left( \nu \nu^{*} \right)_{\Omega, \mathrm{kt}}},
\quad
z_{\pi}(\Omega) = \frac{\left( \pi \pi^{*} \right)_{\Omega, \mathrm{kt}} - \left( \pi \pi^{*} \right)_{\Omega, \mathrm{IS}}}{\left( \pi \pi^{*} \right)_{\Omega, \mathrm{kt}}}
\end{aligned}    
\end{equation}
where $\left( \nu \nu^{*} \right)_{\Omega, \mathrm{IS}}$ and $\left( \pi \pi^{*} \right)_{\Omega, \mathrm{IS}}$ denote Israel-Stewart correlator amplitudes, 
\begin{equation}
\begin{aligned}
&
\left( \nu \nu^{*} \right)_{\Omega, \mathrm{IS}} 
=
\frac{2 \kappa \beta^{2}}{1 + \tau_{n}^{2} \Omega^{2}},
\\
&
\left( \pi \pi^{*} \right)_{\Omega, \mathrm{IS}} 
=
\frac{4 \eta \beta^{3}}{1 + \tau_{\pi}^{2} \Omega^{2}},
\end{aligned}    
\end{equation}
which can be obtained from the results of Ref.~\cite{Mullins:2023tjg}. For the values of the transport coefficients, we consider $\kappa = 3/(g \beta^{2})$, $\tau_{n} = 60/(g n_{\mathrm{eq}} \beta^{2})$, $\eta = 48/(g \beta^{3})$, and $\tau_{\pi} = 72/(g n_{\mathrm{eq}} \beta^{2})$ computed in Ref.~\cite{Rocha:2023hts} for the $\varphi^{4}$ theory employing the IReD truncation scheme. Then, we see in Fig.~\ref{fig:phi4-IS-vs-kin} that for that for $\widehat{\Omega} = 0$, $z_{n}(\Omega) = 0 = z_{\pi}(\Omega)$, thus hydrodynamic fluctuations coincide with the kinetic theory fluctuations in this regime. This result is sensible given the fact that the hydrodynamic behavior should emerge in the limit of small frequencies. Indeed, in Ref.~\cite{Rocha:2024cge} the shear viscosity and shear relaxation time displayed above for $\varphi^{4}$ theory emerge from the full linear response function by expanding it in the  $\widehat{\Omega} \to 0$ limit. For larger values of $|\widehat{\Omega}|$, we see that both $z_{n}(\Omega)$ and $z_{\pi}(\Omega)$ increase, but  they always range values below $1$, as it happened in subsection ~\ref{subsec:fluct-trans-hydro} for fixed time fluctuations. But here they reach values that are larger than the values $z_{n} = 10 \%$ and $z_{\pi} = 17 \%$ reported there. We also remark that $z_{n}(\Omega) \approx z_{\pi}(\Omega)$ in a region around $\widehat{\Omega} = 0$, but, in general, $z_{n}(\Omega) > z_{\pi}(\Omega)$ and thus the particle diffusion current hydrodynamic fluctuations departs farther from the kinetic theory fluctuations when compared with the shear-stress fluctuations. 

\begin{figure}[!h]
\begin{center}
\includegraphics[width=0.4\textwidth]{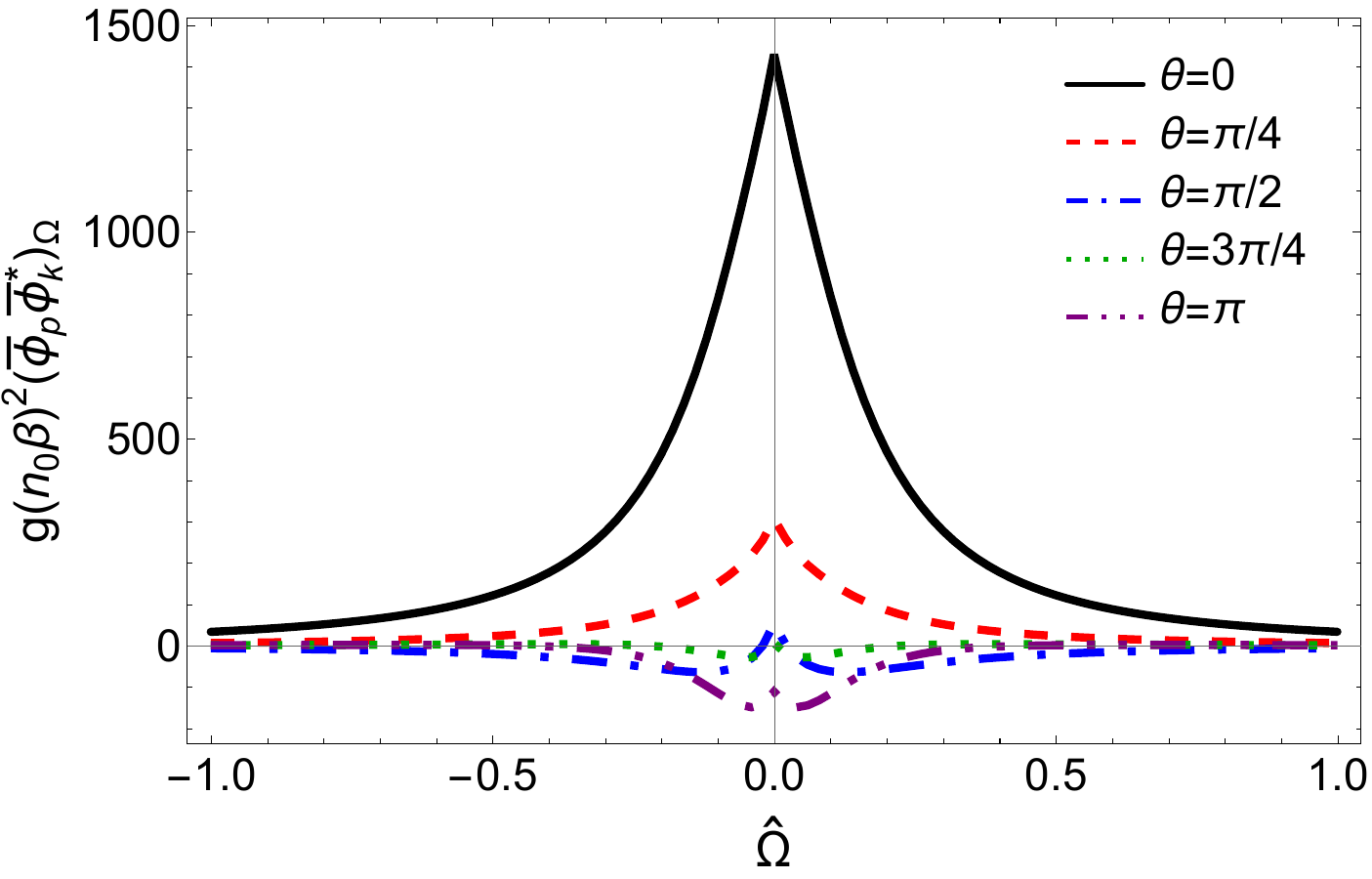}
\includegraphics[width=0.4\textwidth]{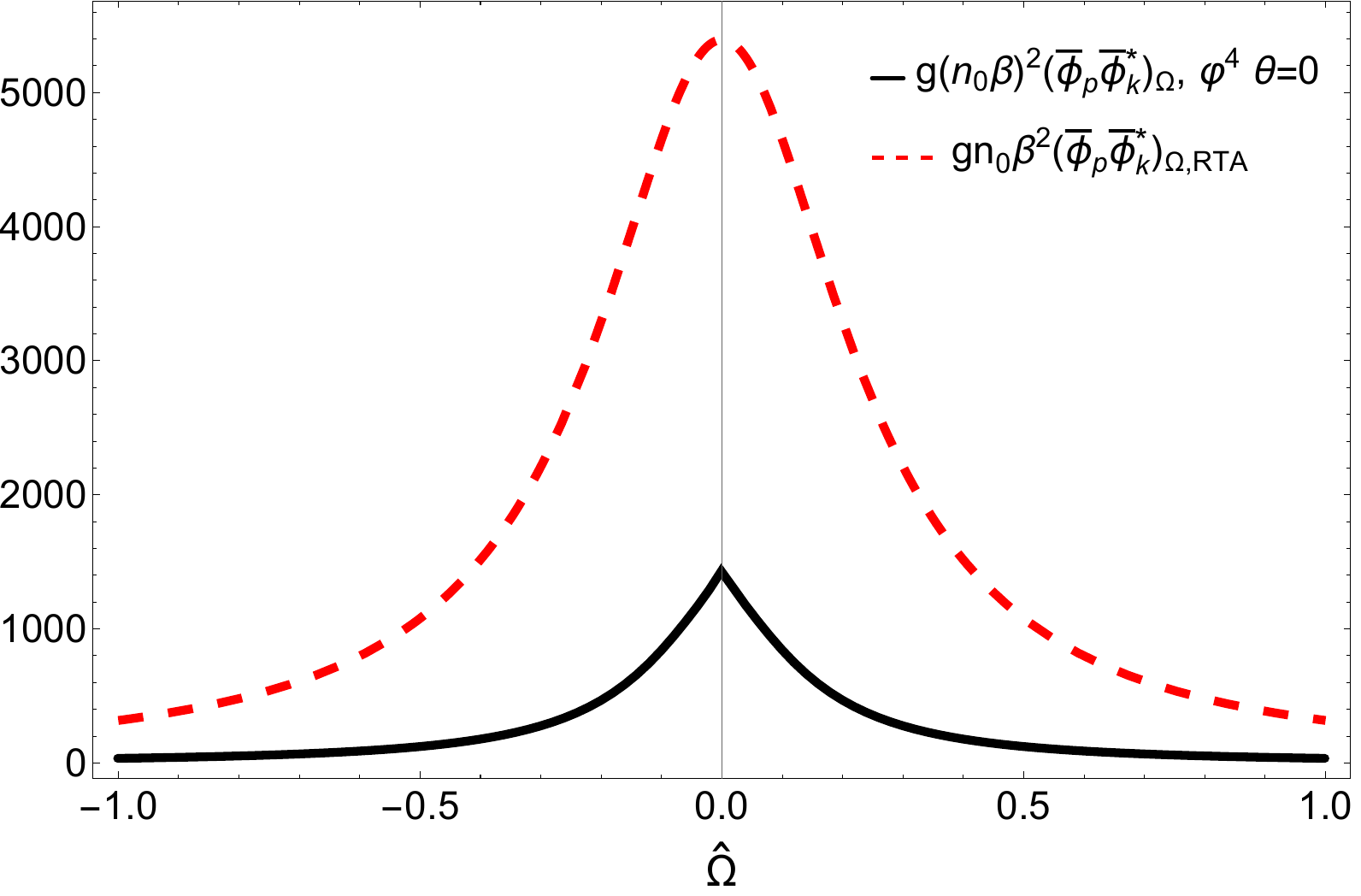}
 \caption{(Left-hand panel) Fractional deviation correlator amplitudes as a function of $\widehat{\Omega}$  for some values of the angle $\theta \equiv \arccos[(\mathbf{p} \cdot \mathbf{k})/(\vert \mathbf{p} \vert \vert \mathbf{k} \vert)]$ for $\beta\vert \mathbf{p} \vert = 1$, $\beta\vert \mathbf{k} \vert = 1$. (Right-hand panel) Comparison with normalized RTA correlation amplitudes.}
	\label{fig:phi4-frac-dev-corr}
	\end{center}
\end{figure}

Having the correlators $\left\langle \widetilde{\Phi}^{\mu_{1} \cdots \mu_{\ell}}_{n}(\Omega)
\widetilde{\Phi}_{n'}^{*\mu_{1} \cdots \mu_{m}}(\Omega')
 \right\rangle$, one can also recover the full fractional deviation function correlators $\langle \Bar{\phi}_{p}(q) \Bar{\phi}^{*}_{k}(q')  \rangle$. As a matter of fact, using Eqs.~\eqref{eq:phi-phi-correlators} and \eqref{eq:lag-fou-phi4-corr-final} we obtain (details also in Appendix \ref{sec:F-moms})
\begin{equation}
\label{eq:phi-phi-correlators-2}
\begin{aligned}
&
\langle \Bar{\phi}_{p}(q) \Bar{\phi}^{*}_{k}(q')  \rangle 
=
\frac{(2 \pi)^{4}}
{n_{0}^{2}\beta^{2}}
\sum_{n,n',\ell=0}^{\infty}  \frac{1}{\widehat{A}_{n}^{(\ell)} \widehat{A}_{n'}^{(\ell)} } 
\left( \widetilde{\Phi}_{n\ell}
\widetilde{\Phi}_{n' \ell}^{*}
 \right)_{\Omega}   L^{(2 \ell + 1)}_{n, p}
L^{(2 \ell + 1)}_{n', k}
 (\beta \vert \mathbf{p} \vert)^{\ell}
 (\beta \vert \mathbf{k} \vert)^{\ell} 
 \frac{\ell!}{(2\ell -1)!!}
P_{\ell}(\cos \theta)
 \delta(\Omega - \Omega')
\\
&
\equiv
(2 \pi)^{4}
\left( \Bar{\phi}_{p} \Bar{\phi}^{*}_{k}  \right)_{\Omega}  \delta(\Omega - \Omega'),
\end{aligned}    
\end{equation}
where $P_{\ell}(x)$ are the Legendre polynomials and $\cos \theta \equiv (\mathbf{p} \cdot \mathbf{k})/(\vert \mathbf{p} \vert \vert \mathbf{k} \vert)$. In the left-hand panel of Fig.~\ref{fig:phi4-frac-dev-corr}, we display the behavior of the above-defined correlator amplitude with $\beta\vert \mathbf{p} \vert = 1$, $\beta\vert \mathbf{k} \vert = 1$ and truncate the above sum in $n$ and $n'$ at $N=15$ and the sum in $\ell$ at $\ell_{\mathrm{max}} = 10$. There, we see that, as expected, the amplitude is largest for $\theta = 0$ for all values of $\widehat\Omega$. Then, it decreases for larger angles until for $\theta = \pi$ the amplitude becomes fully negative. The amplitude becomes very small for $\theta = 3\pi/4$ in comparison with the other angles displayed.

Finally, we compare the $\lambda \varphi^{4}$ correlator amplitudes with the amplitude computed in Sec.~\ref{sec:corr-fourier-RTA} for diluted solution RTA model. To this end, we express the result in Eq.~\eqref{eq:phi-corr-RTA-final} as (in the homogeneous limit, $q^{\langle \mu \rangle} = (0,\textbf{q}) = (0, \mathbf{0})$ and $\Omega = q^{0}$ in the local rest frame) 
\begin{equation}
\label{eq:phi-corr-RTA-final-2}
    (\phi_p\phi_k)_q \equiv (\phi_p\phi_k)_{\Omega,\mathrm{RTA}}
    \delta^3(\textbf{p}-\textbf{k})
    =
    \dfrac{(2\pi)^3}{f_{\text{eq},p}} \, \dfrac{2\tau_p \delta^3(\textbf{p}-\textbf{k})}{1+\tau^2_p \Omega^2}  \, ,
\end{equation}
where we employ $\tau_p = [4/(g n_{\mathrm{eq}} \beta^{2})] E_{p}$, which is the momentum-dependent RTA timescale associated to the $\lambda \varphi^{4}$ interaction, computed in Ref.~\cite{Denicol:2022bsq}. By comparing the above equation with Eq.~\eqref{eq:phi-phi-correlators-2}, we already see that the RTA amplitude strictly enforces that $\mathbf{p}$ and $\mathbf{k}$ are identical, thus $\theta = 0$, while the full $\lambda \varphi^{4}$ amplitude entails a non-trivial distribution both in $\theta$ and in the vector magnitudes $\vert \mathbf{p} \vert$ and $\vert \mathbf{k} \vert$. Besides, in the left-hand panel of Fig.~\ref{fig:phi4-frac-dev-corr}, we see that the normalized fluctuation amplitude of RTA is qualitatively different from the one of the full $\lambda \varphi^{4}$ interaction in the linear regime. Thus, we see that the RTA model cannot accurately represent the complete $\varphi^{4}$ fluctuations in this regime. This discrepancy is increased by the fact that the RTA model employed to compute Eq.~\eqref{eq:phi-corr-RTA-final} does not conserve the solute particle four-current nor their energy momentum tensor, while the $\lambda \varphi^{4}$ model conserves these quantities (at leading order in the coupling constant). Indeed, the RTA collision kernel obtained in Ref.~\cite{Denicol:2022bsq} from the spectral expansion of $\Hat{L}$ contains counter-terms that are essential for the local conservation laws \cite{Rocha:2021zcw}.

\section{Conclusions}

It is well known that any dissipative system will also experience fluctuations around the equilibrium state. While these fluctuations are most often studied for macroscopic systems such as Brownian particles \cite{Dunkel:2008ngc} and viscous fluids \cite{landau_statistical_part_II}, they will also appear in any dissipative microscopic theory. This leads naturally to kinetic theory, which produces entropy according to Boltzmann's H-theorem. Using thermodynamic arguments, made consistent with relativity through the so-called information current \cite{Gavassino:2021kjm}, we constructed both the equal-time and dynamical correlators for the distribution function of relativistic kinetic theory. It is found that the linearized relativistic Boltzmann equation can be obtained using only the information current (or equivalently the entropy current and conserved currents) and entropy production rate, as in theorem 1 of \cite{Gavassino:2023odx} when the antisymmetric term vanishes. 
This means that, at a linear level, the correlators stemming from kinetic theory arise from purely thermodynamic considerations, just like the ones in Israel-Stewart theory.

The fully nonlinear information current of equation \eqref{phuiummum} is manifestly timelike future-directed, vanishes only in equilibrium, and has non-positive divergence due to the H-theorem. This implies that the resulting free energy is thermodynamically consistent, and it plays the role of a Lyapunov functional for the Boltzmann dynamics. Therefore, kinetic theory is stable \cite{Gavassino:2021kjm} in the fully non-linear regime. Furthermore, following the argument of \cite{Gavassino:2021kjm}, we conclude that non-linear deviations from equilibrium cannot expand faster than light (which in the linear regime implies causality). This provides a simple physical perspective on existing stability and causality proofs for the relativistic Boltzmann equation \cite{1973CMaPh..33...83B}. Due to the symmetric hyperbolic nature of the linear Boltzmann equation (when truncated to a finite but arbitrarily large number of moments), these results also imply the existence of solutions for certain initial data \cite{Kato1975TheCP}.

Since the information current can be constructed directly from the entropy current and conserved currents, without any reference to the collision operator, spacelike-separated correlation functions do not depend on the interactions. This allows for several important universal observations to be made about fluctuations in relativistic kinetic systems. By studying the properties of the nonlinear equilibrium probability distribution as the number of particles is reduced, we found that fluctuations become large when there are $\sim$25 particles in a given fluid cell. This indicates that fluctuations should be significant for small systems in heavy-ion collisions when there can be as few as $\sim$5 charged particles in a given fluid cell \cite{Noronha:2024dtq}. Working from equal-time correlation functions in the distribution function, correlation functions involving a conserved current and energy-momentum tensor were determined. By demanding that the equal-time correlation functions of Israel-Stewart match those of kinetic theory, the relaxation times can be identified \emph{in a way that is independent of the interactions}. This is the choice of relaxation time that maximizes the fluctuations of Israel-Stewart, while in general kinetic theory has a more complete high-frequency spectrum allowing it to fluctuate more. 

At non-equal times (or more precisely, for timelike separated spacetime points) we presented two methods of determining correlation functions. The first is to solve the equations of motion using the equal-time correlation functions as initial conditions. This approach was used to study fluctuations of a dilute solution in the relaxation-time approximation, in which case the correlation function of the distribution function was determined analytically. It is found that the correlation function propagates in time with an exact velocity given by the momentum. Using this result, we solved for the correlation function of the density of the solute. With a constant relaxation time, this correlation function decays exponentially with time; whereas with a relaxation time that is linear in $p^0$, the correlation function decays with the exponential of the square root of the time. This subexponential decay appears due to gapless modes that appear in the spectrum of the momentum dependent relaxation time. The other approach for determining correlation functions is the relativistic Fox-Uhlenbeck procedure used in \cite{Mullins:2023tjg, Mullins:2023ott, Gavassino:2024vyu}. In this procedure, a stochastic source is included in the equations of motion, and the distribution of this source is determined such that the system relaxes to the correct equilibrium probability distribution. In the case of the linearized Boltzmann equation, the correlator of this stochastic source is covariantly white and proportional to the kernel of the collision operator.

Using the exact spectrum of the linearized collision operator for weakly interacting ultra-relativistic scalar particles with a $\lambda \varphi^4$ interaction \cite{Mullins:2022fbx, Denicol:2022bsq, Rocha:2023hts, deBrito:2023vzv}, dynamical correlation functions of the linearized Boltzmann equation were studied. These correlation functions can be expanded in terms of the eigenfunctions of the linearized collision operator, reducing the problem to that of inverting a tensor equation for the coefficients of this expansion. In the homogenous case, this becomes a matrix expression that can be inverted numerically. The self-correlation functions are sharply peaked at zero frequency and decay as the magnitude of the frequency increases. Correlation functions involving higher-order moments decay at a faster rate near zero, but at large $|\Omega|$ the higher order moments dominate. Unlike in the relaxation-time approximation, correlation functions of the distribution function, $\langle \bar{\phi}_p(q) \bar{\phi}_k^*(q') \rangle$, have a three dimensional structure, depending on the angle between $\mathbf{p}$ and $\mathbf{k}$. The correlation functions are maximized when this angle is zero; however, as the momenta become anti-parallel, it was found that the correlation function eventually becomes negative. It follows that the magnitude of correlations is not minimized when the momenta are anti-parallel, but rather at some intermediate point. At zero frequency the correlation functions of conserved currents exactly match those of Israel-Stewart theory, while at non-zero frequencies Israel-Stewart under-predicts the magnitude of correlations (using transport coefficients obtained in the IReD scheme in \cite{Rocha:2023hts}). 

These results highlight the utility of the information current. By formulating the Boltzmann equation through its relativistic free energy, causality, stability, and symmetric hyperbolicity are guaranteed (when truncated to finite moments). The Gibbs stability implied by the existence of a well-defined information current further ensures that the system is stable against small perturbations such as those generated by thermal fluctuations. In this sense then, using the information current as the basis for a stochastic theory ensures that the resulting dynamics are well-posed. By simply using natural definitions of the entropy current and conserved current from a distribution function, and assuming the H-theorem is satisfied, the information current can be used to recover the linearized Boltzmann equation. Since the linearized Boltzmann equation can be obtained directly from the information current, it follows that the correlation functions of the distribution function do not depend on the definition of equal-time \cite{Mullins:2023tjg}. We can then think of the linearized Boltzmann equation as being the most general well-posed equation of motion for a distribution function, in the linear regime. The only piece of information that is missing in this construction is the form of the kernel, which is determined by the interactions. Unlike the standard derivation of the Boltzmann equation, starting from the information current also fully defines the corresponding stochastic theory in a natural way. 

For applications to heavy-ion collisions, these results highlight the importance of fluctuations for modeling small-systems such as the ones emerging e.g.~from proton-gold collisions \cite{Noronha:2024dtq}. There have been some recent efforts to include fluctuations in numerical simulations of hydrodynamic systems through 
Monte Carlo techniques \cite{Florio:2021jlx, Florio:2023kmy, Chattopadhyay:2023jfm, Basar:2024qxd}, but state-of-the-art simulations do not include these effects. In small systems, where the system lifetime is short, it is important that such modeling accounts for far-from-equilibrium behavior. Fluctuating kinetic theory provides an exemplary means to understand the behavior of thermal noise far-from-equilibrium and the corresponding impact in small systems, which can be explored in further works.

The same ideas that we have used to include fluctuations in kinetic theory have also been used to study fluctuating hydrodynamics \cite{Mullins:2023tjg, Mullins:2023ott, Gavassino:2024vyu}. However, the connection between fluctuating kinetic theory and hydrodynamics has not yet been explored from this perspective. Do the standard procedures of deriving hydrodynamics from kinetic theory also yield a well-defined stochastic theory? We intend to answer this question in future works.

\section*{Acknowledgements}

G.S.R. is supported by Vanderbilt University and in part by the U.S. Department of Energy, Office of Science under Award Number DE-SC-002434. NM is partially supported by the U.S. Department of Energy, Office of Science, Office for Nuclear Physics under Award No. DE-SC0023861. L.G. is partially supported by a Vanderbilt Seeding Success Grant.

\appendix






\section{Information current for quantum statistics and massive particles}
\label{apn:information-current-FD-BE}

For the sake of completeness, in this Appendix, we shall generalize the discussion put forward in Sec.~\ref{sec:info-current}. Namely, we shall consider here a non-co-moving reference frame and that the particles of the system may be either bosons or fermions. As in Sec.~\ref{sec:info-current}, the probability of the realization of a given macrostate described by the single-particle distribution function profile $f(x,p)$ is given by the functional (see also Eq.~\eqref{PofF})
\begin{equation}
\label{eq:PofF-apn}
    \mathcal{P}[f_p] = Z^{-1} \exp \int_{\Sigma} \big(s^\mu+\alpha^\star J^\mu-\beta_\nu^\star T^{\mu \nu} \big)d\Sigma_\mu \, ,
\end{equation}
In the present case, the entropy four-current and the conserved currents are expressed in terms of the distribution function by
\begin{equation}
\label{eq:Thedefinitions-qstat}
\begin{split}
    s^{\mu} ={}& - \int \frac{d^3p}{(2\pi)^3 p^0} \, p^{\mu} \left( f_p \ln f_p + \frac{\Tilde{f}_p}{a} \ln \Tilde{f}_p \right) , \\
    J^{\mu} ={}& \int \frac{d^3p}{(2\pi)^3 p^0} \, p^{\mu} f_p ,\\
    T^{\mu\nu} ={}& \int \frac{d^3p}{(2\pi)^3 p^0} \, p^{\mu} p^{\nu} f_p,
    \end{split}
\end{equation}
where we made use of the definition
$\Tilde{f}_p = 1 - a f_p$, and $a = +1$ $(-1)$ if the particles are fermions (bosons). The corresponding expressions for classical particles are recovered if one sets $a \to 0$. It is noted that only the definition of $s^{\mu}$ changed with respect to Eq.~\eqref{Thedefinitions}. In this setting, the deviations around global equilibrium are more adequately parametrized by the deviation function 
\begin{equation}
\begin{aligned}
&
\phi_{p} = \frac{f_{p} - f_{\mathrm{eq},p}}{f_{\mathrm{eq},p} \Tilde{f}_{\mathrm{eq},p}},
\end{aligned}    
\end{equation}
which implies that one expresses $f_{p}$ and $\Tilde{f}_{p}$, respectively, as $f_{p} = f_{\mathrm{eq},p}(1 + \Tilde{f}_{\mathrm{eq},p} \phi_{p})$ and
$\Tilde{f}_{p} = \Tilde{f}_{\mathrm{eq},p}(1 - a f_{\mathrm{eq},p} \phi_{p}) $. As before, the information current can be identified by considering the ratio between the realization of a configuration with finite deviation from global equilibrium and the realization of equilibrium itself, 
\begin{equation}
\begin{aligned}
& \dfrac{\mathcal{P}[\phi_p]}{\mathcal{P}[\phi_p{=}0]} = e^{-\int_\Sigma E^\mu d\Sigma_\mu},
\\
&
E^\mu = - \int \dfrac{d^3 p}{(2\pi)^3 p^0} p^\mu   \left\{  
f_{\text{eq},p}\Tilde{f}_{\text{eq},p} \phi_{p}
\ln(f_{\text{eq},p}) 
+
f_{\text{eq},p}(1 + \Tilde{f}_{\text{eq},p} \phi_{p})
\ln(1 + \Tilde{f}_{\text{eq},p} \phi_{p}) 
\right.
\\
&
\left.
-
\Tilde{f}_{\text{eq},p} f_{\text{eq},p} \phi_{p}
\ln(\Tilde{f}_{\text{eq},p}) 
+
\frac{1}{a}\Tilde{f}_{\text{eq},p}(1 - a f_{\text{eq},p} \phi_{p})
\ln(1 - a f_{\text{eq},p} \phi_{p})
+
(-
\alpha
+
\beta_{\nu} p^{\nu}) f_{\text{eq},p} \Tilde{f}_{\text{eq},p} \phi_{p}
\right\}.
\end{aligned}    
\end{equation}
From the imposition that equilibrium is the most probable macrostate, we have that $\delta E^{\mu}/\delta \phi_{p} \vert_{\phi_{p} = 0} = 0$, that leads to $
\ln(f_{\text{eq},p}/\Tilde{f}_{\text{eq},p}) = 
-\alpha
+
\beta_{\nu} p^{\nu}
$ which implies
\begin{equation}
\begin{aligned}
&
f_{\text{eq},p} = \frac{1}{e^{-\alpha
+
\beta_{\nu} p^{\nu}} + a},
\end{aligned}
\end{equation}
which is the Bose-Einstein (Fermi-Dirac) distribution for $a = -1$ ($a = 1$). Hence, the information current in the present case is given by
\begin{equation}
\label{eq:phuiummum-BE-FD}
\begin{aligned}
&
E^\mu = \int \dfrac{d^3 p}{(2\pi)^3 p^0} p^\mu f_{\text{eq},p} \Tilde{f}_{\text{eq},p}\left\{  
\frac{(1 + \Tilde{f}_{\text{eq},p} \phi_{p})}{\Tilde{f}_{\text{eq},p}}
\ln(1 + \Tilde{f}_{\text{eq},p} \phi_{p}) 
+
\frac{(1 - a f_{\text{eq},p} \phi_{p})}{a f_{\text{eq},p}}
\ln(1 - a f_{\text{eq},p} \phi_{p})
\right\}
 \, .
\end{aligned}
\end{equation}
and the probability of realization of a fluctuation with deviation $\phi_{p}$ is given by
\begin{equation}
\label{eq:gavagringo-BE-FD}
\begin{aligned}
&
  \dfrac{\mathcal{P}[\phi_p]}{\mathcal{P}[\phi_p{=}0]} = \exp \left\{- \int_{\mathbb{R}^6} \dfrac{d^3 x \, d^3p}{(2\pi)^3} f_{\text{eq},p} \Tilde{f}_{\text{eq},p}\left[ 
\frac{(1 + \Tilde{f}_{\text{eq},p} \phi_{p})}{\Tilde{f}_{\text{eq},p}}
\ln(1 + \Tilde{f}_{\text{eq},p} \phi_{p}) 
+
\frac{(1 - a f_{\text{eq},p} \phi_{p})}{a f_{\text{eq},p}}
\ln(1 - a f_{\text{eq},p} \phi_{p})
\right]
 \right\},
\end{aligned}
\end{equation}
which reduces to Eq.~\eqref{gavagringo}, as $a \to 0$. Considering a box-like perturbation as in Eq.~\eqref{eq:box-pert}
leads to a considerably more complicated distribution than that in Eq.~\eqref{sao}, which in the present case cannot be expressed in terms of the equilibrium number of particles in the region enclosed by the box because of the $f_{\text{eq},p}$ and $\Tilde{f}_{\text{eq},p}$  factors in the logarithms. On the other hand, in the Gaussian regime, the corresponding distribution for the box perturbation is 
\begin{equation}
\label{eq:gavagringoGauss-FD-BE}    
\begin{aligned}
 \mathcal{P}[\phi] = \sqrt{\frac{1}{2 \pi}\frac{\partial N_{\mathrm{eq}}}{\partial \alpha}} \exp \left\{ - \frac{1}{2} \frac{\partial N_{\mathrm{eq}}}{\partial \alpha}   \phi^2 \right\} \, ,
\end{aligned}
\end{equation}
which is to be contrasted with Eq.~\eqref{eq:prob-dist-gauss-box}. There, the variance of the distribution is given in terms of $N_{\mathrm{eq}}$ and allowed to estimate that the Boltzmann equation should not be adequate to be employed for systems with less than 25 particles, given the threshold for errors to be 20 \%. In the present case, a similar criterion leads to $\partial N_{\mathrm{eq}}/\partial \alpha \lesssim  25$, which means that the Boltzmann equation should not be applicable for systems with quantum statistics in regimes where the typical number of particles varies drastically with the thermal potential. In particular, for bosons ($a=-1$), since the Bose-Einstein distribution diverges at a finite value of  $\vert \mathbf{p} \vert$ if $\alpha > m/T$ (and so does $\partial N_{\mathrm{eq}}/\partial \alpha$), the Boltzmann equation should not be applicable when the chemical potential is close to the mass of the particle.

Now we turn our attention to gaussian fluctuations in generic configurations. Expanding the integrand in the exponential of Eq.~\eqref{eq:phuiummum-BE-FD} up to quadratic order, we have
\begin{equation}
\label{eq:gavagringoGauss-FD-BE}    
\begin{aligned}
  \dfrac{\mathcal{P}[\phi_p]}{\mathcal{P}[\phi_p{=}0]} = \exp \left\{ -\int_{\mathbb{R}^6}  \dfrac{1}{2}\dfrac{d^3 x \, d^3p}{(2\pi)^3} f_{\text{eq},p} \Tilde{f}_{\text{eq},p}  \phi^2_p \right\} \, .
\end{aligned}
\end{equation}
from which the equal-time $\phi$-$\phi$ correlator can be computed to be
\begin{equation}
\label{eq:equaltimecorreltor!}
    \langle \phi_p(\textbf{x})\phi_{p'}(\textbf{x}')\rangle = \dfrac{(2\pi)^3}{f_{\text{eq},p}\Tilde{f}_{\text{eq},p}} \delta^3(\textbf{x}-\textbf{x}')\delta^3(\textbf{p}-\textbf{p}') \, .
\end{equation}
From that, we can compute correlations between fluctuations in deviations of the particle four-current and energy-momentum tensor 
\begin{equation}
\label{eq:ThedefinitionsFluttuanti-FD-BE}
\begin{split}
    \delta J^{\mu} ={}& \int \frac{d^3p}{(2\pi)^3 p^0} \, p^{\mu} f_{\text{eq},p} \Tilde{f}_{\text{eq},p}\phi_p ,\\
    \delta T^{\mu\nu} ={}& \int \frac{d^3p}{(2\pi)^3 p^0} \, p^{\mu} p^{\nu} f_{\text{eq},p} \Tilde{f}_{\text{eq},p}\phi_p .\\
    \end{split}
\end{equation}
as it was done in the main text. These correlations can be expressed as 
\begin{equation}
\label{eq:JJ-TT}
\begin{split}
\langle \delta J^\mu(\textbf{x})\delta J^\nu(\textbf{x}')\rangle ={}& \delta^3(\textbf{x}{-}\textbf{x}') \overline{J^\mu J^\nu}  \, , \\
\langle \delta J^\mu(\textbf{x})\delta T^{\nu\sigma}(\textbf{x}')\rangle ={}& \delta^3(\textbf{x}{-}\textbf{x}') \overline{J^\mu T^{\nu \sigma} }  , \\
\langle \delta T^{\mu\nu}(\textbf{x})\delta T^{\sigma \lambda}(\textbf{x}')\rangle ={}& \delta^3(\textbf{x}{-}\textbf{x}') \overline{T^{\mu \nu} T^{\sigma \lambda} }  \, . \\
\overline{J^\mu J^\nu} = & \int \dfrac{d^3p}{(2\pi)^3(p^0)^2} p^\mu p^\nu f_{\text{eq},p} \Tilde{f}_{\text{eq},p} ,
\\
\overline{J^\mu T^{\nu \sigma} } 
&=
\int \dfrac{d^3p}{(2\pi)^3(p^0)^2} p^\mu p^\nu p^\sigma f_{\text{eq},p} \Tilde{f}_{\text{eq},p} , 
\\
\overline{T^{\mu \nu} T^{\sigma \lambda} } 
&=
\int \dfrac{d^3p}{(2\pi)^3(p^0)^2} p^\mu p^\nu p^\sigma p^\lambda f_{\text{eq},p} \Tilde{f}_{\text{eq},p},
\end{split}
\end{equation}
which can be expressed in terms of the generic thermodynamic integrals
\begin{equation}
\begin{aligned}
&
\mathcal{J}_{nq} = \frac{1}{(2q+1)!!} \int dP \left( - \Delta_{\mu \nu} p^{\mu} p^{\nu} \right)^{q} E_{p}^{n-2q} f_{\text{eq},p} \Tilde{f}_{\text{eq},p} 
\end{aligned}    
\end{equation}
and whose tensor structure must be constructed from $u^{\mu}$ and $g^{\mu \nu}$, which are the only tensors present after integration in momentum space. As a matter of fact, irreducible expressions for Eq.~\eqref{eq:JJ-TT} can be derived using the identities
\begin{equation}
\begin{aligned}
p^{\mu} p^{\nu} &= E_p^{2} u^{\mu} u^{\nu} 
+
\frac{1}{3} \left(\Delta^{\alpha \beta} p_{\alpha} p_{\beta} \right)
\Delta^{\mu \nu}
+
2 E_p u^{(\mu} p^{\langle \nu \rangle)}
+
p^{\langle \mu} p^{\nu \rangle},
\\
p^\mu p^\nu p^\sigma & = E_p^{3} u^{\mu} u^{\nu} u^{\sigma}
+
3 E_p^{2} u^{(\mu} u^{\nu} p^{\langle \sigma \rangle)}
+
3 E_p u^{(\mu} p^{\langle \nu} p^{\sigma \rangle)}
+ 
 E_p  \left(\Delta^{\alpha \beta} p_{\alpha} p_{\beta} \right) u^{(\mu} \Delta^{\nu \sigma )}
+
p^{\langle \mu} p^\nu p^{\sigma \rangle}
\\
&
+
\frac{3}{5}  \left(\Delta^{\alpha \beta} p_{\alpha} p_{\beta} \right) \
p^{( \langle \mu \rangle} \Delta^{\nu \sigma )},
\\
p^\mu p^\nu p^\sigma p^{\lambda} 
& 
=
E_p^{4} u^{\mu} u^{\nu} u^{\sigma} u^{\lambda}
+
4 E_p^{3} u^{(\mu} u^{\nu} u^{\sigma} p^{\langle \lambda \rangle)}
+
6 E_p^{2} u^{(\mu} u^{\nu} p^{\langle \sigma} p^{\lambda \rangle)}
+
2 E_p^{2} \left(\Delta^{\alpha \beta} p_{\alpha} p_{\beta} \right) u^{(\mu} u^{\nu} \Delta^{\sigma \lambda)}
\\
&
+
4 E_p u^{(\mu} p^{\langle \nu} p^\sigma p^{\lambda \rangle )}
+
\frac{12}{5} E_p  \left(\Delta^{\alpha \beta} p_{\alpha} p_{\beta} \right)
\ u^{(\mu} \Delta^{\nu \sigma} p^{\langle \lambda \rangle )}
+
p^{\langle \mu} p^\nu p^\sigma p^{\lambda \rangle}
+
\frac{6}{7}
\left(\Delta^{\alpha \beta} p_{\alpha} p_{\beta} \right) \ 
\Delta^{(\mu \nu} p^{\langle \sigma } p^{ \lambda \rangle)}
\\
&
+ 
\frac{2}{15} \left( \Delta^{\alpha \beta} p_{\alpha} p_{\beta} \right)^{2} \Delta^{\mu \nu \sigma \lambda} 
+ 
\frac{1}{9}\left( \Delta^{\alpha \beta} p_{\alpha} p_{\beta} \right)^{2} \Delta^{\mu \nu} \Delta^{\sigma \lambda},
\end{aligned}    
\end{equation}
where the parentheses denote the symmetrization in the enclosed indices; taking into account the trivial permutations among momentum and the $u^{\mu}$ indices, explicitly, we have
\begin{subequations}
\label{eq:u-p-delta-expressions}
\begin{align}
&    
\label{eq:u-p-delta-expressions-1}
2 u^{(\mu} p^{\langle \nu \rangle)} = u^{\mu} p^{\langle \nu \rangle}
+ 
u^{\nu} p^{\langle \mu \rangle},
\\
&
\label{eq:u-p-delta-expressions-2}
3 u^{(\mu} u^{\nu} p^{\langle \sigma \rangle)} = 
u^{\mu} u^{\nu} p^{\langle \sigma \rangle}
+
u^{\mu} u^{\sigma} p^{\langle \nu \rangle}
+
u^{\sigma} u^{\nu} p^{\langle \mu \rangle},
\\
&
\label{eq:u-p-delta-expressions-3}
4  u^{(\mu} u^{\nu} u^{\sigma} p^{\langle \lambda \rangle)}
=
u^{\mu} u^{\nu} u^{\sigma} p^{\langle \lambda \rangle}
+
u^{\mu} u^{\nu} u^{\lambda} p^{\langle \sigma \rangle}
+
u^{\mu} u^{\lambda} u^{\sigma} p^{\langle \nu \rangle}
+
u^{\lambda} u^{\nu} u^{\sigma} p^{\langle \mu \rangle},
\\
&
\label{eq:u-p-delta-expressions-4}
6 u^{(\mu} u^{\nu} p^{\langle \sigma} p^{\lambda \rangle)}
=
u^{\mu} u^{\nu} p^{\langle \sigma} p^{\lambda \rangle}
+
u^{\sigma} u^{\nu} p^{\langle \mu} p^{\lambda \rangle}
+
u^{\lambda} u^{\nu} p^{\langle \sigma} p^{\mu \rangle}
+
u^{\mu} u^{\sigma} p^{\langle \nu} p^{\lambda \rangle}
+
u^{\mu} u^{\lambda} p^{\langle \sigma} p^{\nu \rangle}
\notag
\\
&
+
u^{\sigma} u^{\lambda} p^{\langle \mu} p^{\nu \rangle}
,
\end{align}    
\end{subequations}
where analogous expressions are valid for $u^{(\mu} p^{\langle \nu} p^{\sigma \rangle)}$, $u^{(\mu} \Delta^{\nu \sigma )}$, $p^{( \langle \mu \rangle} \Delta^{\nu \sigma )}$ (with respect to Eq.~\eqref{eq:u-p-delta-expressions-2}), $u^{(\mu} u^{\nu} \Delta^{\sigma \lambda)}$, $u^{(\mu} p^{\langle \nu} p^\sigma p^{\lambda \rangle )}$, $u^{(\mu} \Delta^{\nu \sigma} p^{\langle \lambda \rangle )}$ (wrt Eq.~\eqref{eq:u-p-delta-expressions-3}), $\Delta^{(\mu \nu} p^{\langle \sigma } p^{ \lambda \rangle)}$ (wrt Eq.~\eqref{eq:u-p-delta-expressions-4}), respectively. Besides that, we employ the fact that 
\begin{equation}
\begin{aligned}
&
\int dP\,
E_p^{n} p^{\langle \mu_{1}} \cdots p^{\mu_{\ell} \rangle}
f_{\text{eq},p} \Tilde{f}_{\text{eq},p} = 0,  \end{aligned}   
\end{equation}
for $\ell \neq 0$, which is a consequence of Eq.~\eqref{eq:main-property-irred-tens}. Then, we finally obtain
\begin{subequations}
\begin{align}
&
\overline{J^\mu J^\nu}= \mathcal{J}_{10} u^{\mu}u^{\nu}
-
\mathcal{J}_{11} \Delta^{\mu \nu},
\\
&
\overline{J^\mu T^{\nu \sigma} } 
=
\mathcal{J}_{20} u^{\mu}u^{\nu}u^{\sigma}
-
\mathcal{J}_{21} u^{(\mu}\Delta^{\nu \sigma)},
\\
&
\overline{T^{\mu \nu} T^{\sigma \lambda} } 
=
\mathcal{J}_{30} u^{\mu}u^{\nu}u^{\sigma} u^{\lambda}
-
6 \mathcal{J}_{31} u^{(\mu} u^{\nu}\Delta^{\sigma \lambda)}
+
2 \mathcal{J}_{32} \Delta^{\mu \nu \sigma \lambda}
+
\frac{5}{3} \mathcal{J}_{32} \Delta^{\mu \nu}\Delta^{\sigma \lambda}.
\end{align}    
\end{subequations}

\section{Properties of the z-parameters}\label{thezetaarecosi}

In the linear regime, we can classify all kinetic states using two scalar variables, $H$ (``Hydrodynamic'') and $N$ (``Not hydrodynamic''). The number $H$ quantifies the cumulative amplitude of the excited modes that can be described using the Israel-Stewart theory, while $N$ quantifies the cumulative amplitude of excited modes that escape the Israel-Stewart description. We can treat each point of the $\{H,N\}$ plane as a larger macrostate, which has a probability $e^{-E(H,N)}$ of being occupied. In the Gaussian limit, the most general function $E(H,N)$ takes the form
\begin{equation}\label{Eooohohoh}
    E = \dfrac{1}{2} (H,N) 
    \begin{bmatrix}
        1 & a \\
        a & 1 \\
    \end{bmatrix} 
    \begin{pmatrix}
        H \\
        N \\
    \end{pmatrix} \, ,
\end{equation}
provided that we adopt some natural units for $H$ and $N$. Thermodynamic stability requires the cross-term $a$ to belong to the interval $(-1,1)$. Explicit evaluation of the Gaussian integrals gives
\begin{equation}
    \begin{bmatrix}
        \langle HH \rangle & \langle HN \rangle \\
        \langle NH \rangle & \langle NN \rangle \\
    \end{bmatrix} = \dfrac{1}{1-a^2}
    \begin{bmatrix}
        1 & -a \\
        -a & 1 \\
    \end{bmatrix} \, .
\end{equation}
Thus, the kinetic-theory prediction for the $H$ correlator is $\langle HH\rangle_{\text{kt}}=(1{-}a^2)^{-1}$. Instead, computing $\langle HH\rangle$ within the Israel-Stewart theory corresponds to constraining $N$ to zero in equation \eqref{Eooohohoh} (i.e. summing only over the $H$ states), giving the hydrodynamic correlator $\langle HH \rangle_{IS} =1$. 

Now, let $A$ be the linear perturbation to an arbitrary physical observable. Within the two-dimensional space of macrostates $(H,N)$, it should be possible to express $A$ in the form $A=A_H H+A_N N$, for some linear-combination coefficients $A_H$ and $A_N$ (not necessarily positive). Then, we find that
\begin{equation}
    \begin{split}
\langle AA \rangle_{\text{kt}} ={}& \dfrac{A_H^2-2aA_H A_N + A_N^2}{1-a^2} \, , \\
\langle AA \rangle_{\text{IS}} ={}& A_H^2 \, . \\
    \end{split}
\end{equation}
Therefore, the z-parameter associated with $A$ is
\begin{equation}
    z_A = \dfrac{\langle AA\rangle_{\text{kt}}-\langle AA\rangle_{\text{IS}}}{\langle AA\rangle_{\text{kt}}} = \dfrac{(A_N{-}aA_H)^2}{(A_N{-}aA_H)^2+(1{-}a^2)A_H^2} \in [0,1] \, ,
\end{equation}
which is what we wanted to prove.

The above proof relies on a coarse-graining, where we replace all the detailed degrees of freedom of the system with two scalars $\{H,N\}$. There is no loss of generality in this procedure, as long as $E$ is appropriately increased to account for the degrees that are lost in the process. However, for completeness, let us also provide the proof that $z_A \in[0,1]$ when $H$ and $Z$ are column vectors (i.e. list of variables). In this case, we have
\begin{equation}\label{Eooohohoh2}
\begin{split}
    E ={}& \dfrac{1}{2} (H^T,N^T) 
    \begin{bmatrix}
        P & Q \\
        Q^T & R \\
    \end{bmatrix} 
    \begin{pmatrix}
        H \\
        N \\
    \end{pmatrix} \, , \\
    A={}& A_H^T H + A_N^T N \, , \\
\end{split}
\end{equation}
where $P$ and $R$ are symmetric positive definite matrices, $Q$ is a rectangular matrix, and $A_H$ and $A_N$ are column vectors. The correlators of the fundamental variables now read
\begin{equation}
     \begin{bmatrix}
        \langle HH^T \rangle & \langle HN^T \rangle \\
        \langle NH^T \rangle & \langle NN^T \rangle \\
    \end{bmatrix} =   
\begin{bmatrix}
    P^{-1}+P^{-1}QZQ^T P^{-1} & -P^{-1}QZ \\
    -ZQ^T P^{-1} & Z \\
\end{bmatrix} \, ,
\end{equation}
with $Z=(R{-}Q^T P^{-1}Q)^{-1}$. Therefore, the $A$ correlators read
\begin{equation}
\begin{split}
\langle AA\rangle_{\text{kt}}={}& A_H^T P^{-1} A_H + (A_N{-}Q^T P^{-1} A_H)^T Z (A_N{-}Q^T P^{-1} A_H) \, , \\
\langle AA\rangle_{\text{IS}}={}& A_H^T P^{-1} A_H \, , \\
\end{split}
\end{equation}
As before, the Israel-Stewart correlators are computed assuming that $N$ does not fluctuate (i.e. forcing $N=0$ in the Gaussian integral). Note that, since the matrix $Z$ coincides with the correlator $\langle N N^T \rangle_{\text{kt}}$, it must be positive definite. Clearly, $P^{-1}$ is positive definite too. Thus, the $z$-parameter of $A$ reads
\begin{equation}
    z_A = \dfrac{(A_N{-}Q^T P^{-1} A_H)^T Z (A_N{-}Q^T P^{-1} A_H) }{A_H^T P^{-1} A_H + (A_N{-}Q^T P^{-1} A_H)^T Z (A_N{-}Q^T P^{-1} A_H) } \in [0,1] \, ,
\end{equation}
completing our proof.

\section{Properties of the linearized collision operator and the collision kernel $K_{p\Tilde{p}}$}
\label{eq:K-kernel}

In the present Appendix, we shall display the details of the derivation of Eq.~\eqref{eq:kernel-K}, which expresses the linearized collision term as a symmetric kernel $K_{p \widetilde{p}}$. In fact, linearizing Eq.~\eqref{eq:Boltzmann}, we have
\begin{equation}
\label{Lhat}
\begin{aligned}
&
f_{\mathrm{eq},p}
p^{\mu}\partial_{\mu}\phi_p
=
f_{\mathrm{eq},p}
\hat{L}[\phi_{p}] \\
&
\equiv
\frac{1}{2}\int \frac{d^{3}k}{(2 \pi)^{3} k^{0}} \ \frac{d^{3}k'}{(2 \pi)^{3} k'^{0}} \ \frac{d^{3}p'}{(2 \pi)^{3} p'^{0}}  W_{pp' \leftrightarrow kk'} f_{\mathrm{eq},p} f_{\mathrm{eq},p'} (  \phi_{k} + \phi_{k'} - \phi_{p} - \phi_{p'}  ),
\end{aligned}    
\end{equation}
In order to express the linearized Boltzmann equation in the form of Eq.~\eqref{sprintone}, we must find an integral kernel such that 
\begin{equation}
\label{eq:fund-K-prop}
\begin{aligned}
&
f_{\text{eq},p} \int \dfrac{d^3 \Bar{p}}{(2\pi)^3 \Bar{p}^0} K_{p\Bar{p}} \phi_{\Bar{p}}
=
f_{\text{eq},p}
\Hat{L}[\phi_{p}].
\end{aligned}    
\end{equation}
To this end, first, from the definition \eqref{Lhat}, we can derive that
\begin{equation}
\label{eq:apn-K-1}
\begin{aligned}
f_{\mathrm{eq},p}
\Hat{L}[\phi_{p}] 
&= \frac{(2\pi)^3}{2} \int\frac{d^3\widetilde{p}}{(2\pi)^3 \widetilde{p}^0} \int \frac{d^3k}{(2\pi)^3 k^0}  \ \frac{d^{3}k'}{(2 \pi)^{3} k'^{0}} \ \frac{d^{3}p'}{(2 \pi)^{3} p'^{0}} \, \widetilde{p}^0   W_{pp' \rightarrow kk'} f_{\mathrm{eq},p} f_{\mathrm{eq},p'} 
\\
&
\times
\Big[ \delta^{3}(\textbf{k}-\widetilde{\textbf{p}}) + \delta^{3}(\textbf{k}'-\widetilde{\textbf{p}}) 
- \delta^{3}(\textbf{p}-\widetilde{\textbf{p}}) - \delta^{3}(\textbf{p}'-\widetilde{\textbf{p}}) \Big] \phi_{\widetilde{p}},
\end{aligned}
\end{equation}
where we have used the filtering property of the Dirac delta. On the other hand, using more elaborate properties of the linearized collision, we derive
\begin{equation}
\label{eq:apn-K-2}
\begin{aligned}
&
f_{\mathrm{eq},p}
\Hat{L}[\phi_{p}]
\\
&
=
\frac{(2 \pi)^{3}}{2}\left[\int \frac{d^{3}\widetilde{p}}{(2 \pi)^{3}\widetilde{p}^{0}} p^{0}\ \delta^{3}(\textbf{p} - \widetilde{\textbf{p}}) \right] \int \frac{d^{3}k}{(2 \pi)^{3} k^{0}} \ \frac{d^{3}k'}{(2 \pi)^{3} k'^{0}} \ \frac{d^{3}p'}{(2 \pi)^{3} p'^{0}}  W_{pp' \leftrightarrow kk'} f_{\mathrm{eq},\widetilde{p}} f_{\mathrm{eq},\textbf{p}'} (  \phi_{k} + \phi_{k'} - \phi_{p} - \phi_{p'}  ),
\\
&
=
\frac{(2 \pi)^{3}}{2} 
\int \frac{d^{3}\widetilde{p}}{(2 \pi)^{3}\widetilde{p}^{0}} 
\frac{d^{3}k}{(2 \pi)^{3} k^{0}} \ \frac{d^{3}k'}{(2 \pi)^{3} k'^{0}} \ \frac{d^{3}p'}{(2 \pi)^{3} p'^{0}} p^{0}
\delta^{3}(\textbf{p} - \widetilde{\textbf{p}})  
W_{\widetilde{p}p' \leftrightarrow kk'} 
f_{\mathrm{eq},\widetilde{p}}
f_{\mathrm{eq},p'} (  \phi_{k} + \phi_{k'} - \phi_{\widetilde{p}} - \phi_{p'}  ),
\\
&
=
\frac{(2 \pi)^{3}}{2} 
\int \frac{d^{3}\widetilde{p}}{(2 \pi)^{3}\Tilde{p}^{0}} 
\frac{d^{3}k}{(2 \pi)^{3} k^{0}} \ \frac{d^{3}k'}{(2 \pi)^{3} k'^{0}} \ \frac{d^{3}p'}{(2 \pi)^{3} p'^{0}}
p^{0} W_{\widetilde{p}p' \leftrightarrow kk'} f_{\mathrm{eq},\widetilde{p}} f_{\mathrm{eq},p'}  
\\
&
\times \left[\delta^{3}(\textbf{p} - \textbf{k}) 
+ 
 \delta^{3}(\textbf{p} - \textbf{k}') 
- 
\delta^{3}(\textbf{p} - \widetilde{\textbf{p}}) 
- 
\delta^{3}(\textbf{p} - \textbf{p}')  \right]  \phi_{\widetilde{p}},
\end{aligned}    
\end{equation}
where, in the final equality, the self-adjoint property of the linearized collision term has been employed, i.e., for given arbitrary functions of the momentum $A_{\widetilde{p}}$ and $B_{\widetilde{p}}$, 
\begin{equation}
\begin{aligned}
\int \frac{d^{3}\widetilde{p}}{(2 \pi)^{3}\Tilde{p}^{0}}
A_{\widetilde{p}}
f_{\mathrm{eq},\widetilde{p}}
\Hat{L}[B_{\widetilde{p}}]
=
\int \frac{d^{3}\widetilde{p}}{(2 \pi)^{3}\Tilde{p}^{0}}
B_{\widetilde{p}}
f_{\mathrm{eq},\widetilde{p}}
\Hat{L}[A_{\widetilde{p}}]
\end{aligned}    
\end{equation}
Then, we can express the linearized collision term as a manifestly symmetric integral kernel by taking the average of Eqs.~\eqref{eq:apn-K-1} and \eqref{eq:apn-K-2}, 
\begin{equation}
\begin{aligned}
f_{\mathrm{eq},p}
\Hat{L}[\phi_{p}] 
&
=
\frac{(2\pi)^3}{4} \int 
\frac{d^3\widetilde{p}}{(2\pi)^3 \widetilde{p}^0} \frac{d^3k}{(2\pi)^3 k^0}  \ \frac{d^{3}k'}{(2 \pi)^{3} k'^{0}} \ \frac{d^{3}p'}{(2 \pi)^{3} p'^{0}} \,  
\\
&
\times
\left\{ 
\tilde{p}^0   W_{pp' \rightarrow kk'} f_{\mathrm{eq},p} f_{\mathrm{eq},p'}
\left[ \delta^{3}(\textbf{k}-\widetilde{\textbf{p}}) + \delta^{3}(\textbf{k}'-\widetilde{\textbf{p}}) 
- \delta^{3}(\textbf{p}-\widetilde{\textbf{p}}) - \delta^{3}(\textbf{p}'-\widetilde{\textbf{p}}) \right]
\right.
\\
&
\left.
+
p^{0} W_{\widetilde{p}p' \leftrightarrow kk'} f_{\mathrm{eq},\widetilde{p}} f_{\mathrm{eq},p'}  
\left[\delta^{3}(\textbf{k} - \textbf{p} ) 
+ 
 \delta^{3}(\textbf{k}' - \textbf{p}) 
- 
\delta^{3}(\widetilde{\textbf{p}} - \textbf{p}) 
- 
\delta^{3}(\textbf{p}' - \textbf{p})  \right]
\right\}
\phi_{\widetilde{p}},
\end{aligned}    
\end{equation}
from which $K_{p \widetilde{p}}$ can be readily identified.

\section{Details of the obtention of the Fourier-Laguerre component equations}
\label{sec:F-moms}

Given the role played by the linearized collision term in dynamical fluctuations in Kinetic Theory (see e.g.~\eqref{sprintone}), it is convenient to perform expansions in the eigenbasis of this linear operator. This basis is given in terms of projected tensors, $p^{\langle \mu_{1}} \cdots p^{\mu_{\ell} \rangle} \equiv \Delta^{\mu_{1} \cdots \mu_{\ell}}_{\nu_{1} \cdots \nu_{\ell}} p^{\nu_{1}} \cdots p^{\nu_{\ell}}$. The $2 \ell$-rank tensor $\Delta^{\mu_{1} \cdots \mu_{\ell}}_{\nu_{1} \cdots \nu_{\ell}} p^{\nu_{1}} \cdots p^{\nu_{\ell}}$ is constructed from the $\Delta^{\mu \nu} = g^{\mu \nu} - u^{\mu} u^{\nu}$ projectors. Explicitly, it reads
\cite{DeGroot:1980dk}
\begin{equation}
\label{eq:def-deltao-1}
\begin{aligned}
&   \Delta^{\mu_{1} \cdots \mu_{\ell} \nu_{1} \cdots \nu_{\ell}} = \sum_{k = 0}^{\lceil \ell / 2 \rceil} \frac{C_{\ell k}}{\mathcal{N}_{\ell k}}\sum_{\wp_{\mu} \wp_{\nu}} \Delta^{\mu_{1} \mu_{2}} \cdots \Delta^{\mu_{2k-1} \mu_{2k}} \Delta^{\nu_{1} \nu_{2}} \cdots \Delta^{\nu_{2k-1} \nu_{2k}} \Delta^{\mu_{2k+1} \nu_{2k+1}} \cdots \Delta^{\mu_{\ell} \nu_{\ell}}\;, 
\end{aligned}    
\end{equation}
where $\lceil \ell / 2 \rceil$ is the largest integer not exceeding $\ell / 2$, and we define the combinatorial factors
\begin{subequations}
\label{eq:def-deltao-2}
\begin{align}
 &
 C_{\ell k} = (-1)^{k} \frac{(\ell!)^{2}}{(2 \ell)!} \frac{(2 \ell - 2k)!}{k!(\ell-k)!(\ell-2k)!}\;,
\\
&
\mathcal{N}_{\ell k} = \frac{1}{(\ell - 2k)!} \left( \frac{\ell!}{2^{k}k!} \right)^{2} \;.
\end{align}    
\end{subequations}
which guarantees that it is symmetric with respect to permutations in any of the indices $\mu_{1} \cdots \mu_{\ell}$ and $\nu_{1} \cdots \nu_{\ell}$, separately, and also $\Delta^{\mu_{1} \cdots \mu_{\ell}}_{\nu_{1} \cdots \nu_{\ell}} g^{\nu_{j} \nu_{k}} = 0$, $\forall \; j,k =1, \ldots \ell$ \cite{DeGroot:1980dk,Denicol:2021,Rocha:2023ilf}. 

The eigenbasis is also given in terms of Laguerre polynomials (see Eq.~\eqref{eq:eigenvalues-lin-col}). Then, it is convenient to express the fluctuating equations of motion \eqref{eq:lin-BEq-Fou} in terms of the Fourier-Laguerre moments $\widetilde{\Phi}_{n}^{\mu_{1} \cdots \mu_{\ell}}(q) \equiv \int (d^{3}p/[(2 \pi)^{3}p^{0}]) f_{\mathrm{eq},p} L^{(2 \ell + 1)}_{n p} p^{\langle \mu_{1}} \cdots p^{ \mu_{\ell} \rangle} \Bar{\phi}_p$. These can be obtained by integrating Eq.~\eqref{eq:lin-BEq-Fou} with $L^{(2 \ell + 1)}_{n p} p_{\langle \mu_{1}} \cdots p_{ \mu_{\ell} \rangle}$. From this first step, we obtain
\begin{equation}
\label{eq:lin-BEq-Fou-mom-Lag}
\begin{aligned}
&
i \Omega \int \frac{d^{3}p}{(2 \pi)^{3} p^{0}} f_{\mathrm{eq},p} E_p L^{(2 \ell + 1)}_{n, p} p^{\langle \mu_{1}} \cdots p^{ \mu_{\ell} \rangle}   \Bar{\phi}_p 
+
i q_{\langle \mu \rangle} \int \frac{d^{3}p}{(2 \pi)^{3} p^{0}} f_{\mathrm{eq},p} L^{(2 \ell + 1)}_{n p} p^{\langle \mu_{1}} \cdots p^{ \mu_{\ell} \rangle} p^{\langle \mu \rangle} \Bar{\phi}_p  
\\
&
-
\chi_{n\ell}
\widetilde{\Phi}_{n}^{\mu_{1} \cdots \mu_{\ell}}
   = 
\widetilde{\Xi}_{n}^{\mu_{1} \cdots \mu_{\ell}},
\end{aligned}
\end{equation}
In order to express the above equation solely in terms of the $\widetilde{\Phi}_{n}^{\mu_{1} \cdots \mu_{\ell}}$-moments themselves, we shall make use of the following identities
\begin{subequations}
\label{eq:ids-Lag-tensor}
\begin{align}
&
\label{eq:ids-Lag-tensor-1}
\beta E_p L_{n, p}^{(2 \ell +1)} 
=
-
(n+1) L_{n+1, p}^{(2 \ell +1)}
+
2(n+ \ell +1) L_{n, p}^{(2 \ell +1)}
-
(n+2 \ell +1) L_{n-1, p}^{(2 \ell +1)},
\\
&
\label{eq:ids-Lag-tensor-2}
p^{\langle \mu_{1}} \cdots p^{ \mu_{\ell} \rangle} 
 p^{\langle \mu \rangle} 
 =
p^{\langle \mu_{1}} \cdots p^{ \mu_{\ell} } 
 p^{\mu \rangle} 
 +
 \frac{\ell}{2 \ell + 1}
 (\Delta_{\alpha \beta} p^{\alpha} p^{\beta})
\Delta^{\mu_{1} \cdots \mu_{\ell}}_{ \ \ \ \ \ \ \ \nu_{1} \nu_{2} \cdots \nu_{\ell}} \Delta^{\nu_{1} \mu}
p^{\langle \nu_{2}} \cdots p^{ \nu_{\ell} \rangle} ,
\\
&
\label{eq:ids-Lag-tensor-2.5}
L_{n, p}^{(2 \ell +1)} 
=
L_{n, p}^{(2 \ell + 3)}
-
2 L_{n-1, p}^{(2 \ell +3)}
+
L_{n-2, p}^{(2 \ell +3)},
\\
&
\label{eq:ids-Lag-tensor-3}
(\beta E_p)^{2} L_{n, p}^{(2 \ell +1)}
=
(n+1)(n+2) L_{n+2, p}^{(2 \ell -1)}
-
2(n+1)(n+2\ell+1) L_{n, p}^{(2 \ell -1)}
+
(n+2\ell+1)(n+2\ell) L_{n, p}^{(2 \ell -1)},
\end{align}    
\end{subequations}
where Eqs.~\eqref{eq:ids-Lag-tensor-1},
\eqref{eq:ids-Lag-tensor-2.5},
and \eqref{eq:ids-Lag-tensor-3} can be derived from the basic properties,  $x L_{n}^{(\alpha)} = -(n+1) L_{n+1}^{(\alpha-1)} + (n + \alpha) L_{n}^{(\alpha-1)}$, $L_{n}^{(\alpha)} = L_{n}^{(\alpha + 1)} - L_{n-1}^{(\alpha+1)}$, of the Laguerre polynomials \cite{NIST:DLMF,gradshteyn2014table}. In its turn, Eq.~\eqref{eq:ids-Lag-tensor-2} can be derived from the symmetry properties of the irreducible tensors. Namely, we know that it is symmetric, traceless, orthogonal to $u^{\mu}$ in all of the indexes $\mu_{1} \cdots \mu_{\ell}$ of the tensor $p^{\langle \mu_{1}} \cdots p^{ \mu_{\ell} \rangle} p^{\langle \mu \rangle}$ and it is orthogonal with respect to $u_{\mu}$ in index $\mu$. This tensor can then only obey this symmetries if it is expressed as a linear combination of $p^{\langle \mu_{1}} \cdots p^{ \mu_{\ell} } p^{\mu \rangle}$ and $\Delta^{\mu_{1} \cdots \mu_{\ell} \mu}_{ \ \ \ \ \ \ \ \ \nu_{2} \cdots \nu_{\ell}} 
p^{\langle \nu_{2}} \cdots p^{ \nu_{\ell} \rangle}$, 
\begin{equation}
\begin{aligned}
&
p^{\langle \mu_{1}} \cdots p^{ \mu_{\ell} \rangle} 
 p^{\langle \mu \rangle} 
 =
A p^{\langle \mu_{1}} \cdots p^{ \mu_{\ell} } 
 p^{\mu \rangle} 
 +
B 
\Delta^{\mu_{1} \cdots \mu_{\ell}}_{ \ \ \ \ \ \ \ \ \nu_{1} \cdots \nu_{\ell}} \Delta^{\nu_{1} \mu} 
p^{\langle \nu_{2}} \cdots p^{ \nu_{\ell} \rangle} .
\end{aligned}    
\end{equation}
Contracting this equation on both sides with $\Delta^{\lambda_{1} \cdots \lambda_{\ell + 1}}_{ \ \ \ \ \ \ \ \ \mu_{1} \cdots \mu_{\ell} \mu}$
yields $p^{\langle \lambda_{1}} \cdots p^{ \lambda_{\ell + 1} \rangle} 
 =
A p^{\langle \lambda_{1}} \cdots p^{ \lambda_{\ell + 1} \rangle} $, which follows from the properties $\Delta^{\lambda_{1} \cdots \lambda_{\ell + 1}}_{ \ \ \ \ \ \ \ \ \mu_{1} \cdots \mu_{\ell} \mu} \Delta^{\mu_{1} \cdots \mu_{\ell} \mu}_{ \ \ \ \ \ \ \ \ \nu_{1} \cdots \nu_{\ell + 1}} =
\Delta^{\lambda_{1} \cdots \lambda_{\ell + 1}}_{ \ \ \ \ \ \ \ \ \nu_{1} \cdots \nu_{\ell + 1}}$ and $\Delta^{\lambda_{1} \cdots \lambda_{\ell + 1}}_{ \ \ \ \ \ \ \ \ \mu_{1} \cdots \mu_{\ell} \mu} \Delta^{\mu_{1} \cdots \mu_{\ell} \mu}_{ \ \ \ \ \ \ \ \ \nu_{1} \cdots \nu_{\ell}} = 0$ \cite{deBrito:2024vhm}. The former property means that when the projector is acted twice this is the same as acting once, whereas the latter is derived from the definition of the projector, Eq.~\eqref{eq:def-deltao-1}, in which every term in the summation has at least one fundamental $\Delta^{\mu \nu}$ projector fully contracted with the $2 \ell + 2$-rank projector. 

Alternatively, contracting the linear combination with $p_{\mu_{1}} \cdots p_{\mu_{\ell-1}} g_{\mu_{\ell} \mu}$, we have $[\ell!/(2\ell-1)!!] = [(2\ell + 1)/(2\ell - 1)][(\ell-1)!/(2\ell-3)!!] B$, which follows from $p^{\langle \mu_{1}} \cdots p^{\mu_{\ell} \rangle} p_{\langle \mu_{1}} \cdots p_{\mu_{\ell} \rangle}
=
[\ell!/(2 \ell - 1)!!] 
(\Delta_{\mu \nu} p^{\mu} p^{\nu})^{\ell}$, $\Delta^{\mu_{1} \cdots \mu_{\ell} \mu}_{ \ \ \ \ \ \ \ \ \nu_{1} \cdots \nu_{\ell + 1}} g_{\mu_{\ell} \mu} = 0$, and 
$\Delta^{\mu_{1} \cdots \mu_{\ell}}_{ \ \ \ \ \ \ \ \ \nu_{1} \cdots \nu_{\ell}} \Delta^{\nu_{1} \mu} g_{\mu_{\ell} \mu} = [(2\ell + 1)/(2\ell - 1)] \Delta^{\mu_{1} \cdots \mu_{\ell -1}}_{ \ \ \ \ \ \ \ \ \nu_{1} \cdots \nu_{\ell-1}} \Delta^{\nu_{1} \mu}$. Hence, $B = [\ell / (2\ell + 1)](\Delta_{\mu \nu} p^{\mu} p^{\nu})$. Equation \eqref{eq:ids-Lag-tensor-2} is thus established. Afterwards, this equation is combined with the identity $\Delta_{\mu \nu} p^{\mu} p^{\nu} = - E_p^{2}$, which is valid for massless particles, and we derive 
\begin{equation}
\label{eq:moments-lag}
\begin{aligned}
&
i \frac{\Omega}{\beta} \left[
-(n+1) \widetilde{\Phi}_{n+1}^{\mu_{1} \cdots \mu_{\ell}}
+
2(n+\ell+1)
\widetilde{\Phi}_{n}^{\mu_{1} \cdots \mu_{\ell}}
-
(n+2\ell+1)
\widetilde{\Phi}_{n-1}^{\mu_{1} \cdots \mu_{\ell}}
\right]
+ 
i q_{\langle \mu \rangle} \left[ \widetilde{\Phi}_{n}^{\mu_{1} \cdots \mu_{\ell} \mu}
-
2 \widetilde{\Phi}_{n-1}^{\mu_{1} \cdots \mu_{\ell} \mu}
+
\widetilde{\Phi}_{n-2}^{\mu_{1} \cdots \mu_{\ell} \mu}
\right]
\\
&
-
\frac{i}{\beta^{2}}  
 \frac{\ell}{2 \ell + 1}
q^{\langle \mu_{1}} 
\left[
(n+1)(n+2)
\widetilde{\Phi}_{n+2}^{\mu_{2} \cdots \mu_{\ell} \rangle}
-
2(n+1)(n+2\ell+1)
\widetilde{\Phi}_{n+1}^{\mu_{2} \cdots \mu_{\ell} \rangle}
+
(n+2\ell+1)(n+2\ell)
\widetilde{\Phi}_{n}^{\mu_{2} \cdots \mu_{\ell} \rangle}
\right]
\\
&
-
\chi_{n\ell}
\widetilde{\Phi}_{n}^{\mu_{1} \cdots \mu_{\ell}}
   = 
\widetilde{\Xi}_{n}^{\mu_{1} \cdots \mu_{\ell}}.
\end{aligned}
\end{equation}

Now we turn our attention to the computation of the correlators of the stochastic components $\widetilde{\Xi}_{n}^{\mu_{1} \cdots \mu_{\ell}}(q)$. First, we substitute the expansion \eqref{eq:phi-xi-expn-1b} in Eq.~\eqref{eq:Fourier-correlators-Xi} for the Fourier correlators and integrate with the basis elements $L^{(2 \ell + 1)}_{n p} p_{\langle \mu_{1}} \cdots p_{ \mu_{\ell} \rangle}$ and $L^{(2 \ell + 1)}_{n, p'} p'_{\langle \mu_{1}} \cdots p'_{ \mu_{\ell} \rangle}$
\begin{equation}
\label{eq:XiXi-corr-apn}
\begin{aligned}
&
\left\langle \widetilde{\Xi}_{n}^{\mu_{1} \cdots \mu_{\ell}}(q) \widetilde{\Xi}_{n'}^{*\nu_{1} \cdots \nu_{m}}(q')
\right\rangle
=
\int \dfrac{d^3 p}{(2\pi)^3 p^0} \dfrac{d^3 \Bar{p}}{(2\pi)^3 \Bar{p}^0}  L^{(2 \ell + 1)}_{n p} p^{\langle \mu_{1}} \cdots p^{ \mu_{\ell} \rangle}
\left\langle \Bar{\xi}^{*}_{\Tilde{p}}(q)
 \Bar{\xi}_p(q')
 \right\rangle
 L^{(2 \ell + 1)}_{n \tilde{p}} \tilde{p}^{\langle \nu_{1}} \cdots \tilde{p}^{ \nu_{m} \rangle}
\\
&
=
- 2 (2 \pi)^{4} \delta^{4}(q-q')
\int \dfrac{d^3 p}{(2\pi)^3 p^0} \dfrac{d^3 \Bar{p}}{(2\pi)^3 \Bar{p}^0} L^{(2 \ell + 1)}_{n,p} p^{\langle \mu_{1}} \cdots p^{ \mu_{\ell} \rangle}
f_{\mathrm{eq},p}K_{p \tilde{p}}
 L^{(2 \ell + 1)}_{n, \tilde{p} } \tilde{p}^{\langle \nu_{1}} \cdots \tilde{p}^{ \nu_{m} \rangle}
\\
&
=
- 2 (2 \pi)^{4} \delta^{4}(q-q')
\int \dfrac{d^3 p}{(2\pi)^3 p^0} L^{(2 \ell + 1)}_{n,p} p^{\langle \mu_{1}} \cdots p^{ \mu_{\ell} \rangle}
f_{\mathrm{eq},p} \Hat{L}[L^{(2 \ell + 1)}_{n, p} p^{\langle \nu_{1}} \cdots p^{ \nu_{m} \rangle}]
\\
&
= 
-2 (2 \pi)^{4} A_{n}^{(\ell)} \chi_{n,\ell}  \delta^{4}(q-q') \delta_{\ell m} \delta_{nn'} \Delta^{\mu_{1} \cdots \mu_{\ell}\nu_{1} \cdots \nu_{\ell}},
\end{aligned}    
\end{equation}
where from the first to the second equality we employed Eq.~\eqref{eq:Fourier-correlators-Xi}, and from the second to the third equality, we employed the property \eqref{eq:fund-K-prop}, which defines the kernel $K_{p \Tilde{p}}$. We note that the above equation is independent of the interaction if one ``forgets'' for a moment that $L^{(2 \ell + 1)}_{n,p}$ denote Laguerre polynomials. In deriving Eq.~\eqref{eq:XiXi-corr-apn}, we have only used the fact that the linearized collision term possess eigenvalues $\chi_{n,\ell}$ whose corresponding eigenvectors possess finite norm $A_{n}^{(\ell)}$.

Finally, from Eqs.~\eqref{eq:phi-phi-correlators} and \eqref{eq:lag-fou-phi4-corr-final}, we compute the following expression for the correlators of the fractional deviation from equilibrium in Fourier space
\begin{equation}
\label{eq:phi-phi-correlators-apn}
\begin{aligned}
&
\langle \Bar{\phi}_{p}(q) \Bar{\phi}^{*}_{k}(q')  \rangle 
=
\sum_{n,\ell, n',m = 0}^{\infty} \frac{1}{A_{n}^{(\ell)} A_{n'}^{(m)} } 
\left\langle \widetilde{\Phi}_{n}^{\mu_{1} \cdots \mu_{\ell}}(q) \widetilde{\Phi}_{n'}^{*\nu_{1} \cdots \nu_{m}}(q') \right\rangle
L^{(2 \ell + 1)}_{n, p}
L^{(2 \ell + 1)}_{n', k}
 p_{\langle \mu_{1}} \cdots p_{ \mu_{\ell}
 \rangle}
  k_{\langle \mu_{1}} \cdots k_{ \mu_{m}
 \rangle}
\\
&
=
(2 \pi)^{4}
\sum_{n,n',\ell}^{\infty} \frac{(-1)^{\ell}}{\beta^{2\ell}} \frac{1}{A_{n}^{(\ell)} A_{n'}^{(\ell)} } 
\left( \widetilde{\Phi}_{n\ell}
\widetilde{\Phi}_{n' \ell}^{*}
 \right)_{\Omega}   L^{(2 \ell + 1)}_{n, p}
L^{(2 \ell + 1)}_{n', k}
 p_{\langle \mu_{1}} \cdots p_{ \mu_{\ell}
 \rangle}
  k^{\langle \mu_{1}} \cdots k^{ \mu_{\ell}
 \rangle}
 \delta(\Omega - \Omega').
\\
&
=
\frac{(2 \pi)^{4}}
{n_{0}^{2}\beta^{2}}
\sum_{n,n',\ell}^{\infty}  \frac{1}{\widehat{A}_{n}^{(\ell)} \widehat{A}_{n'}^{(\ell)} } 
\left( \widetilde{\Phi}_{n\ell}
\widetilde{\Phi}_{n' \ell}^{*}
 \right)_{\Omega}   L^{(2 \ell + 1)}_{n, p}
L^{(2 \ell + 1)}_{n', k}
 (\beta \vert \mathbf{p} \vert)^{\ell}
 (\beta \vert \mathbf{k} \vert)^{\ell} 
 \frac{\ell!}{(2\ell -1)!!}
P_{\ell}(\cos \theta)
 \delta(\Omega - \Omega'),
\end{aligned}    
\end{equation}
where $P_{\ell}(x)$ are the Legendre polynomials and $\cos \theta \equiv (\mathbf{p} \cdot \mathbf{k})/(\vert \mathbf{p} \vert \vert \mathbf{k} \vert)$. In the last step of the above derivation, we have employed the fact that 
\begin{equation}
\begin{aligned}
&
p_{\langle \mu_{1}} \cdots p_{\mu_{\ell} \rangle}
k^{\langle \mu_{1}} \cdots k^{\mu_{\ell} \rangle} 
= 
\sum_{s=0}^{\lceil \ell / 2 \rceil} C_{\ell s} \left(\Delta^{\mu \nu} k_{\mu} k_{\nu}\right)^{s} \left(\Delta^{\mu \nu} p_{\mu} p_{\nu}\right)^{s} 
\left(\Delta^{\mu \nu} k_{\mu} p_{\nu}\right)^{\ell - 2s}
\\
&
= 
(-1)^{\ell}
\vert \mathbf{p} \vert^{\ell} \vert \mathbf{k} \vert^{\ell}
\sum_{s=0}^{\lceil \ell / 2 \rceil}(-1)^{s} \frac{(\ell!)^{2}}{(2 \ell)!} \frac{(2 \ell - 2s)!}{s!(\ell-s)!(\ell-2s)!} \left(\cos \theta\right)^{\ell - 2s}
=
(-1)^{\ell}\frac{\ell!}{(2\ell -1)!!}
\vert \mathbf{p} \vert^{\ell} \vert \mathbf{k} \vert^{\ell} P_{\ell}(\cos \theta)
,
\end{aligned}    
\end{equation}
which follows from definition \eqref{eq:def-deltao-1} (see also 
Eq.~(7.28) from Ref.~\cite{Denicol:2021}) and the following representation of Legendre polynomials
\begin{equation}
\begin{aligned}
&
P_{\ell}(x) = \frac{1}{2^{\ell}}\sum_{s=0}^{\lceil \ell / 2 \rceil} 
(-1)^{s} \binom{\ell}{s}
\binom{2\ell-2s}{\ell} x^{\ell - 2s}
\end{aligned}    
\end{equation}
and the identity $(2 \ell)! = 2^{\ell} \ell! (2 \ell - 1)!!$.

\bibliography{references}

\begin{thebibliography}{67}%
\makeatletter
\providecommand \@ifxundefined [1]{%
 \@ifx{#1\undefined}
}%
\providecommand \@ifnum [1]{%
 \ifnum #1\expandafter \@firstoftwo
 \else \expandafter \@secondoftwo
 \fi
}%
\providecommand \@ifx [1]{%
 \ifx #1\expandafter \@firstoftwo
 \else \expandafter \@secondoftwo
 \fi
}%
\providecommand \natexlab [1]{#1}%
\providecommand \enquote  [1]{``#1''}%
\providecommand \bibnamefont  [1]{#1}%
\providecommand \bibfnamefont [1]{#1}%
\providecommand \citenamefont [1]{#1}%
\providecommand \href@noop [0]{\@secondoftwo}%
\providecommand \href [0]{\begingroup \@sanitize@url \@href}%
\providecommand \@href[1]{\@@startlink{#1}\@@href}%
\providecommand \@@href[1]{\endgroup#1\@@endlink}%
\providecommand \@sanitize@url [0]{\catcode `\\12\catcode `\$12\catcode `\&12\catcode `\#12\catcode `\^12\catcode `\_12\catcode `\%12\relax}%
\providecommand \@@startlink[1]{}%
\providecommand \@@endlink[0]{}%
\providecommand \url  [0]{\begingroup\@sanitize@url \@url }%
\providecommand \@url [1]{\endgroup\@href {#1}{\urlprefix }}%
\providecommand \urlprefix  [0]{URL }%
\providecommand \Eprint [0]{\href }%
\providecommand \doibase [0]{http://dx.doi.org/}%
\providecommand \selectlanguage [0]{\@gobble}%
\providecommand \bibinfo  [0]{\@secondoftwo}%
\providecommand \bibfield  [0]{\@secondoftwo}%
\providecommand \translation [1]{[#1]}%
\providecommand \BibitemOpen [0]{}%
\providecommand \bibitemStop [0]{}%
\providecommand \bibitemNoStop [0]{.\EOS\space}%
\providecommand \EOS [0]{\spacefactor3000\relax}%
\providecommand \BibitemShut  [1]{\csname bibitem#1\endcsname}%
\let\auto@bib@innerbib\@empty
\bibitem [{\citenamefont {Huang}(1987)}]{huang_book}%
  \BibitemOpen
  \bibfield  {author} {\bibinfo {author} {\bibfnamefont {K.}~\bibnamefont {Huang}},\ }\href@noop {} {\emph {\bibinfo {title} {Statistical Mechanics}}},\ \bibinfo {edition} {2nd}\ ed.\ (\bibinfo  {publisher} {John Wiley \& Sons},\ \bibinfo {year} {1987})\BibitemShut {NoStop}%
\bibitem [{\citenamefont {{Rezzolla}}\ and\ \citenamefont {{Zanotti}}(2013)}]{rezzolla_book}%
  \BibitemOpen
  \bibfield  {author} {\bibinfo {author} {\bibfnamefont {L.}~\bibnamefont {{Rezzolla}}}\ and\ \bibinfo {author} {\bibfnamefont {O.}~\bibnamefont {{Zanotti}}},\ }\href@noop {} {\emph {\bibinfo {title} {Relativistic Hydrodynamics, by L.~Rezzolla and O.~Zanotti.~Oxford University Press, 2013.~ISBN-10: 0198528906; ISBN-13: 978-0198528906}}}\ (\bibinfo {year} {2013})\BibitemShut {NoStop}%
\bibitem [{\citenamefont {{Hakim}}(2011)}]{Hakim2011}%
  \BibitemOpen
  \bibfield  {author} {\bibinfo {author} {\bibfnamefont {R{\'e}mi~Joel}\ \bibnamefont {{Hakim}}},\ }\href@noop {} {\emph {\bibinfo {title} {{Introduction to Relativistic Statistical Mechanics: Classical and Quantum}}}}\ (\bibinfo  {publisher} {World Scientific},\ \bibinfo {year} {2011})\BibitemShut {NoStop}%
\bibitem [{\citenamefont {{Florkowski}}\ \emph {et~al.}(2018)\citenamefont {{Florkowski}}, \citenamefont {{Heller}},\ and\ \citenamefont {{Spali{\'n}ski}}}]{FlorkowskiReview2018}%
  \BibitemOpen
  \bibfield  {author} {\bibinfo {author} {\bibfnamefont {Wojciech}\ \bibnamefont {{Florkowski}}}, \bibinfo {author} {\bibfnamefont {Michal~P.}\ \bibnamefont {{Heller}}}, \ and\ \bibinfo {author} {\bibfnamefont {Micha{\l}}\ \bibnamefont {{Spali{\'n}ski}}},\ }\bibfield  {title} {\enquote {\bibinfo {title} {{New theories of relativistic hydrodynamics in the LHC era}},}\ }\href {\doibase 10.1088/1361-6633/aaa091} {\bibfield  {journal} {\bibinfo  {journal} {Reports on Progress in Physics}\ }\textbf {\bibinfo {volume} {81}},\ \bibinfo {eid} {046001} (\bibinfo {year} {2018})},\ \Eprint {http://arxiv.org/abs/1707.02282} {arXiv:1707.02282 [hep-ph]} \BibitemShut {NoStop}%
\bibitem [{\citenamefont {{Romatschke}}(2010)}]{Romatschke2010}%
  \BibitemOpen
  \bibfield  {author} {\bibinfo {author} {\bibfnamefont {Paul}\ \bibnamefont {{Romatschke}}},\ }\bibfield  {title} {\enquote {\bibinfo {title} {{New Developments in Relativistic Viscous Hydrodynamics}},}\ }\href {\doibase 10.1142/S0218301310014613} {\bibfield  {journal} {\bibinfo  {journal} {International Journal of Modern Physics E}\ }\textbf {\bibinfo {volume} {19}},\ \bibinfo {pages} {1--53} (\bibinfo {year} {2010})},\ \Eprint {http://arxiv.org/abs/0902.3663} {arXiv:0902.3663 [hep-ph]} \BibitemShut {NoStop}%
\bibitem [{\citenamefont {Rocha}\ \emph {et~al.}(2023)\citenamefont {Rocha}, \citenamefont {de~Brito},\ and\ \citenamefont {Denicol}}]{Rocha:2023hts}%
  \BibitemOpen
  \bibfield  {author} {\bibinfo {author} {\bibfnamefont {Gabriel~S.}\ \bibnamefont {Rocha}}, \bibinfo {author} {\bibfnamefont {Caio V.~P.}\ \bibnamefont {de~Brito}}, \ and\ \bibinfo {author} {\bibfnamefont {Gabriel~S.}\ \bibnamefont {Denicol}},\ }\bibfield  {title} {\enquote {\bibinfo {title} {{Hydrodynamic theories for a system of weakly self-interacting classical ultrarelativistic scalar particles: Microscopic derivations and attractors}},}\ }\href {\doibase 10.1103/PhysRevD.108.036017} {\bibfield  {journal} {\bibinfo  {journal} {Phys. Rev. D}\ }\textbf {\bibinfo {volume} {108}},\ \bibinfo {pages} {036017} (\bibinfo {year} {2023})},\ \Eprint {http://arxiv.org/abs/2306.07423} {arXiv:2306.07423 [nucl-th]} \BibitemShut {NoStop}%
\bibitem [{\citenamefont {Rocha}\ \emph {et~al.}(2024{\natexlab{a}})\citenamefont {Rocha}, \citenamefont {Wagner}, \citenamefont {Denicol}, \citenamefont {Noronha},\ and\ \citenamefont {Rischke}}]{Rocha:2023ilf}%
  \BibitemOpen
  \bibfield  {author} {\bibinfo {author} {\bibfnamefont {Gabriel~S.}\ \bibnamefont {Rocha}}, \bibinfo {author} {\bibfnamefont {David}\ \bibnamefont {Wagner}}, \bibinfo {author} {\bibfnamefont {Gabriel~S.}\ \bibnamefont {Denicol}}, \bibinfo {author} {\bibfnamefont {Jorge}\ \bibnamefont {Noronha}}, \ and\ \bibinfo {author} {\bibfnamefont {Dirk~H.}\ \bibnamefont {Rischke}},\ }\bibfield  {title} {\enquote {\bibinfo {title} {{Theories of Relativistic Dissipative Fluid Dynamics}},}\ }\href {\doibase 10.3390/e26030189} {\bibfield  {journal} {\bibinfo  {journal} {Entropy}\ }\textbf {\bibinfo {volume} {26}},\ \bibinfo {pages} {189} (\bibinfo {year} {2024}{\natexlab{a}})},\ \Eprint {http://arxiv.org/abs/2311.15063} {arXiv:2311.15063 [nucl-th]} \BibitemShut {NoStop}%
\bibitem [{\citenamefont {De~Groot}(1980)}]{DeGroot:1980dk}%
  \BibitemOpen
  \bibfield  {author} {\bibinfo {author} {\bibfnamefont {S.~R.}\ \bibnamefont {De~Groot}},\ }\href@noop {} {\emph {\bibinfo {title} {{Relativistic Kinetic Theory. Principles and Applications}}}},\ edited by\ \bibinfo {editor} {\bibfnamefont {W.~A.}\ \bibnamefont {Van~Leeuwen}}\ and\ \bibinfo {editor} {\bibfnamefont {C.~G.}\ \bibnamefont {Van~Weert}}\ (\bibinfo {year} {1980})\BibitemShut {NoStop}%
\bibitem [{\citenamefont {Cercignani}(1990)}]{cercignani:90mathematical}%
  \BibitemOpen
  \bibfield  {author} {\bibinfo {author} {\bibfnamefont {C.}~\bibnamefont {Cercignani}},\ }\href@noop {} {\emph {\bibinfo {title} {Mathematical methods in kinetic theory}}}\ (\bibinfo  {publisher} {Springer},\ \bibinfo {year} {1990})\BibitemShut {NoStop}%
\bibitem [{\citenamefont {{Jaynes}}(1965)}]{Jaynes1965}%
  \BibitemOpen
  \bibfield  {author} {\bibinfo {author} {\bibfnamefont {E.~T.}\ \bibnamefont {{Jaynes}}},\ }\bibfield  {title} {\enquote {\bibinfo {title} {{Gibbs vs Boltzmann Entropies}},}\ }\href {\doibase 10.1119/1.1971557} {\bibfield  {journal} {\bibinfo  {journal} {American Journal of Physics}\ }\textbf {\bibinfo {volume} {33}},\ \bibinfo {pages} {391--398} (\bibinfo {year} {1965})}\BibitemShut {NoStop}%
\bibitem [{\citenamefont {{Sevick}}\ \emph {et~al.}(2008)\citenamefont {{Sevick}}, \citenamefont {{Prabhakar}}, \citenamefont {{Williams}},\ and\ \citenamefont {{Searles}}}]{Sevick2008}%
  \BibitemOpen
  \bibfield  {author} {\bibinfo {author} {\bibfnamefont {E.~M.}\ \bibnamefont {{Sevick}}}, \bibinfo {author} {\bibfnamefont {R.}~\bibnamefont {{Prabhakar}}}, \bibinfo {author} {\bibfnamefont {Stephen~R.}\ \bibnamefont {{Williams}}}, \ and\ \bibinfo {author} {\bibfnamefont {Debra~J.}\ \bibnamefont {{Searles}}},\ }\bibfield  {title} {\enquote {\bibinfo {title} {{Fluctuation Theorems}},}\ }\href {\doibase 10.1146/annurev.physchem.58.032806.104555} {\bibfield  {journal} {\bibinfo  {journal} {Annual Review of Physical Chemistry}\ }\textbf {\bibinfo {volume} {59}},\ \bibinfo {pages} {603--633} (\bibinfo {year} {2008})},\ \Eprint {http://arxiv.org/abs/0709.3888} {arXiv:0709.3888 [cond-mat.stat-mech]} \BibitemShut {NoStop}%
\bibitem [{\citenamefont {Pitaevskii}\ and\ \citenamefont {Lifshitz}(2012)}]{landau10}%
  \BibitemOpen
  \bibfield  {author} {\bibinfo {author} {\bibfnamefont {L.P.}\ \bibnamefont {Pitaevskii}}\ and\ \bibinfo {author} {\bibfnamefont {E.M.}\ \bibnamefont {Lifshitz}},\ }\href {https://books.google.pl/books?id=DTHxPDfV0fQC} {\emph {\bibinfo {title} {Physical Kinetics}}},\ \bibinfo {number} {v. 10}\ (\bibinfo  {publisher} {Elsevier Science},\ \bibinfo {year} {2012})\BibitemShut {NoStop}%
\bibitem [{\citenamefont {Fox}\ and\ \citenamefont {Uhlenbeck}(1970{\natexlab{a}})}]{fox1970contributions}%
  \BibitemOpen
  \bibfield  {author} {\bibinfo {author} {\bibfnamefont {Ronald~Forrest}\ \bibnamefont {Fox}}\ and\ \bibinfo {author} {\bibfnamefont {George~E}\ \bibnamefont {Uhlenbeck}},\ }\bibfield  {title} {\enquote {\bibinfo {title} {Contributions to nonequilibrium thermodynamics. ii. fluctuation theory for the boltzmann equation},}\ }\href@noop {} {\bibfield  {journal} {\bibinfo  {journal} {The Physics of Fluids}\ }\textbf {\bibinfo {volume} {13}},\ \bibinfo {pages} {2881--2890} (\bibinfo {year} {1970}{\natexlab{a}})}\BibitemShut {NoStop}%
\bibitem [{\citenamefont {Bixon}\ and\ \citenamefont {Zwanzig}(1969)}]{Bixon:1969zz}%
  \BibitemOpen
  \bibfield  {author} {\bibinfo {author} {\bibfnamefont {Mordechai}\ \bibnamefont {Bixon}}\ and\ \bibinfo {author} {\bibfnamefont {Robert}\ \bibnamefont {Zwanzig}},\ }\bibfield  {title} {\enquote {\bibinfo {title} {{Boltzmann-Langevin Equation and Hydrodynamic Fluctuations}},}\ }\href {\doibase 10.1103/PhysRev.187.267} {\bibfield  {journal} {\bibinfo  {journal} {Phys. Rev.}\ }\textbf {\bibinfo {volume} {187}},\ \bibinfo {pages} {267--272} (\bibinfo {year} {1969})}\BibitemShut {NoStop}%
\bibitem [{\citenamefont {Gavin}\ \emph {et~al.}(2017)\citenamefont {Gavin}, \citenamefont {Moschelli},\ and\ \citenamefont {Zin}}]{Gavin:2016nir}%
  \BibitemOpen
  \bibfield  {author} {\bibinfo {author} {\bibfnamefont {Sean}\ \bibnamefont {Gavin}}, \bibinfo {author} {\bibfnamefont {George}\ \bibnamefont {Moschelli}}, \ and\ \bibinfo {author} {\bibfnamefont {Christopher}\ \bibnamefont {Zin}},\ }\bibfield  {title} {\enquote {\bibinfo {title} {{Boltzmann-Langevin Approach to Pre-equilibrium Correlations in Nuclear Collisions}},}\ }\href {\doibase 10.1103/PhysRevC.95.064901} {\bibfield  {journal} {\bibinfo  {journal} {Phys. Rev. C}\ }\textbf {\bibinfo {volume} {95}},\ \bibinfo {pages} {064901} (\bibinfo {year} {2017})},\ \Eprint {http://arxiv.org/abs/1612.07856} {arXiv:1612.07856 [nucl-th]} \BibitemShut {NoStop}%
\bibitem [{\citenamefont {Mir\'on-Granese}\ \emph {et~al.}(2020)\citenamefont {Mir\'on-Granese}, \citenamefont {Kandus},\ and\ \citenamefont {Calzetta}}]{Miron-Granese:2020mbf}%
  \BibitemOpen
  \bibfield  {author} {\bibinfo {author} {\bibfnamefont {Nahuel}\ \bibnamefont {Mir\'on-Granese}}, \bibinfo {author} {\bibfnamefont {Alejandra}\ \bibnamefont {Kandus}}, \ and\ \bibinfo {author} {\bibfnamefont {Esteban}\ \bibnamefont {Calzetta}},\ }\bibfield  {title} {\enquote {\bibinfo {title} {{Nonlinear Fluctuations in Relativistic Causal Fluids}},}\ }\href {\doibase 10.1007/JHEP07(2020)064} {\bibfield  {journal} {\bibinfo  {journal} {JHEP}\ }\textbf {\bibinfo {volume} {07}},\ \bibinfo {pages} {064} (\bibinfo {year} {2020})},\ \Eprint {http://arxiv.org/abs/2002.08323} {arXiv:2002.08323 [hep-th]} \BibitemShut {NoStop}%
\bibitem [{\citenamefont {Torrieri}(2024)}]{Torrieri:2023thk}%
  \BibitemOpen
  \bibfield  {author} {\bibinfo {author} {\bibfnamefont {Giorgio}\ \bibnamefont {Torrieri}},\ }\bibfield  {title} {\enquote {\bibinfo {title} {{The functional generalization of the Boltzmann-Vlasov equation and its Gauge-like symmetry}},}\ }\href {\doibase 10.21468/SciPostPhys.16.3.070} {\bibfield  {journal} {\bibinfo  {journal} {SciPost Phys.}\ }\textbf {\bibinfo {volume} {16}},\ \bibinfo {pages} {070} (\bibinfo {year} {2024})},\ \Eprint {http://arxiv.org/abs/2309.05154} {arXiv:2309.05154 [cond-mat.stat-mech]} \BibitemShut {NoStop}%
\bibitem [{\citenamefont {Gavassino}\ \emph {et~al.}(2022{\natexlab{a}})\citenamefont {Gavassino}, \citenamefont {Antonelli},\ and\ \citenamefont {Haskell}}]{Gavassino:2021kjm}%
  \BibitemOpen
  \bibfield  {author} {\bibinfo {author} {\bibfnamefont {Lorenzo}\ \bibnamefont {Gavassino}}, \bibinfo {author} {\bibfnamefont {Marco}\ \bibnamefont {Antonelli}}, \ and\ \bibinfo {author} {\bibfnamefont {Brynmor}\ \bibnamefont {Haskell}},\ }\bibfield  {title} {\enquote {\bibinfo {title} {{Thermodynamic Stability Implies Causality}},}\ }\href {\doibase 10.1103/PhysRevLett.128.010606} {\bibfield  {journal} {\bibinfo  {journal} {Phys. Rev. Lett.}\ }\textbf {\bibinfo {volume} {128}},\ \bibinfo {pages} {010606} (\bibinfo {year} {2022}{\natexlab{a}})},\ \Eprint {http://arxiv.org/abs/2105.14621} {arXiv:2105.14621 [gr-qc]} \BibitemShut {NoStop}%
\bibitem [{\citenamefont {van Kampen}(1968)}]{vanKampen1968}%
  \BibitemOpen
  \bibfield  {author} {\bibinfo {author} {\bibfnamefont {N.~G.}\ \bibnamefont {van Kampen}},\ }\bibfield  {title} {\enquote {\bibinfo {title} {Relativistic thermodynamics of moving systems},}\ }\href {\doibase 10.1103/PhysRev.173.295} {\bibfield  {journal} {\bibinfo  {journal} {Phys. Rev.}\ }\textbf {\bibinfo {volume} {173}},\ \bibinfo {pages} {295--301} (\bibinfo {year} {1968})}\BibitemShut {NoStop}%
\bibitem [{\citenamefont {Israel}(1989)}]{Noto_full}%
  \BibitemOpen
  \bibfield  {author} {\bibinfo {author} {\bibfnamefont {Werner}\ \bibnamefont {Israel}},\ }\bibfield  {title} {\enquote {\bibinfo {title} {Covariant fluid mechanics and thermodynamics: An introduction},}\ }in\ \href@noop {} {\emph {\bibinfo {booktitle} {Relativistic Fluid Dynamics}}},\ \bibinfo {editor} {edited by\ \bibinfo {editor} {\bibfnamefont {Angelo~M.}\ \bibnamefont {Anile}}\ and\ \bibinfo {editor} {\bibfnamefont {Yvonne}\ \bibnamefont {Choquet-Bruhat}}}\ (\bibinfo  {publisher} {Springer Berlin Heidelberg},\ \bibinfo {address} {Berlin, Heidelberg},\ \bibinfo {year} {1989})\ pp.\ \bibinfo {pages} {152--210}\BibitemShut {NoStop}%
\bibitem [{\citenamefont {Israel}(1981)}]{Israel_1981_review}%
  \BibitemOpen
  \bibfield  {author} {\bibinfo {author} {\bibfnamefont {W.}~\bibnamefont {Israel}},\ }\bibfield  {title} {\enquote {\bibinfo {title} {Thermodynamics of relativistic systems},}\ }\href {\doibase https://doi.org/10.1016/0378-4371(81)90220-X} {\bibfield  {journal} {\bibinfo  {journal} {Physica A: Statistical Mechanics and its Applications}\ }\textbf {\bibinfo {volume} {106}},\ \bibinfo {pages} {204 -- 214} (\bibinfo {year} {1981})}\BibitemShut {NoStop}%
\bibitem [{\citenamefont {Mullins}\ \emph {et~al.}(2023{\natexlab{a}})\citenamefont {Mullins}, \citenamefont {Hippert},\ and\ \citenamefont {Noronha}}]{Mullins:2023tjg}%
  \BibitemOpen
  \bibfield  {author} {\bibinfo {author} {\bibfnamefont {Nicki}\ \bibnamefont {Mullins}}, \bibinfo {author} {\bibfnamefont {Mauricio}\ \bibnamefont {Hippert}}, \ and\ \bibinfo {author} {\bibfnamefont {Jorge}\ \bibnamefont {Noronha}},\ }\bibfield  {title} {\enquote {\bibinfo {title} {{Stochastic fluctuations in relativistic fluids: causality, stability, and the information current}},}\ }\href@noop {} {\  (\bibinfo {year} {2023}{\natexlab{a}})},\ \Eprint {http://arxiv.org/abs/2306.08635} {arXiv:2306.08635 [nucl-th]} \BibitemShut {NoStop}%
\bibitem [{\citenamefont {Mullins}\ \emph {et~al.}(2023{\natexlab{b}})\citenamefont {Mullins}, \citenamefont {Hippert}, \citenamefont {Gavassino},\ and\ \citenamefont {Noronha}}]{Mullins:2023ott}%
  \BibitemOpen
  \bibfield  {author} {\bibinfo {author} {\bibfnamefont {Nicki}\ \bibnamefont {Mullins}}, \bibinfo {author} {\bibfnamefont {Mauricio}\ \bibnamefont {Hippert}}, \bibinfo {author} {\bibfnamefont {Lorenzo}\ \bibnamefont {Gavassino}}, \ and\ \bibinfo {author} {\bibfnamefont {Jorge}\ \bibnamefont {Noronha}},\ }\bibfield  {title} {\enquote {\bibinfo {title} {{Relativistic hydrodynamic fluctuations from an effective action: causality, stability, and the information current}},}\ }\href@noop {} {\  (\bibinfo {year} {2023}{\natexlab{b}})},\ \Eprint {http://arxiv.org/abs/2309.00512} {arXiv:2309.00512 [hep-th]} \BibitemShut {NoStop}%
\bibitem [{\citenamefont {Romatschke}(2016)}]{Romatschke:2015gic}%
  \BibitemOpen
  \bibfield  {author} {\bibinfo {author} {\bibfnamefont {Paul}\ \bibnamefont {Romatschke}},\ }\bibfield  {title} {\enquote {\bibinfo {title} {{Retarded correlators in kinetic theory: branch cuts, poles and hydrodynamic onset transitions}},}\ }\href {\doibase 10.1140/epjc/s10052-016-4169-7} {\bibfield  {journal} {\bibinfo  {journal} {Eur. Phys. J. C}\ }\textbf {\bibinfo {volume} {76}},\ \bibinfo {pages} {352} (\bibinfo {year} {2016})},\ \Eprint {http://arxiv.org/abs/1512.02641} {arXiv:1512.02641 [hep-th]} \BibitemShut {NoStop}%
\bibitem [{\citenamefont {Kurkela}\ and\ \citenamefont {Wiedemann}(2019)}]{Kurkela:2017xis}%
  \BibitemOpen
  \bibfield  {author} {\bibinfo {author} {\bibfnamefont {Aleksi}\ \bibnamefont {Kurkela}}\ and\ \bibinfo {author} {\bibfnamefont {Urs~Achim}\ \bibnamefont {Wiedemann}},\ }\bibfield  {title} {\enquote {\bibinfo {title} {{Analytic structure of nonhydrodynamic modes in kinetic theory}},}\ }\href {\doibase 10.1140/epjc/s10052-019-7271-9} {\bibfield  {journal} {\bibinfo  {journal} {Eur. Phys. J. C}\ }\textbf {\bibinfo {volume} {79}},\ \bibinfo {pages} {776} (\bibinfo {year} {2019})},\ \Eprint {http://arxiv.org/abs/1712.04376} {arXiv:1712.04376 [hep-ph]} \BibitemShut {NoStop}%
\bibitem [{\citenamefont {Bajec}\ \emph {et~al.}(2024)\citenamefont {Bajec}, \citenamefont {Grozdanov},\ and\ \citenamefont {Soloviev}}]{Bajec:2024jez}%
  \BibitemOpen
  \bibfield  {author} {\bibinfo {author} {\bibfnamefont {Matej}\ \bibnamefont {Bajec}}, \bibinfo {author} {\bibfnamefont {Sa\v{s}o}\ \bibnamefont {Grozdanov}}, \ and\ \bibinfo {author} {\bibfnamefont {Alexander}\ \bibnamefont {Soloviev}},\ }\bibfield  {title} {\enquote {\bibinfo {title} {{Spectra of correlators in the relaxation time approximation of kinetic theory}},}\ }\href@noop {} {\  (\bibinfo {year} {2024})},\ \Eprint {http://arxiv.org/abs/2403.17769} {arXiv:2403.17769 [hep-th]} \BibitemShut {NoStop}%
\bibitem [{\citenamefont {Denicol}\ and\ \citenamefont {Noronha}(2024)}]{Denicol:2022bsq}%
  \BibitemOpen
  \bibfield  {author} {\bibinfo {author} {\bibfnamefont {Gabriel~S.}\ \bibnamefont {Denicol}}\ and\ \bibinfo {author} {\bibfnamefont {Jorge}\ \bibnamefont {Noronha}},\ }\bibfield  {title} {\enquote {\bibinfo {title} {{Spectrum of the Boltzmann collision operator for \ensuremath{\lambda}\ensuremath{\phi}4 theory in the classical regime}},}\ }\href {\doibase 10.1016/j.physletb.2024.138487} {\bibfield  {journal} {\bibinfo  {journal} {Phys. Lett. B}\ }\textbf {\bibinfo {volume} {850}},\ \bibinfo {pages} {138487} (\bibinfo {year} {2024})},\ \Eprint {http://arxiv.org/abs/2209.10370} {arXiv:2209.10370 [nucl-th]} \BibitemShut {NoStop}%
\bibitem [{\citenamefont {Gavassino}(2020)}]{GavassinoTermometri}%
  \BibitemOpen
  \bibfield  {author} {\bibinfo {author} {\bibfnamefont {Lorenzo}\ \bibnamefont {Gavassino}},\ }\bibfield  {title} {\enquote {\bibinfo {title} {{The zeroth law of thermodynamics in special relativity}},}\ }\href {\doibase 10.1007/s10701-020-00393-x} {\bibfield  {journal} {\bibinfo  {journal} {Found. Phys.}\ }\textbf {\bibinfo {volume} {50}},\ \bibinfo {pages} {1554--1586} (\bibinfo {year} {2020})},\ \Eprint {http://arxiv.org/abs/2005.06396} {arXiv:2005.06396 [gr-qc]} \BibitemShut {NoStop}%
\bibitem [{\citenamefont {Hakim}(2011)}]{HakimBook}%
  \BibitemOpen
  \bibfield  {author} {\bibinfo {author} {\bibfnamefont {Rémi}\ \bibnamefont {Hakim}},\ }\href {\doibase 10.1142/7881} {\emph {\bibinfo {title} {Introduction to Relativistic Statistical Mechanics: Classical and Quantum}}}\ (\bibinfo  {publisher} {World Scientific},\ \bibinfo {address} {New Jersey},\ \bibinfo {year} {2011})\ pp.\ \bibinfo {pages} {1--538}\BibitemShut {NoStop}%
\bibitem [{\citenamefont {Landau}\ and\ \citenamefont {Lifshitz}(1980)}]{landau_statistical_1980}%
  \BibitemOpen
  \bibfield  {author} {\bibinfo {author} {\bibfnamefont {L.~D.}\ \bibnamefont {Landau}}\ and\ \bibinfo {author} {\bibfnamefont {E.~M.}\ \bibnamefont {Lifshitz}},\ }\href@noop {} {\emph {\bibinfo {title} {Statistical Physics Part I - Volume 5 (Course of Theoretical Physics)}}},\ \bibinfo {edition} {3rd}\ ed.\ (\bibinfo  {publisher} {Butterworth-Heinemann},\ \bibinfo {address} {Oxford, UK},\ \bibinfo {year} {1980})\BibitemShut {NoStop}%
\bibitem [{\citenamefont {Gavassino}(2021)}]{Gavassino:2021cli}%
  \BibitemOpen
  \bibfield  {author} {\bibinfo {author} {\bibfnamefont {Lorenzo}\ \bibnamefont {Gavassino}},\ }\bibfield  {title} {\enquote {\bibinfo {title} {{Applying the Gibbs stability criterion to relativistic hydrodynamics}},}\ }\href {\doibase 10.1088/1361-6382/ac2b0e} {\bibfield  {journal} {\bibinfo  {journal} {Class. Quant. Grav.}\ }\textbf {\bibinfo {volume} {38}},\ \bibinfo {pages} {21LT02} (\bibinfo {year} {2021})},\ \Eprint {http://arxiv.org/abs/2104.09142} {arXiv:2104.09142 [gr-qc]} \BibitemShut {NoStop}%
\bibitem [{\citenamefont {Gavassino}\ \emph {et~al.}(2024)\citenamefont {Gavassino}, \citenamefont {Mullins},\ and\ \citenamefont {Hippert}}]{Gavassino:2024vyu}%
  \BibitemOpen
  \bibfield  {author} {\bibinfo {author} {\bibfnamefont {Lorenzo}\ \bibnamefont {Gavassino}}, \bibinfo {author} {\bibfnamefont {Nicki}\ \bibnamefont {Mullins}}, \ and\ \bibinfo {author} {\bibfnamefont {Mauricio}\ \bibnamefont {Hippert}},\ }\bibfield  {title} {\enquote {\bibinfo {title} {{Consistent inclusion of fluctuations in first-order causal and stable relativistic hydrodynamics}},}\ }\href@noop {} {\  (\bibinfo {year} {2024})},\ \Eprint {http://arxiv.org/abs/2402.06776} {arXiv:2402.06776 [nucl-th]} \BibitemShut {NoStop}%
\bibitem [{\citenamefont {Olson}(1990)}]{OLSON199018}%
  \BibitemOpen
  \bibfield  {author} {\bibinfo {author} {\bibfnamefont {Timothy~S}\ \bibnamefont {Olson}},\ }\bibfield  {title} {\enquote {\bibinfo {title} {Stability and causality in the israel-stewart energy frame theory},}\ }\href {\doibase https://doi.org/10.1016/0003-4916(90)90366-V} {\bibfield  {journal} {\bibinfo  {journal} {Annals of Physics}\ }\textbf {\bibinfo {volume} {199}},\ \bibinfo {pages} {18--36} (\bibinfo {year} {1990})}\BibitemShut {NoStop}%
\bibitem [{\citenamefont {Gavassino}\ \emph {et~al.}(2022{\natexlab{b}})\citenamefont {Gavassino}, \citenamefont {Antonelli},\ and\ \citenamefont {Haskell}}]{GavassinoSymmetricQuasi2022roi}%
  \BibitemOpen
  \bibfield  {author} {\bibinfo {author} {\bibfnamefont {Lorenzo}\ \bibnamefont {Gavassino}}, \bibinfo {author} {\bibfnamefont {Marco}\ \bibnamefont {Antonelli}}, \ and\ \bibinfo {author} {\bibfnamefont {Brynmor}\ \bibnamefont {Haskell}},\ }\bibfield  {title} {\enquote {\bibinfo {title} {{Symmetric-hyperbolic quasihydrodynamics}},}\ }\href {\doibase 10.1103/PhysRevD.106.056010} {\bibfield  {journal} {\bibinfo  {journal} {Phys. Rev. D}\ }\textbf {\bibinfo {volume} {106}},\ \bibinfo {pages} {056010} (\bibinfo {year} {2022}{\natexlab{b}})},\ \Eprint {http://arxiv.org/abs/2207.14778} {arXiv:2207.14778 [gr-qc]} \BibitemShut {NoStop}%
\bibitem [{\citenamefont {Denicol}\ \emph {et~al.}(2012)\citenamefont {Denicol}, \citenamefont {Niemi}, \citenamefont {Molnar},\ and\ \citenamefont {Rischke}}]{Denicol:2012cn}%
  \BibitemOpen
  \bibfield  {author} {\bibinfo {author} {\bibfnamefont {G.~S.}\ \bibnamefont {Denicol}}, \bibinfo {author} {\bibfnamefont {H.}~\bibnamefont {Niemi}}, \bibinfo {author} {\bibfnamefont {E.}~\bibnamefont {Molnar}}, \ and\ \bibinfo {author} {\bibfnamefont {D.~H.}\ \bibnamefont {Rischke}},\ }\bibfield  {title} {\enquote {\bibinfo {title} {{Derivation of transient relativistic fluid dynamics from the Boltzmann equation}},}\ }\href {\doibase 10.1103/PhysRevD.85.114047} {\bibfield  {journal} {\bibinfo  {journal} {Phys. Rev. D}\ }\textbf {\bibinfo {volume} {85}},\ \bibinfo {pages} {114047} (\bibinfo {year} {2012})},\ \bibinfo {note} {[Erratum: Phys.Rev.D 91, 039902 (2015)]},\ \Eprint {http://arxiv.org/abs/1202.4551} {arXiv:1202.4551 [nucl-th]} \BibitemShut {NoStop}%
\bibitem [{\citenamefont {Wagner}\ \emph {et~al.}(2022)\citenamefont {Wagner}, \citenamefont {Palermo},\ and\ \citenamefont {Ambru\c{s}}}]{Wagner:2022ayd}%
  \BibitemOpen
  \bibfield  {author} {\bibinfo {author} {\bibfnamefont {David}\ \bibnamefont {Wagner}}, \bibinfo {author} {\bibfnamefont {Andrea}\ \bibnamefont {Palermo}}, \ and\ \bibinfo {author} {\bibfnamefont {Victor~E.}\ \bibnamefont {Ambru\c{s}}},\ }\bibfield  {title} {\enquote {\bibinfo {title} {{Inverse-Reynolds-dominance approach to transient fluid dynamics}},}\ }\href {\doibase 10.1103/PhysRevD.106.016013} {\bibfield  {journal} {\bibinfo  {journal} {Phys. Rev. D}\ }\textbf {\bibinfo {volume} {106}},\ \bibinfo {pages} {016013} (\bibinfo {year} {2022})},\ \Eprint {http://arxiv.org/abs/2203.12608} {arXiv:2203.12608 [nucl-th]} \BibitemShut {NoStop}%
\bibitem [{\citenamefont {Gavassino}\ \emph {et~al.}(2022{\natexlab{c}})\citenamefont {Gavassino}, \citenamefont {Antonelli},\ and\ \citenamefont {Haskell}}]{Gavassino:2022roi}%
  \BibitemOpen
  \bibfield  {author} {\bibinfo {author} {\bibfnamefont {Lorenzo}\ \bibnamefont {Gavassino}}, \bibinfo {author} {\bibfnamefont {Marco}\ \bibnamefont {Antonelli}}, \ and\ \bibinfo {author} {\bibfnamefont {Brynmor}\ \bibnamefont {Haskell}},\ }\bibfield  {title} {\enquote {\bibinfo {title} {{Symmetric-hyperbolic quasihydrodynamics}},}\ }\href {\doibase 10.1103/PhysRevD.106.056010} {\bibfield  {journal} {\bibinfo  {journal} {Phys. Rev. D}\ }\textbf {\bibinfo {volume} {106}},\ \bibinfo {pages} {056010} (\bibinfo {year} {2022}{\natexlab{c}})},\ \Eprint {http://arxiv.org/abs/2207.14778} {arXiv:2207.14778 [gr-qc]} \BibitemShut {NoStop}%
\bibitem [{\citenamefont {Lifshitz}\ and\ \citenamefont {Pitaevskii}(1980)}]{landau_statistical_part_II}%
  \BibitemOpen
  \bibfield  {author} {\bibinfo {author} {\bibfnamefont {E.~M.}\ \bibnamefont {Lifshitz}}\ and\ \bibinfo {author} {\bibfnamefont {L.~P.}\ \bibnamefont {Pitaevskii}},\ }\href@noop {} {\emph {\bibinfo {title} {Statistical Physics Part II - Volume 9 (Course of Theoretical Physics)}}}\ (\bibinfo  {publisher} {Butterworth-Heinemann},\ \bibinfo {address} {Oxford, UK},\ \bibinfo {year} {1980})\BibitemShut {NoStop}%
\bibitem [{\citenamefont {Ambrus}\ \emph {et~al.}(2022)\citenamefont {Ambrus}, \citenamefont {Moln\'ar},\ and\ \citenamefont {Rischke}}]{Ambrus:2022vif}%
  \BibitemOpen
  \bibfield  {author} {\bibinfo {author} {\bibfnamefont {Victor~E.}\ \bibnamefont {Ambrus}}, \bibinfo {author} {\bibfnamefont {Etele}\ \bibnamefont {Moln\'ar}}, \ and\ \bibinfo {author} {\bibfnamefont {Dirk~H.}\ \bibnamefont {Rischke}},\ }\bibfield  {title} {\enquote {\bibinfo {title} {{Transport coefficients of second-order relativistic fluid dynamics in the relaxation-time approximation}},}\ }\href {\doibase 10.1103/PhysRevD.106.076005} {\bibfield  {journal} {\bibinfo  {journal} {Phys. Rev. D}\ }\textbf {\bibinfo {volume} {106}},\ \bibinfo {pages} {076005} (\bibinfo {year} {2022})},\ \Eprint {http://arxiv.org/abs/2207.05670} {arXiv:2207.05670 [nucl-th]} \BibitemShut {NoStop}%
\bibitem [{\citenamefont {Wagner}\ and\ \citenamefont {Gavassino}(2023)}]{Wagner:2023jgq}%
  \BibitemOpen
  \bibfield  {author} {\bibinfo {author} {\bibfnamefont {David}\ \bibnamefont {Wagner}}\ and\ \bibinfo {author} {\bibfnamefont {Lorenzo}\ \bibnamefont {Gavassino}},\ }\bibfield  {title} {\enquote {\bibinfo {title} {{The regime of applicability of Israel-Stewart hydrodynamics}},}\ }\href@noop {} {\  (\bibinfo {year} {2023})},\ \Eprint {http://arxiv.org/abs/2309.14828} {arXiv:2309.14828 [nucl-th]} \BibitemShut {NoStop}%
\bibitem [{\citenamefont {Petrosyan}\ and\ \citenamefont {Zaccone}(2022)}]{Petrosyan:2021lqi}%
  \BibitemOpen
  \bibfield  {author} {\bibinfo {author} {\bibfnamefont {Aleksandr}\ \bibnamefont {Petrosyan}}\ and\ \bibinfo {author} {\bibfnamefont {Alessio}\ \bibnamefont {Zaccone}},\ }\bibfield  {title} {\enquote {\bibinfo {title} {{Relativistic Langevin equation derived from a particle-bath Lagrangian}},}\ }\href {\doibase 10.1088/1751-8121/ac3a33} {\bibfield  {journal} {\bibinfo  {journal} {J. Phys. A}\ }\textbf {\bibinfo {volume} {55}},\ \bibinfo {pages} {015001} (\bibinfo {year} {2022})},\ \Eprint {http://arxiv.org/abs/2107.07205} {arXiv:2107.07205 [cond-mat.stat-mech]} \BibitemShut {NoStop}%
\bibitem [{\citenamefont {Peskin}\ and\ \citenamefont {Schroeder}(1995)}]{Peskin_book}%
  \BibitemOpen
  \bibfield  {author} {\bibinfo {author} {\bibfnamefont {Michael~E.}\ \bibnamefont {Peskin}}\ and\ \bibinfo {author} {\bibfnamefont {Daniel~V.}\ \bibnamefont {Schroeder}},\ }\href {http://www.slac.stanford.edu/~mpeskin/QFT.html} {\emph {\bibinfo {title} {{An introduction to quantum field theory}}}}\ (\bibinfo  {publisher} {Addison-Wesley},\ \bibinfo {address} {Reading, USA},\ \bibinfo {year} {1995})\BibitemShut {NoStop}%
\bibitem [{\citenamefont {Rocha}\ \emph {et~al.}(2021)\citenamefont {Rocha}, \citenamefont {Denicol},\ and\ \citenamefont {Noronha}}]{Rocha:2021zcw}%
  \BibitemOpen
  \bibfield  {author} {\bibinfo {author} {\bibfnamefont {Gabriel~S.}\ \bibnamefont {Rocha}}, \bibinfo {author} {\bibfnamefont {Gabriel~S.}\ \bibnamefont {Denicol}}, \ and\ \bibinfo {author} {\bibfnamefont {Jorge}\ \bibnamefont {Noronha}},\ }\bibfield  {title} {\enquote {\bibinfo {title} {{Novel Relaxation Time Approximation to the Relativistic Boltzmann Equation}},}\ }\href {\doibase 10.1103/PhysRevLett.127.042301} {\bibfield  {journal} {\bibinfo  {journal} {Phys. Rev. Lett.}\ }\textbf {\bibinfo {volume} {127}},\ \bibinfo {pages} {042301} (\bibinfo {year} {2021})},\ \Eprint {http://arxiv.org/abs/2103.07489} {arXiv:2103.07489 [nucl-th]} \BibitemShut {NoStop}%
\bibitem [{\citenamefont {Rocha}\ \emph {et~al.}(2022)\citenamefont {Rocha}, \citenamefont {Ferreira}, \citenamefont {Denicol},\ and\ \citenamefont {Noronha}}]{Rocha:2022fqz}%
  \BibitemOpen
  \bibfield  {author} {\bibinfo {author} {\bibfnamefont {Gabriel~S.}\ \bibnamefont {Rocha}}, \bibinfo {author} {\bibfnamefont {Maur\'\i{}cio~N.}\ \bibnamefont {Ferreira}}, \bibinfo {author} {\bibfnamefont {Gabriel~S.}\ \bibnamefont {Denicol}}, \ and\ \bibinfo {author} {\bibfnamefont {Jorge}\ \bibnamefont {Noronha}},\ }\bibfield  {title} {\enquote {\bibinfo {title} {{Transport coefficients of quasiparticle models within a new relaxation time approximation of the Boltzmann equation}},}\ }\href {\doibase 10.1103/PhysRevD.106.036022} {\bibfield  {journal} {\bibinfo  {journal} {Phys. Rev. D}\ }\textbf {\bibinfo {volume} {106}},\ \bibinfo {pages} {036022} (\bibinfo {year} {2022})},\ \Eprint {http://arxiv.org/abs/2203.15571} {arXiv:2203.15571 [nucl-th]} \BibitemShut {NoStop}%
\bibitem [{\citenamefont {Gavassino}(2024{\natexlab{a}})}]{Gavassino:2024pgl}%
  \BibitemOpen
  \bibfield  {author} {\bibinfo {author} {\bibfnamefont {Lorenzo}\ \bibnamefont {Gavassino}},\ }\bibfield  {title} {\enquote {\bibinfo {title} {{Infinite Order Hydrodynamics: an Analytical Example}},}\ }\href@noop {} {\  (\bibinfo {year} {2024}{\natexlab{a}})},\ \Eprint {http://arxiv.org/abs/2402.19343} {arXiv:2402.19343 [nucl-th]} \BibitemShut {NoStop}%
\bibitem [{\citenamefont {Gavassino}(2024{\natexlab{b}})}]{Gavassino:2024rck}%
  \BibitemOpen
  \bibfield  {author} {\bibinfo {author} {\bibfnamefont {Lorenzo}\ \bibnamefont {Gavassino}},\ }\bibfield  {title} {\enquote {\bibinfo {title} {{Gapless non-hydrodynamic modes in relativistic kinetic theory}},}\ }\href@noop {} {\  (\bibinfo {year} {2024}{\natexlab{b}})},\ \Eprint {http://arxiv.org/abs/2404.12327} {arXiv:2404.12327 [nucl-th]} \BibitemShut {NoStop}%
\bibitem [{\citenamefont {Rocha}\ \emph {et~al.}(2024{\natexlab{b}})\citenamefont {Rocha}, \citenamefont {Danhoni}, \citenamefont {Ingles}, \citenamefont {Denicol},\ and\ \citenamefont {Noronha}}]{Rocha:2024cge}%
  \BibitemOpen
  \bibfield  {author} {\bibinfo {author} {\bibfnamefont {Gabriel~S.}\ \bibnamefont {Rocha}}, \bibinfo {author} {\bibfnamefont {Isabella}\ \bibnamefont {Danhoni}}, \bibinfo {author} {\bibfnamefont {Kevin}\ \bibnamefont {Ingles}}, \bibinfo {author} {\bibfnamefont {Gabriel~S.}\ \bibnamefont {Denicol}}, \ and\ \bibinfo {author} {\bibfnamefont {Jorge}\ \bibnamefont {Noronha}},\ }\bibfield  {title} {\enquote {\bibinfo {title} {{Branch-cut in the shear-stress response function of massless $\lambda \varphi^4$ with Boltzmann statistics}},}\ }\href@noop {} {\  (\bibinfo {year} {2024}{\natexlab{b}})},\ \Eprint {http://arxiv.org/abs/2404.04679} {arXiv:2404.04679 [nucl-th]} \BibitemShut {NoStop}%
\bibitem [{\citenamefont {Fox}\ and\ \citenamefont {Uhlenbeck}(1970{\natexlab{b}})}]{doi:10.1063/1.1693183}%
  \BibitemOpen
  \bibfield  {author} {\bibinfo {author} {\bibfnamefont {Ronald~Forrest}\ \bibnamefont {Fox}}\ and\ \bibinfo {author} {\bibfnamefont {George~E.}\ \bibnamefont {Uhlenbeck}},\ }\bibfield  {title} {\enquote {\bibinfo {title} {Contributions to non‐equilibrium thermodynamics. i. theory of hydrodynamical fluctuations},}\ }\href {\doibase 10.1063/1.1693183} {\bibfield  {journal} {\bibinfo  {journal} {The Physics of Fluids}\ }\textbf {\bibinfo {volume} {13}},\ \bibinfo {pages} {1893--1902} (\bibinfo {year} {1970}{\natexlab{b}})},\ \Eprint {http://arxiv.org/abs/https://aip.scitation.org/doi/pdf/10.1063/1.1693183} {https://aip.scitation.org/doi/pdf/10.1063/1.1693183} \BibitemShut {NoStop}%
\bibitem [{\citenamefont {Gavassino}\ \emph {et~al.}(2023{\natexlab{a}})\citenamefont {Gavassino}, \citenamefont {Disconzi},\ and\ \citenamefont {Noronha}}]{Gavassino:2023odx}%
  \BibitemOpen
  \bibfield  {author} {\bibinfo {author} {\bibfnamefont {Lorenzo}\ \bibnamefont {Gavassino}}, \bibinfo {author} {\bibfnamefont {Marcelo~M.}\ \bibnamefont {Disconzi}}, \ and\ \bibinfo {author} {\bibfnamefont {Jorge}\ \bibnamefont {Noronha}},\ }\bibfield  {title} {\enquote {\bibinfo {title} {{Universality Classes of Relativistic Fluid Dynamics I: Foundations}},}\ }\href@noop {} {\  (\bibinfo {year} {2023}{\natexlab{a}})},\ \Eprint {http://arxiv.org/abs/2302.03478} {arXiv:2302.03478 [nucl-th]} \BibitemShut {NoStop}%
\bibitem [{\citenamefont {Gavassino}\ \emph {et~al.}(2023{\natexlab{b}})\citenamefont {Gavassino}, \citenamefont {Disconzi},\ and\ \citenamefont {Noronha}}]{Gavassino:2023qwl}%
  \BibitemOpen
  \bibfield  {author} {\bibinfo {author} {\bibfnamefont {L.}~\bibnamefont {Gavassino}}, \bibinfo {author} {\bibfnamefont {M.~M.}\ \bibnamefont {Disconzi}}, \ and\ \bibinfo {author} {\bibfnamefont {J.}~\bibnamefont {Noronha}},\ }\bibfield  {title} {\enquote {\bibinfo {title} {{Universality Classes of Relativistic Fluid Dynamics II: Applications}},}\ }\href@noop {} {\  (\bibinfo {year} {2023}{\natexlab{b}})},\ \Eprint {http://arxiv.org/abs/2302.05332} {arXiv:2302.05332 [nucl-th]} \BibitemShut {NoStop}%
\bibitem [{\citenamefont {Cercignani}\ and\ \citenamefont {Kremer}(2002)}]{cercignani:02relativistic}%
  \BibitemOpen
  \bibfield  {author} {\bibinfo {author} {\bibfnamefont {C.}~\bibnamefont {Cercignani}}\ and\ \bibinfo {author} {\bibfnamefont {G.~M.}\ \bibnamefont {Kremer}},\ }\href@noop {} {\emph {\bibinfo {title} {The Relativistic {B}oltzmann Equation: Theory and Applications}}}\ (\bibinfo  {publisher} {Springer},\ \bibinfo {year} {2002})\BibitemShut {NoStop}%
\bibitem [{\citenamefont {Chang}(1952)}]{chang1952propagation}%
  \BibitemOpen
  \bibfield  {author} {\bibinfo {author} {\bibfnamefont {C.S.~Wang}\ \bibnamefont {Chang}},\ }\href@noop {} {\emph {\bibinfo {title} {On the propagation of sound in monoatomic gases}}}\ (\bibinfo {year} {1952})\BibitemShut {NoStop}%
\bibitem [{{\relax DLMF}()}]{NIST:DLMF}%
  \BibitemOpen
  {\relax DLMF},\ \href {http://dlmf.nist.gov/} {\enquote {\bibinfo {title} {{\it NIST Digital Library of Mathematical Functions}},}\ }\bibinfo {howpublished} {http://dlmf.nist.gov/, Release 1.1.3 of 2021-09-15},\ \bibinfo {note} {f.~W.~J. Olver, A.~B. {Olde Daalhuis}, D.~W. Lozier, B.~I. Schneider, R.~F. Boisvert, C.~W. Clark, B.~R. Miller, B.~V. Saunders, H.~S. Cohl, and M.~A. McClain, eds.}\BibitemShut {Stop}%
\bibitem [{\citenamefont {Gradshteyn}\ and\ \citenamefont {Ryzhik}(2014)}]{gradshteyn2014table}%
  \BibitemOpen
  \bibfield  {author} {\bibinfo {author} {\bibfnamefont {I.S.}\ \bibnamefont {Gradshteyn}}\ and\ \bibinfo {author} {\bibfnamefont {I.M.}\ \bibnamefont {Ryzhik}},\ }\href@noop {} {\emph {\bibinfo {title} {Table of integrals, series, and products}}}\ (\bibinfo  {publisher} {Academic press},\ \bibinfo {year} {2014})\BibitemShut {NoStop}%
\bibitem [{\citenamefont {Denicol}\ and\ \citenamefont {Rischke}(2021)}]{Denicol:2021}%
  \BibitemOpen
  \bibfield  {author} {\bibinfo {author} {\bibfnamefont {G.~S.}\ \bibnamefont {Denicol}}\ and\ \bibinfo {author} {\bibfnamefont {D.~H.}\ \bibnamefont {Rischke}},\ }\href@noop {} {\emph {\bibinfo {title} {Microscopic Foundations of Relativistic Fluid Dynamics}}}\ (\bibinfo  {publisher} {Springer},\ \bibinfo {year} {2021})\BibitemShut {NoStop}%
\bibitem [{\citenamefont {Huang}\ and\ \citenamefont {McColl}(1997)}]{huang1997analytical}%
  \BibitemOpen
  \bibfield  {author} {\bibinfo {author} {\bibfnamefont {Y}~\bibnamefont {Huang}}\ and\ \bibinfo {author} {\bibfnamefont {WF}~\bibnamefont {McColl}},\ }\bibfield  {title} {\enquote {\bibinfo {title} {Analytical inversion of general tridiagonal matrices},}\ }\href@noop {} {\bibfield  {journal} {\bibinfo  {journal} {Journal of Physics A: Mathematical and General}\ }\textbf {\bibinfo {volume} {30}},\ \bibinfo {pages} {7919} (\bibinfo {year} {1997})}\BibitemShut {NoStop}%
\bibitem [{\citenamefont {Dunkel}\ and\ \citenamefont {H\"anggi}(2009)}]{Dunkel:2008ngc}%
  \BibitemOpen
  \bibfield  {author} {\bibinfo {author} {\bibfnamefont {J\"orn}\ \bibnamefont {Dunkel}}\ and\ \bibinfo {author} {\bibfnamefont {Peter}\ \bibnamefont {H\"anggi}},\ }\bibfield  {title} {\enquote {\bibinfo {title} {{Relativistic Brownian motion}},}\ }\href {\doibase 10.1016/j.physrep.2008.12.001} {\bibfield  {journal} {\bibinfo  {journal} {Phys. Rept.}\ }\textbf {\bibinfo {volume} {471}},\ \bibinfo {pages} {1--73} (\bibinfo {year} {2009})},\ \Eprint {http://arxiv.org/abs/0812.1996} {arXiv:0812.1996 [cond-mat.stat-mech]} \BibitemShut {NoStop}%
\bibitem [{\citenamefont {{Bancel}}\ and\ \citenamefont {{Choquet-Bruhat}}(1973)}]{1973CMaPh..33...83B}%
  \BibitemOpen
  \bibfield  {author} {\bibinfo {author} {\bibfnamefont {Daniel}\ \bibnamefont {{Bancel}}}\ and\ \bibinfo {author} {\bibfnamefont {Yvonne}\ \bibnamefont {{Choquet-Bruhat}}},\ }\bibfield  {title} {\enquote {\bibinfo {title} {{Existence, uniqueness, and local stability for the Einstein-Maxwell-Boltzman system}},}\ }\href {\doibase 10.1007/BF01645621} {\bibfield  {journal} {\bibinfo  {journal} {Communications in Mathematical Physics}\ }\textbf {\bibinfo {volume} {33}},\ \bibinfo {pages} {83--96} (\bibinfo {year} {1973})}\BibitemShut {NoStop}%
\bibitem [{\citenamefont {Kato}(1975)}]{Kato1975TheCP}%
  \BibitemOpen
  \bibfield  {author} {\bibinfo {author} {\bibfnamefont {Tosio}\ \bibnamefont {Kato}},\ }\bibfield  {title} {\enquote {\bibinfo {title} {The cauchy problem for quasi-linear symmetric hyperbolic systems},}\ }\href {https://api.semanticscholar.org/CorpusID:120479390} {\bibfield  {journal} {\bibinfo  {journal} {Archive for Rational Mechanics and Analysis}\ }\textbf {\bibinfo {volume} {58}},\ \bibinfo {pages} {181--205} (\bibinfo {year} {1975})}\BibitemShut {NoStop}%
\bibitem [{\citenamefont {Noronha}\ \emph {et~al.}(2024)\citenamefont {Noronha}, \citenamefont {Schenke}, \citenamefont {Shen},\ and\ \citenamefont {Zhao}}]{Noronha:2024dtq}%
  \BibitemOpen
  \bibfield  {author} {\bibinfo {author} {\bibfnamefont {Jorge}\ \bibnamefont {Noronha}}, \bibinfo {author} {\bibfnamefont {Bj\"orn}\ \bibnamefont {Schenke}}, \bibinfo {author} {\bibfnamefont {Chun}\ \bibnamefont {Shen}}, \ and\ \bibinfo {author} {\bibfnamefont {Wenbin}\ \bibnamefont {Zhao}},\ }\bibfield  {title} {\enquote {\bibinfo {title} {{Progress and Challenges in Small Systems}},}\ \ }(\bibinfo {year} {2024})\ \Eprint {http://arxiv.org/abs/2401.09208} {arXiv:2401.09208 [nucl-th]} \BibitemShut {NoStop}%
\bibitem [{\citenamefont {Mullins}\ \emph {et~al.}(2022)\citenamefont {Mullins}, \citenamefont {Denicol},\ and\ \citenamefont {Noronha}}]{Mullins:2022fbx}%
  \BibitemOpen
  \bibfield  {author} {\bibinfo {author} {\bibfnamefont {Nicki}\ \bibnamefont {Mullins}}, \bibinfo {author} {\bibfnamefont {Gabriel~S.}\ \bibnamefont {Denicol}}, \ and\ \bibinfo {author} {\bibfnamefont {Jorge}\ \bibnamefont {Noronha}},\ }\bibfield  {title} {\enquote {\bibinfo {title} {{Far-from-equilibrium kinetic dynamics of \ensuremath{\lambda}\ensuremath{\phi}4 theory in an expanding universe}},}\ }\href {\doibase 10.1103/PhysRevD.106.056024} {\bibfield  {journal} {\bibinfo  {journal} {Phys. Rev. D}\ }\textbf {\bibinfo {volume} {106}},\ \bibinfo {pages} {056024} (\bibinfo {year} {2022})},\ \Eprint {http://arxiv.org/abs/2207.07786} {arXiv:2207.07786 [hep-ph]} \BibitemShut {NoStop}%
\bibitem [{\citenamefont {de~Brito}\ \emph {et~al.}(2023)\citenamefont {de~Brito}, \citenamefont {Rocha},\ and\ \citenamefont {Denicol}}]{deBrito:2023vzv}%
  \BibitemOpen
  \bibfield  {author} {\bibinfo {author} {\bibfnamefont {Caio V.~P.}\ \bibnamefont {de~Brito}}, \bibinfo {author} {\bibfnamefont {Gabriel~S.}\ \bibnamefont {Rocha}}, \ and\ \bibinfo {author} {\bibfnamefont {Gabriel~S.}\ \bibnamefont {Denicol}},\ }\bibfield  {title} {\enquote {\bibinfo {title} {{Hydrodynamic theories for a system of weakly self-interacting classical ultra-relativistic scalar particles: causality and stability}},}\ }\href@noop {} {\  (\bibinfo {year} {2023})},\ \Eprint {http://arxiv.org/abs/2311.07272} {arXiv:2311.07272 [nucl-th]} \BibitemShut {NoStop}%
\bibitem [{\citenamefont {Florio}\ \emph {et~al.}(2022)\citenamefont {Florio}, \citenamefont {Grossi}, \citenamefont {Soloviev},\ and\ \citenamefont {Teaney}}]{Florio:2021jlx}%
  \BibitemOpen
  \bibfield  {author} {\bibinfo {author} {\bibfnamefont {Adrien}\ \bibnamefont {Florio}}, \bibinfo {author} {\bibfnamefont {Eduardo}\ \bibnamefont {Grossi}}, \bibinfo {author} {\bibfnamefont {Alexander}\ \bibnamefont {Soloviev}}, \ and\ \bibinfo {author} {\bibfnamefont {Derek}\ \bibnamefont {Teaney}},\ }\bibfield  {title} {\enquote {\bibinfo {title} {{Dynamics of the $O(4)$ critical point in QCD}},}\ }\href {\doibase 10.1103/PhysRevD.105.054512} {\bibfield  {journal} {\bibinfo  {journal} {Phys. Rev. D}\ }\textbf {\bibinfo {volume} {105}},\ \bibinfo {pages} {054512} (\bibinfo {year} {2022})},\ \Eprint {http://arxiv.org/abs/2111.03640} {arXiv:2111.03640 [hep-lat]} \BibitemShut {NoStop}%
\bibitem [{\citenamefont {Florio}\ \emph {et~al.}(2024)\citenamefont {Florio}, \citenamefont {Grossi},\ and\ \citenamefont {Teaney}}]{Florio:2023kmy}%
  \BibitemOpen
  \bibfield  {author} {\bibinfo {author} {\bibfnamefont {Adrien}\ \bibnamefont {Florio}}, \bibinfo {author} {\bibfnamefont {Eduardo}\ \bibnamefont {Grossi}}, \ and\ \bibinfo {author} {\bibfnamefont {Derek}\ \bibnamefont {Teaney}},\ }\bibfield  {title} {\enquote {\bibinfo {title} {{Dynamics of the O(4) critical point in QCD: Critical pions and diffusion in model G}},}\ }\href {\doibase 10.1103/PhysRevD.109.054037} {\bibfield  {journal} {\bibinfo  {journal} {Phys. Rev. D}\ }\textbf {\bibinfo {volume} {109}},\ \bibinfo {pages} {054037} (\bibinfo {year} {2024})},\ \Eprint {http://arxiv.org/abs/2306.06887} {arXiv:2306.06887 [hep-lat]} \BibitemShut {NoStop}%
\bibitem [{\citenamefont {Chattopadhyay}\ \emph {et~al.}(2023)\citenamefont {Chattopadhyay}, \citenamefont {Ott}, \citenamefont {Schaefer},\ and\ \citenamefont {Skokov}}]{Chattopadhyay:2023jfm}%
  \BibitemOpen
  \bibfield  {author} {\bibinfo {author} {\bibfnamefont {Chandrodoy}\ \bibnamefont {Chattopadhyay}}, \bibinfo {author} {\bibfnamefont {Josh}\ \bibnamefont {Ott}}, \bibinfo {author} {\bibfnamefont {Thomas}\ \bibnamefont {Schaefer}}, \ and\ \bibinfo {author} {\bibfnamefont {Vladimir}\ \bibnamefont {Skokov}},\ }\bibfield  {title} {\enquote {\bibinfo {title} {{Dynamic scaling of order parameter fluctuations in model B}},}\ }\href {\doibase 10.1103/PhysRevD.108.074004} {\bibfield  {journal} {\bibinfo  {journal} {Phys. Rev. D}\ }\textbf {\bibinfo {volume} {108}},\ \bibinfo {pages} {074004} (\bibinfo {year} {2023})},\ \Eprint {http://arxiv.org/abs/2304.07279} {arXiv:2304.07279 [nucl-th]} \BibitemShut {NoStop}%
\bibitem [{\citenamefont {Ba\c{s}ar}\ \emph {et~al.}(2024)\citenamefont {Ba\c{s}ar}, \citenamefont {Bhambure}, \citenamefont {Singh},\ and\ \citenamefont {Teaney}}]{Basar:2024qxd}%
  \BibitemOpen
  \bibfield  {author} {\bibinfo {author} {\bibfnamefont {G\"ok\c{c}e}\ \bibnamefont {Ba\c{s}ar}}, \bibinfo {author} {\bibfnamefont {Jay}\ \bibnamefont {Bhambure}}, \bibinfo {author} {\bibfnamefont {Rajeev}\ \bibnamefont {Singh}}, \ and\ \bibinfo {author} {\bibfnamefont {Derek}\ \bibnamefont {Teaney}},\ }\bibfield  {title} {\enquote {\bibinfo {title} {{The stochastic relativistic advection diffusion equation from the Metropolis algorithm}},}\ }\href@noop {} {\  (\bibinfo {year} {2024})},\ \Eprint {http://arxiv.org/abs/2403.04185} {arXiv:2403.04185 [nucl-th]} \BibitemShut {NoStop}%
\bibitem [{\citenamefont {de~Brito}\ and\ \citenamefont {Denicol}(2024)}]{deBrito:2024vhm}%
  \BibitemOpen
  \bibfield  {author} {\bibinfo {author} {\bibfnamefont {Caio V.~P.}\ \bibnamefont {de~Brito}}\ and\ \bibinfo {author} {\bibfnamefont {Gabriel~S.}\ \bibnamefont {Denicol}},\ }\bibfield  {title} {\enquote {\bibinfo {title} {{Method of moments for a relativistic single-component gas}},}\ }\href@noop {} {\  (\bibinfo {year} {2024})},\ \Eprint {http://arxiv.org/abs/2401.10098} {arXiv:2401.10098 [nucl-th]} \BibitemShut {NoStop}%
\end{thebibliography}%

\end{document}